\documentclass[11pt,a4paper]{article} 
\pdfoutput=1
\usepackage{jheppub}
\usepackage{multirow}
\usepackage{fix-cm}
\usepackage{epsfig}
\usepackage{amsmath}
\usepackage{color}
\usepackage{graphicx}
\usepackage{dcolumn}
\usepackage{bm}

\def\beq{\begin{equation}}
\def\eeq{\end{equation}}
\def\beqa{\begin{eqnarray}}
\def\eeqa{\end{eqnarray}}

\def \as {\relax\ifmmode\alpha_s\else{$\alpha_s${ }}\fi}

\newcommand{\lsim}{
\mathrel{\hbox{\rlap{\hbox{\lower4pt\hbox{$\sim$}}}\hbox{$<$}}}}
\newcommand{\gsim}{
\mathrel{\hbox{\rlap{\hbox{\lower4pt\hbox{$\sim$}}}\hbox{$>$}}}}

\titlepage


\title{From Webs to Polylogarithms}

\author{Einan Gardi}

\affiliation{Higgs Centre for Theoretical Physics, School of Physics and Astronomy, \\
The University of Edinburgh,  
Edinburgh EH9 3JZ, Scotland, UK}

\emailAdd{Einan.Gardi@ed.ac.uk}

\abstract{We compute a class of diagrams contributing to the multi-leg soft anomalous dimension through three loops, by renormalizing a product of semi-infinite non-lightlike Wilson lines in dimensional regularization. Using non-Abelian exponentiation we directly compute contributions to the exponent in terms of webs.
We develop a general strategy to compute webs with multiple gluon exchanges between Wilson lines in configuration space, and explore their analytic structure in terms of $\alpha_{ij}$, the exponential of the Minkowski cusp angle formed between the lines $i$ and $j$. We show that beyond the obvious inversion symmetry $\alpha_{ij}\to 1/\alpha_{ij}$, at the level of the symbol the result also admits a crossing symmetry $\alpha_{ij}\to -\alpha_{ij}$, relating spacelike and timelike kinematics, and hence argue that in this class of webs the symbol alphabet is restricted to $\alpha_{ij}$ and $1-\alpha_{ij}^2$. 
We carry out the calculation up to three gluons connecting four Wilson lines, finding that the contributions to the soft anomalous dimension are remarkably simple: they involve pure functions of uniform weight, which are written as a sum of products of polylogarithms, each depending on a single cusp angle. We conjecture that this type of factorization extends to all multiple-gluon-exchange contributions to the anomalous dimension.
}
\begin{document}
\begin{flushright}
Edinburgh 2013/28
\vspace*{-25pt}
\end{flushright}
\maketitle
\allowdisplaybreaks 

\section{Introduction\label{sec:introduction}}

Gauge theory scattering amplitudes are known to be infrared singular~\cite{Bloch:1937pw}. Understanding the detailed structure of these singularities is important for collider physics~\cite{Korchemsky:1985xj,Ivanov:1985np,Korchemsky:1987wg,Korchemsky:1988hd,Korchemsky:1988si,Korchemsky:1992xv,Gardi:2005yi,Becher:2005pd,Becher:2006qw,Korchemskaya:1994qp,Botts:1989kf,Contopanagos:1996nh,Kidonakis:1998nf,Kidonakis:1997gm,Kidonakis:2009zc,Catani:1998bh,Sterman:2002qn,Aybat:2006mz,Aybat:2006wq,Laenen:2008ux,Kyrieleis:2005dt,Sjodahl:2008fz,Seymour:2008xr,Kidonakis:2009ev,Mitov:2009sv,Becher:2009kw,Beneke:2009rj,Czakon:2009zw,Ferroglia:2009ep,Ferroglia:2009ii,Chiu:2009mg,Mitov:2010xw,Ferroglia:2010mi,Becher:2009cu,Gardi:2009qi,Becher:2009qa,Dixon:2008gr,Dixon:2009gx,Dixon:2009ur,Bierenbaum:2011gg,Laenen:2008gt}: it is a prior condition to cross-section calculations in which these singularities cancel in the sum of real and virtual corrections. 
In many circumstances these cancellations leave behind parametrically-large logarithms, which can be resummed to all order, see e.g.~refs.~\cite{Sterman:1986aj,Catani:1992ua,Contopanagos:1993yq,Catani:1989ne,Oderda:1999im,Kidonakis:1998bk,Laenen:1998qw,Laenen:2000ij,Bozzi:2007pn,Catani:1996yz,Gardi:2007ma,Andersen:2005mj,Gardi:2002bg,Gardi:2001ny,Cacciari:2002xb,Gardi:2002xm,Becher:2006mr,Becher:2007ty,Becher:2008cf,Ahrens:2008nc,Bozzi:2008bb,Kang:2008zzd,Idilbi:2009cc,Almeida:2009jt,Moch:2009mu,Moch:2009my,Mantry:2009qz,Beenakker:2010fw,Papaefstathiou:2010bw,Chien:2010kc,Beneke:2012eb,Banfi:2012jm,Becher:2013xia}.

The infrared structure of amplitudes is also interesting from a purely field-theoretic perspective. This is underlined by the possibility to explore all-order structures and form a bridge with strong-coupling methods. While progress on this front was largely restricted to (planar) ${\cal N}=4$ supersymmetric Yang-Mills theory, e.g.~\cite{Bern:2005iz,Basso:2007wd,Brandhuber:2007yx,Drummond:2007aua,Alday:2007hr,Gaiotto:2011dt,Correa:2012nk,Henn:2012qz,Henn:2012ia,Henn:2013wfa}, where amplitudes appear to be directly related to certain Wilson loops, infrared singularities are generally similar across all gauge theories and can be computed by considering products of Wilson-line operators. 

Long-distance singularities have been investigated since the early days of QCD, and have been a subject of continuous theoretical interest~\cite{Korchemsky:1985xj,Ivanov:1985np,Korchemsky:1987wg,Korchemsky:1988hd,Korchemsky:1988si,Korchemsky:1992xv,Gardi:2005yi,
Becher:2006qw,Becher:2005pd,Korchemskaya:1994qp,Botts:1989kf,Contopanagos:1996nh,Kidonakis:1998nf,Kidonakis:1997gm,Kidonakis:2009zc,
Sterman:2002qn,Aybat:2006mz,Aybat:2006wq,Laenen:2008ux,Kyrieleis:2005dt,Sjodahl:2008fz,Seymour:2008xr,Kidonakis:2009ev,Mitov:2009sv,
Becher:2009kw,Beneke:2009rj,Czakon:2009zw,Ferroglia:2009ep,Ferroglia:2009ii,Chiu:2009mg,Mitov:2010xw,Ferroglia:2010mi,Becher:2009cu,
Gardi:2009qi,Becher:2009qa,Dixon:2008gr,Dixon:2009gx,Dixon:2009ur,Bierenbaum:2011gg,Laenen:2008gt}. In recent years we have seen significant progress towards determining the singularities of multi-leg amplitudes with general kinematics beyond the planar limit and beyond one loop~\cite{Korchemsky:1987wg,Aybat:2006wq,Aybat:2006mz,Kidonakis:2009ev,Mitov:2009sv,Becher:2009kw,Beneke:2009rj,Czakon:2009zw,Ferroglia:2009ep,Ferroglia:2009ii,Chiu:2009mg,Mitov:2010xw,Ferroglia:2010mi,Bierenbaum:2011gg}.
Complete two-loop results are now available for the soft anomalous dimension in both the massless and massive cases. Furthermore, in the massless case all-order constraints have been deduced from factorization and rescaling symmetry, leading to a minimal ansatz in which the anomalous dimension takes the form of a sum over colour dipoles~\cite{Becher:2009cu,Gardi:2009qi,Becher:2009qa,Dixon:2009gx,Dixon:2009ur}. Corrections to the dipole formula may first appear at three loops, from diagrams involving four Wilson lines. 
Despite recent progress~\cite{Dixon:2008gr,Gardi:2009qi,Dixon:2009gx,Becher:2009cu,Becher:2009qa,Dixon:2009ur,Gardi:2009zv,Gehrmann:2010ue,Bret:2011xm,DelDuca:2011ae,Ahrens:2012qz,Naculich:2013xa,Caron-Huot:2013fea},  
it seems that general considerations fall short of excluding or fixing these corrections\footnote{A very interesting argument has been formulated recently~\cite{Caron-Huot:2013fea} based on the Regge limit, indicating that the dipole formula should be violated at four loops.}; further input from explicit calculations is needed. This motivates the present work, which is part of a larger project aimed at determining the anomalous dimension at three-loop order, and understanding its structure to all orders.

A separate line of investigation where much progress was made recently is soft-gluon exponentiation.
Denoting the soft-gluon amplitude, or equivalently the correlator of Wilson lines, by ${\cal S}$ (see eq.~(\ref{S_def_m}) below), the general observation is that the corresponding exponent~$w$, defined by ${\cal S}= \exp w$, has a simpler perturbative expansion with some remarkable properties. In particular, all colour factors appearing in the exponent correspond to connected graphs~\cite{Gardi:2013ita}.
There are two complementary approaches to exponentiation: one is based on evolution equations~\cite{Korchemsky:1985xj,Ivanov:1985np,Korchemsky:1987wg,Korchemsky:1988hd,Korchemsky:1988si,Korchemsky:1992xv,Gardi:2005yi,Becher:2006qw,Becher:2005pd,Korchemskaya:1994qp,Botts:1989kf,Contopanagos:1996nh,Kidonakis:1997gm,Kidonakis:1998nf,Sterman:2002qn,Aybat:2006mz,Laenen:2008ux,Kyrieleis:2005dt,Sjodahl:2008fz,Seymour:2008xr,Kidonakis:2009ev,Mitov:2009sv,Becher:2009kw,Beneke:2009rj,Czakon:2009zw,Ferroglia:2009ep,Ferroglia:2009ii,Chiu:2009mg,Mitov:2010xw,Ferroglia:2010mi,Becher:2009cu,Gardi:2009qi,Becher:2009qa,Dixon:2008gr,Dixon:2009gx,Dixon:2009ur}, which are ultimately a consequence of multiplicative renormalizability~\cite{Polyakov:1980ca,Arefeva:1980zd,Dotsenko:1979wb,Brandt:1981kf}, and the second is a diagrammatic approach, the direct computation of the exponent in terms of webs~\cite{Sterman:1981jc,Gatheral:1983cz,Frenkel:1984pz,Gardi:2010rn,Mitov:2010rp,Gardi:2011wa,Gardi:2011yz,Gardi:2013ita}. Following ref.~\cite{Gardi:2011yz}, we shall make simultaneous use of both approaches, aiming to compute the anomalous dimension in terms of webs.   

The diagrammatic approach to exponentiation is summarised by the non-Abelian exponentiation theorem. This theorem was established in the 1980's in the context of a Wilson loop~\cite{Gatheral:1983cz,Frenkel:1984pz}, and was recently generalised to the case of multiple Wilson lines in arbitrary representations of the colour group~\cite{Gardi:2010rn,Mitov:2010rp,Gardi:2011wa,Gardi:2011yz,Gardi:2013ita}. In contrast to the Abelian case, where only connected diagrams contribute to the exponent, in a non-Abelian theory certain non-connected diagrams\footnote{In this context `connected' and `non-connected' refers to the diagram after removing the Wilson lines.} contribute as well. The non-Abelian exponentiation theorem states that such a diagram $D$ contributes to the exponent with a modified, Exponentiated Colour Factor (ECF) $\widetilde{C}(D)$, which corresponds to a connected graph~\cite{Gardi:2013ita}.

In the case of multiple Wilson lines it is useful to define webs as sets of diagrams which contribute to the exponent together, rather than as individual diagrams. These sets are formed by taking all possible permutations of the order of gluon attachments to the Wilson lines. 
Each such set of diagrams constitutes a single web $W$. A web will be denoted by $W_{(n_1,\ldots,n_L)}$ or simply $n_1 -  n_2 - \ldots -  n_L$, where $n_i >0$ is the number of gluon attachments on line $i$, and it is assumed that there are $L$ lines in total.
The contribution of these diagrams to the exponent $w$ takes the form~\cite{Gardi:2010rn}:
\begin{equation}
W=\sum_{D}{\cal F}(D)\widetilde{C}(D)=\sum_{D,D'}{\cal F}(D)R_{DD'}C(D').
\label{Wform}
\end{equation}
Here $\{{\cal F}(D)\}$ and $\{C(D)\}$ are, respectively, the sets of kinematic functions and colour factors associated with all diagrams $D\in W$, diagrams which are related to each other by permuting the order of gluon emissions, while $R_{DD'}$ is a matrix of rational numbers called the \emph{web mixing matrix}. Thus each web has an associated web mixing matrix which dictate how colour and kinematic information is entangled. These matrices were further studied in refs.~\cite{Gardi:2011wa,Gardi:2011yz}, where it was shown that web mixing matrices are idempotent, $R^2=R$, namely they act as projection operators, selecting particular linear combinations of colour factors,  those corresponding to unit eigenvalues of~$R$, to appear in the exponent. It was then proven~\cite{Gardi:2013ita} that these combinations always correspond to connected graphs.
\begin{figure}[htb]
\begin{center}
\scalebox{1.0}{\includegraphics{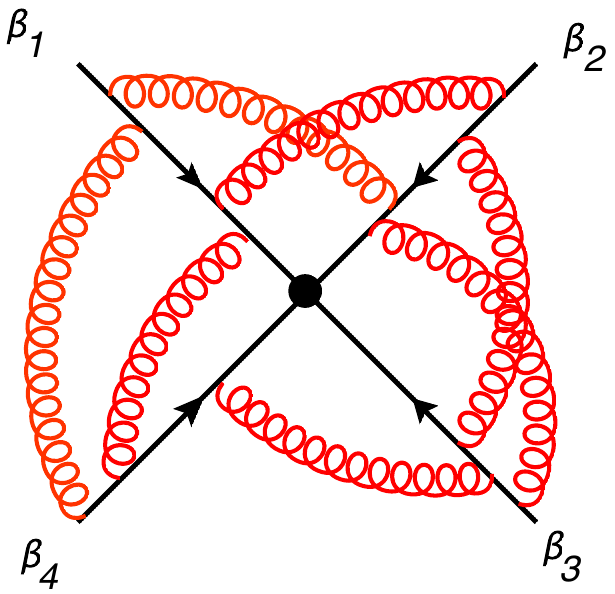}}
\caption{An example multiple-gluon-exchange diagram connecting four semi-infinite Wilson lines at seven loops. The 4-velocities associated with the directions of the Wilson lines in Minkowki space are indicated by~$\beta_i$. The lines all meet at the origin, where there is a local effective vertex representing the hard interaction.  }
\label{multi}
\end{center}
\end{figure}

The mixing matrix can also be viewed as acting on the kinematic factors, generating particular linear combinations of  $\{{\cal F}(D)\}$ in which certain subdivergences cancel, as dictated by the renormalisation properties of the vertex at which the Wilson lines meet~\cite{Gardi:2011yz}. 
It was also shown that the contents of web mixing matrices can be obtained purely from combinatorial reasoning~\cite{Gardi:2011wa}. This has been further used in refs.~\cite{Dukes:2013wa,Dukes:2013gea} to establish a relation with partially ordered sets, and deduce all-order solutions for certain classes of webs.

\begin{figure}[htb]
\begin{center}
\scalebox{1.2}{\includegraphics{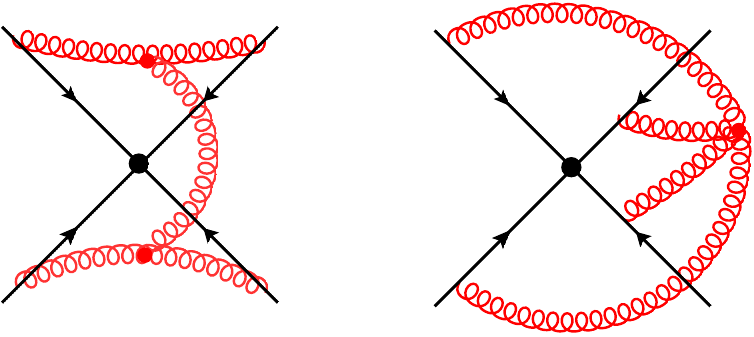}}
\caption{Connected three-loop diagrams with 3- or 4-gluon vertices spanning four Wilson lines.}
\label{Connected_four}
\end{center}
\end{figure}

\begin{figure}[htb]
\begin{center}
\scalebox{1.2}{\includegraphics{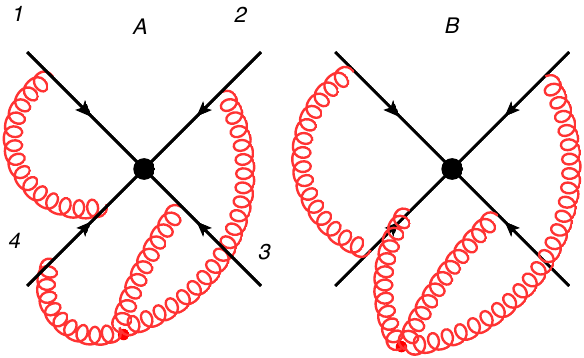}}
\caption{The 1-1-1-2 web. Each of the two diagrams in this web ($A$ and $B$) contains two connected subdiagrams, one of which has a three-gluon vertex.}
\label{two_pieces}
\end{center}
\end{figure}

Our interest in the present paper is to develop techniques to evaluate the integrals associated with webs connecting several non-lightlike Wilson lines and determine the contributions of these webs to the angle-dependent soft anomalous dimension. We consider here webs involving multiple-gluon-exchange diagrams with no three- or four-gluon vertices. An example of a diagram in this class is shown in figure~\ref{multi}.
Within this general class we further focus on these webs that connect the maximal number of Wilson lines at any given order, that is three Wilson lines\footnote{The two-line case is the familiar angle-dependent cusp anomalous dimension, which was computed at two-loops in refs.~\cite{Korchemsky:1987wg,Kidonakis:2009zc}.} at two loops, as in figure~\ref{2loopfig}, where we will reproduce the results of refs.~\cite{Ferroglia:2009ii,Mitov:2009sv,Mitov:2010xw}, and four Wilson lines at three loops, 
as in figures~\ref{3lfour} and~\ref{1113}. 
We emphasise at the outset that evaluating the full soft anomalous dimension at three loops is beyond the scope of the present paper. In particular the computation of webs involving three- or four-gluon vertices, as shown in figures~\ref{Connected_four} and \ref{two_pieces}, requires different techniques; this work is being carried out in parallel and the results will be published separately.\\

In the final part of the introduction we briefly review how correlators of Wilson-lines arise, define the soft anomalous dimension and outline the general strategy for computing it. This closely follows the approach of  ref.~\cite{Gardi:2011yz}  (see also refs.~\cite{Mitov:2010rp,Ferroglia:2009ii}) where detailed derivations can be found.
Our approach uses the very powerful observation~\cite{Korchemsky:1987wg}, that soft singularities in amplitudes can be deduced from the ultraviolet singularities of a product of semi-infinite Wilson lines. Wilson-line operators arise upon factorizing an amplitude into soft and hard modes:
\begin{equation}
\label{soft_hard_fact}
{\cal M}^{I}(p_i,\epsilon_{\rm IR})={\cal S}^{IJ}_{\rm ren.}(\gamma_{ij},\epsilon_{\rm IR}) \, {\cal H}_J(p_i)
\end{equation}
where the hard function ${\cal H}_{J}$, similarly to the amplitude itself, is a vector in colour space, while the soft factor, ${\cal S}^{IJ}$, for which we give an operator definition below, is a matrix in this space (we denoted the collection of colour indices schematically by $I$ and $J$). To identify the soft modes one introduces Wilson lines as follows: a field $\Psi(z)$ in the time-ordered product in ${\cal M}$, corresponding to the annihilation of an outgoing quark with momentum $q$, is replaced by
\begin{align}
\label{field_replcaement}
\begin{split}
\Psi(z)&\,\to\, 
\overline{\cal P} \exp\left(-{\rm i}g_s\int_{z}^{\infty}dx^{\mu}A_{\mu}(x)\right)\, \psi(z)
\end{split}
\end{align}
where the soft gauge field interacts with the Wilson line and decouples from~$\psi(z)$. The integration path in the exponent is along a ray in the direction of quark, and the anti-path-ordering operator $\overline{\cal P}$ places colour fields that are closer to the lower limit of integration to the right; in fig.~\ref{multi} this is indicated by the direction of the arrow along the Wilson lines\footnote{We follow the conventions used in Ref.~\cite{Gardi:2005yi}; Appendix A there summarises some useful properties 
of Wilson lines.}.
In this paper we follow the convention where all momenta $p_i$ are incoming, so the outgoing quark in (\ref{field_replcaement}) has 4-momentum $p_i=-q$.
Using the rescaling symmetry of the Wilson line we can use the 4-velocity $\beta_i$, which is proportional to $p_i$, and parametrize the path as $x^{\mu}=z^{\mu}-t\beta_i^{\mu}$, obtaining\footnote{Gauge fields are still ordered such that the rightmost ones are those closest to the point $z$; the Wilson line is path ordered with regards to $\beta_i$.}:
\begin{align}
\label{Phi_def1}
\begin{split}
\Phi_{\beta_i}(\infty, z)
=
{\cal P} \exp\left({\rm i}g_s \beta_i^{\mu}\int_{0}^{\infty}dt A_{\mu}(z-t\beta_i)\right)\,,
\end{split}
\end{align}
and the replacement of eq.~(\ref{field_replcaement}) takes the form
$\Psi(z)\,\to\,\Phi_{\beta_i}(\infty, z) \, \psi(z)$.
Similar replacements with the same formula for the Wilson line (\ref{Phi_def1}) apply to other partons where the relevant representation of the gauge field is dictated in each case by that of the corresponding parton~\cite{Catani:1996jh,Catani:1996vz}, $A_{\mu}=A_{\mu}^a T_i^a$, where $(T_i^a)_{\alpha\beta} = t^a_{\alpha\beta}$ 
for a final-state quark (or an initial-state anti-quark), 
 $(T_i^a)_{\alpha\beta} = -t^a_{\beta\alpha}$ for an initial-state
quark (or a final-state anti-quark), and for gluons $T$ is in the Adjoint representation with $(T_i^a)_{bc} = -{\rm i}f_{abc}$.

Soft gauge fields correlate the Wilson lines associated with all partons, while they decouple from the remaining hard components. The latter ($\psi(z)$ in eq.~(\ref{field_replcaement})), become part of the hard coefficient function ${\cal H}_J$ in eq.~(\ref{soft_hard_fact}) while the soft function is defined by
\begin{align}
\label{S_def}
{\cal S}\left(\gamma_{ij},\epsilon_{\rm IR}\right)
&\equiv\left<0\left|\Phi_{\beta_1}\otimes\Phi_{\beta_2}\otimes\ldots\otimes\Phi_{\beta_L} \right|0\right>\,,
\end{align}
where $\Phi_{\beta_i}\equiv \Phi_{\beta_i}(\infty,0)$.
${\cal S}\left(\gamma_{ij},\epsilon_{\rm IR}\right)$, as hinted by the notation, captures all infrared singularities.
Note that we ignored here collinear singularities which occur for massless partons. For the purpose of the present paper it is convenient to assume that the Wilson-line velocities are non-lightlike ($\beta_i^2\neq 0$). This will also 
guarantee that the operator in eq.~(\ref{S_def}) is multiplicatively renormalizable (see below). The lightlike limit may of course be taken at the end.

Owing to the rescaling symmetry mentioned above the dependence of the soft function on the kinematics is only through the Minkowski-space angles between the Wilson lines $i$ and $j$. We define
\begin{equation}
\label{gamma_ij}
\gamma_{ij}\equiv\frac{2\beta_i\cdot\beta_j+{\rm i}0}{\sqrt{\beta_i^2-{\rm i0}}\sqrt{\beta_j^2-{\rm i}0}}
=\frac{2p_i\cdot p_j+{\rm i}0}{\sqrt{m_i^2-{\rm i0}}\sqrt{m_j^2-{\rm i}0}}\,,
\end{equation} 
where we indicated the prescription for each Lorentz invariant. In the second expression we restored the dimensionful kinematic variables: $p_{i}$ is the momentum of a heavy parton with squared mass $p_{i}^2=m_{i}^2$.
Related kinematic variables will be defined below (see eq.~(\ref{eq:F1m1})).

In dimensional regularization $S$ presents a rather remarkable relation between the ultraviolet and the infrared singularity structure~\cite{Korchemsky:1987wg} owing to the fact that scaleless integrals vanish identically. Instead of computing infrared singularities we will compute the renormalization of the vertex formed by the Wilson lines in eq.~(\ref{S_def}). This operator renormalizes multiplicatively~\cite{Polyakov:1980ca,Arefeva:1980zd,Dotsenko:1979wb,Brandt:1981kf}:
\begin{equation}
S_{\rm ren.}(\epsilon_{\rm IR},\mu)=S_{{\rm UV}+{\rm IR}}\, Z(\epsilon_{\rm UV},\mu)\,,
\end{equation}
where for lightness of notation we omitted here the dependence on the kinematic variables.
 In the absence of any cutoff all radiative corrections vanish and $S_{{\rm UV}+{\rm IR}}=1$, which implies 
\begin{equation}
\label{IR_UV}
S_{\rm ren.}(\epsilon_{\rm IR},\mu)= Z(\epsilon_{\rm UV},\mu).
\end{equation}
Thus our task is to compute the renormalization factor $Z$. To this end we consider the correlator of eq.~(\ref{S_def}) with an infrared cutoff:
\begin{align}
\label{S_def_m}
{\cal S}\left(\gamma_{ij},\alpha_s(\mu,\epsilon), \epsilon, m\right)
&\equiv\left<0\left|\Phi_{\beta_1}^{(m)}\otimes\Phi_{\beta_2}^{(m)}\otimes\ldots\otimes\Phi_{\beta_L}^{(m)} \right|0\right>\,,
\end{align}
where $m$ is a mass scale associated with the (exponential) damping  we use for the coupling of the gauge field to the Wilson lines as defined in eq.~(\ref{FRmod}) below. Dimensional regularization with $d=4-2\epsilon$ is used in the ultraviolet ($\epsilon>0$). After renormalizing the strong coupling, all remaining ultraviolet singularities should be associated with the multi-Wilson-line vertex $Z$ as follows:
\begin{equation}
\label{S_reg_ren_UV_and_IR}
{\cal S}  \left(\gamma_{ij},\alpha_s(\mu_R^2),\epsilon, m\right)\,
Z\left(\gamma_{ij},\alpha_s(\mu_R^2),\epsilon,\mu\right)\,=
\,
{\cal S}_{\rm ren.}\left(\gamma_{ij},\alpha_s(\mu_R^2),\mu,m\right)
,
\end{equation}
where ${\cal S}_{\rm ren.}$ is finite for $\epsilon\to 0$ and $\mu$ is a renormalization scale introduced in defining $Z$ (this scale is in principle distinct from the renormalization scale of the strong coupling, $\mu_R$). The $Z$ factor does not depend on the infrared regulator. 
The soft anomalous dimension~$\Gamma$, which is itself finite, is defined through
\begin{equation}
\frac{dZ}{d\ln \mu}=-Z\Gamma\,.
\end{equation}
As one expects from this evolution equation, the perturbative expansion of $Z$ takes a particularly simple form upon exponentiation.
The exponent of the $Z$ factor may be written in terms of the coefficients of the anomalous dimension 
$\Gamma=\sum_{n=1}^{\infty} \alpha_s^n \,\Gamma^{(n)}$ (and those of the $\beta$ function, $b_n$) yielding:   
\begin{align}
\label{Z_exp_expanded_1l_multiparton}
\begin{split}
Z=&\,\exp\left\{
\,\frac{1}{2\epsilon}\,\Gamma^{(1)}\,\alpha_s
+\,\left(\frac{1}{4\epsilon}\,\Gamma^{(2)}-\frac{b_0}{4\epsilon^2}\,\Gamma^{(1)}\right)\,\alpha_s^2
\right.\\&
\left.
+\,\left(\frac{1}{6\epsilon}\,\Gamma^{(3)}
+\frac{1}{48\epsilon^2}\left[\Gamma^{(1)},\Gamma^{(2)}\right]-\frac{1}{6\epsilon^2}
\left(b_0\Gamma^{(2)}+b_1\Gamma^{(1)}\right)+\frac{b_0^2}{6\epsilon^3}\Gamma^{(1)}\right)\,\alpha_s^3
\,+\,{\cal O}(\alpha_s^4)\,
 \right\}\,.
\end{split}
\end{align}
Note that in contrast to the Wilson loop case (or, equivalently, the case of two Wilson lines with a colour singlet operator at the cusp), $\Gamma^{(n)}$, similarly to $Z$, are matrices and thus even in a conformal theory where $b_n=0$ the expression for the exponent involves higher-order poles in $\epsilon$ governed by commutators.  
As discussed above, also the Wilson-line correlator ${\cal S}$ itself is most naturally expressed as an exponential
\begin{equation}
\label{w-def}
{\cal S}\left(\gamma_{ij},\alpha_s(\mu,\epsilon), \epsilon, m\right)=\exp\left\{w\right\}=
\exp\left\{\sum_{n=1}^{\infty}w^{(n)} \,\alpha_s^n\right\}
=
\exp\left\{\sum_{n,k \, }w^{(n,k)}\, \alpha_s^n\,\epsilon^{n}\right\}\,.
\end{equation}
where the coefficients $w^{(n)}$ and $w^{(n,k)}$ collect all non-renormalized webs at a given order in $\alpha_s$, i.e.
\begin{equation}
w^{(n)}\,\alpha_s^n\,=\,\sum_{n_1,\ldots,n_L} W_{(n_1,\ldots,n_L)}^{(n)}\,.
\label{wndef}
\end{equation}

Our strategy will be to make use of the non-Abelian exponentiation theorem and compute the exponent $w$ as a sum of webs according to eq.~(\ref{wndef}).  Having $w^{(n,k)}$ at hand, we will determine the coefficients of the soft anomalous dimension using the following relations~\cite{Gardi:2011yz}:
\begin{subequations}
\label{Gamma_n}
\begin{align}
\Gamma^{(1)}&=-2w^{(1,-1)}\\
\Gamma^{(2)}&=-4w^{(2,-1)}-2\left[w^{(1,-1)},w^{(1,0)}\right]\\
\begin{split}
\label{Gamma_3}
\Gamma^{(3)}&=-6w^{(3,-1)}+\frac32b_0\left[w^{(1,-1)},w^{(1,1)}\right]
+3\left[w^{(1,0)},w^{(2,-1)}\right]
+3\left[w^{(2,0)},w^{(1,-1)}\right]
\\&
+\left[w^{(1,0)},\left[w^{(1,-1)},w^{(1,0)}\right]\right]
-\left[w^{(1,-1)},\left[w^{(1,-1)},w^{(1,1)}\right]\right]\,.
\end{split}
\end{align}
\end{subequations}
We emphasise that while individual web coefficients $w^{(n,k)}$ may depend on the infrared regularization, $\Gamma^{(n)}$ are strictly regulator independent. 

In the following sections we explicitly evaluate the integrals corresponding to multiple-gluon-exchange diagrams in the Feynman gauge and determine their contributions to $w^{(n,k)}$ through three loops ($n=3$).  
At each order $n$ the soft anomalous dimension coefficient $\Gamma^{(n)}$ 
involves a sum over terms depending on up to $(n+1)$ Wilson lines, namely
\begin{equation}
\Gamma^{(n)}=\sum_{k=2}^{n+1}\Gamma_k^{(n)}\,,
\end{equation}
where $\Gamma^{(n)}_2$ is the familiar cusp anomalous dimension.
As mentioned above we focus here on these gluon-exchange webs that connect the maximal number of Wilson lines at each order.
Individual webs will be identified by the number of gluon attachments to each of the Wilson lines. We will compute the 1-1 web at one loop (figure~\ref{1loopfig}), which contributes to the leading-order cusp anomalous dimension $\Gamma^{(1)}_2$, the 1-2-1 web at two-loops (upper diagrams in figure~\ref{2loopfig}), which contributes to  $\Gamma^{(2)}_3$, and the 1-2-2-1 and 1-1-1-3 webs at three loops (figures~\ref{3lfour} and~\ref{1113}, respectively) both contributing to  $\Gamma^{(3)}_4$.

The entire calculation is done in configuration space. In case of gluon-exchange diagrams this has an obvious advantage over a momentum-space calculation: one obtains parameter integrals along the Wilson lines instead of $d$-dimensional loop integrals. 
Besides computing individual integrals we develop in this paper a general strategy for evaluating and organising the contributions of multiple-gluon-exchange diagrams.
We will see that it is straightforward to combine integrals corresponding to different diagrams in a given web. We will further see that it is natural to combine the web integrals with the relevant commutators of lower-order webs corresponding to their subdiagrams according to the combinations that contribute to the anomalous dimension coefficients, eqs.~(\ref{Gamma_n}). These combinations, which we refer to as \emph{subtracted webs}, are singled out by the fact that they are associated with a single pole, where the effect of subdivergences associated with the hard-interaction vertex is removed by including the commutators~\cite{Gardi:2011yz}. We will see that certain symmetry properties are only
recovered at the level of these subtracted webs, which provides an important consistency check of the computation. 

The paper is organised as follows:
in section \ref{sec:1_loop} we recall the Feynman rules and go through the calculation of a single gluon exchange between two semi-infinite Wilson lines to all orders in $\epsilon$. In section \ref{sec:2_loop} we compute the contributions to the two-loop anomalous dimension associated with three Wilson lines from the 1-2-1 web ($w^{(2,-1)}$) and then proceed to evaluate the corresponding finite term $w^{(2,0)}$ which is relevant for the three-loop anomalous dimension. In section \ref{sec:functions} we perform a general analysis of the integrals corresponding to multiple-gluon-exchange diagrams and determine a basis of functions by which one may express the contributions to the anomalous dimension from this class of webs. This also prepares the grounds for the more complex three-loop computations.  In~section~\ref{sec:3_loop} we summarise the results for the contributions of the three-loop webs 1-2-2-1 and 1-1-1-3 to the anomalous dimension, where the details are relegated to three appendices, appendix \ref{sec:summary_of_integrals} where we compile the results for all relevant webs, appendix~\ref{sec:3_loop_calc} where the three-loop web diagrams are expressed in a canonical form as parameter integrals, and appendix  
\ref{sec:combining} where the latter are combined with the corresponding commutators of their subdiagrams, obtaining subtracted webs, and these are evaluated explicitly in terms of the basis of functions of section~\ref{sec:functions}. In section~\ref{sec:conclusions} we discuss our results and conclude.

\section{One-loop calculation and the choice of kinematic variables\label{sec:1_loop}}

In this section we recall the one-loop calculation. Although the calculation is essentially the same as that of the cusp anomalous dimension~\cite{Korchemsky:1987wg,Kidonakis:2009zc,Henn:2012qz,Henn:2012ia}, we present it in some detail in order to introduce the infrared regulator, the kinematic variables and some further notation. We will also explain how the integrals are performed, preparing the grounds for similar higher-loop computations.

Throughout this paper we use an infrared regulator which exponentially suppresses the coupling to the Wilson lines at large distances from the vertex as follows~\cite{Gardi:2011wa}:
\begin{equation}
(ig_s)\,\beta_i^\mu\int_0^\infty d\lambda \,\left(\cdots\right) \qquad\,\,\ \longrightarrow\qquad\,\,
(ig_s)\,\beta_i^\mu\int_0^\infty d\lambda\,e^{-{\rm i}m\lambda\sqrt{\beta_i^2-{\rm i}0}} \,\left(\cdots\right)
\label{FRmod}
\end{equation}
where we indicated the relevant prescription\footnote{In the calculation we will eventually rescale the integration variable to absorb the factor $\sqrt{\beta_i^2-{\rm i}0}$. In this case the prescription will be carried by the regulator $m\to m-{\rm i}0$.} for $\beta_i^2\to \beta_i^2-i0$, the same prescription as in the denominator of eq.~(\ref{gamma_ij}). This is in keeping with the fact that this variable takes the role of a squared mass, rather than a squared off-shell momentum.
Note that thanks to this prescription the convergence of the $\lambda$ integral at infinity is guaranteed\footnote{To verify this note that the analytic continuation reads $\beta_i^2-{\rm i}0=|\beta_i^2| \exp(-{\rm i}\theta)$ where $\theta>0$. One has $\theta=\varepsilon\to 0^+$ for time-like Wilson lines, while $\theta=\pi$ for space-like lines. In the former case the exponent in eq.~(\ref{FRmod})  is
$-{\rm i}m\lambda\sqrt{|\beta_i^2|}(1-{\rm i}\varepsilon/2)$
and in the latter it is
$-{\rm i}m\lambda\sqrt{|\beta_i^2|}(-{\rm i})$, so in both cases one finds exponential suppression in the $\lambda\to \infty$ limit.} for both space-like and time-like Wilson lines.
\begin{figure}[htb]
\begin{center}
\scalebox{1.1}{\includegraphics{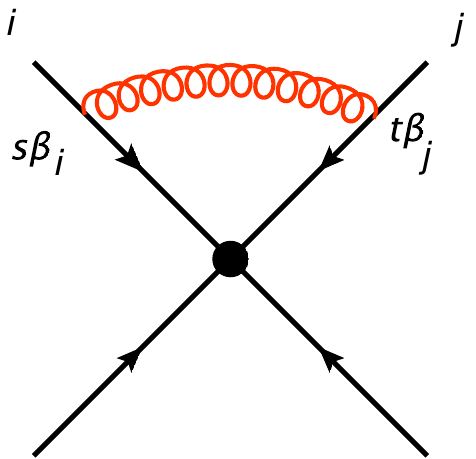}}
\caption{One loop web, where the gluon is emitted between partons $i$ and $j$,
whose kinematic part is given by eq.~(\ref{eq:Fijoneloop2}).}
\label{1loopfig}
\end{center}
\end{figure}

Let us consider the calculation of the one-loop diagram in figure~\ref{1loopfig} with this regulator in place. We follow the conventions described in the introduction, where the Wilson lines are defined in eq.~(\ref{Phi_def1}) and the velocities are incoming.
We denote the one-loop diagram with a gluon exchanged between legs $i$ and $j$ by 
\begin{equation}
\label{one_loop_diag}
W_{(i,j)}^{(1)}\,=\,T_{i}\cdot T_{j}\,\,{\cal F}^{(1)}_{ij}(\gamma_{ij},\mu^2/m^2,\epsilon)\,,
\end{equation}
where we have written the colour factor in terms of the generators of the two lines, defining $T_{i}\cdot T_{j} =\sum_a T_i^{(a)} T_j^{(a)}$, where $T_i$ and $T_j$ may belong to different representations.
The Feynman rule for the configuration-space gluon propagator in the Feynman gauge in $d=4-2\epsilon$ dimensions is 
({\it cf.} appendix A in Ref.~\cite{Korchemsky:1992xv}) 
\begin{align}
\label{propdef}
\begin{split}
&D_{\mu\nu}(x)=\mu^{2\epsilon}\, \int\frac{d^dk}{(2\pi)^d}{\rm e}^{-{\rm i}kx} \frac{-{\rm i}\,g_{\mu\nu}}{k^2+{\rm i}\varepsilon}=-{\cal N}\,g_{\mu\nu}\,
\left(-x^2+{\rm i}\varepsilon\right)^{\epsilon-1}\,,
\\&\text{where}\qquad
{\cal N}\equiv
\mu^{2\epsilon}\, \frac{\Gamma(1-\epsilon)}{4\pi^{2-\epsilon}}\,,\qquad \varepsilon>0\,.
\end{split}
\end{align}
Note that we included the dimensional regularization scale in the propagator, making it dimensionless. 
Parametrizing the positions along the two Wilson lines $i$ and $j$ by $s$ and $t$ respectively, the one-loop calculation reduces to a parameter integral over the gluon propagator 
(\ref{propdef}) between the two points $s\beta_i$ and $t\beta_j$,
with exponential damping associated with the infrared-regulated interaction vertices (\ref{FRmod}):
\begin{align}
  \label{eq:Fijoneloop1}
  \begin{split}
  {\cal F}^{(1)}_{ij}\left(\gamma_{ij},\frac{\mu^2}{m^2},\epsilon\right)
  &= g_s^2\,{\cal N}\,\beta_i\cdot\beta_j\!
\int_0^{\infty}\!\!\! ds\! \int_0^{\infty} \!\!\! dt
  \, \Big(-(s\beta_i-t\beta_j)^2+{\rm i}\varepsilon\Big)^{\epsilon-1}
\!{\rm e}^{-{\rm i}m s\sqrt{\beta_i^2-{\rm i}0}-{\rm i}mt\sqrt{\beta_j^2-{\rm i}0}
}
\\
&= \frac{g_s^2}{2}\,{\cal N}\,\gamma_{ij}\,
\int_0^{\infty} d\sigma \int_0^{\infty}  d\tau
  \Big(-\sigma^2-\tau^2+\gamma_{ij} \sigma \tau+{\rm i}\varepsilon\Big)^{\epsilon-1}
\,{\rm e}^{-{\rm i}(\sigma+\tau)(m-{\rm i}0)}
\\
 &=
\kappa 
 \,\Gamma(2\epsilon) \,\gamma_{ij}\int_0^1dx \, P(x,\gamma_{ij})\,,
  \end{split}
\end{align}
where in the second line we rescaled\footnote{Note that for \emph{spacelike} Wilson lines, $\sqrt{\beta_i^2}=-{\rm i} \sqrt{|\beta_i^2|}$, one may simply define $\sigma=s\sqrt{|\beta_i^2|}$ and $\tau=t\sqrt{|\beta_j^2|}$, and then identify $(2\beta_i\cdot\beta_j+{\rm i}\varepsilon)/\sqrt{|\beta_i^2||\beta_j^2|}=-\gamma_{ij}$, consistently with the final result in eq.~(\ref{eq:Fijoneloop1}).} the integration variables such that 
$\sigma=s\sqrt{\beta_i^2-{\rm i}0}$ and $\tau=t\sqrt{\beta_i^2-{\rm i}0}$ and in the third line we defined 
\begin{equation}
\label{kappa1}
\kappa\equiv -\left(\frac{\mu^2}{m^2}\right)^{\epsilon} \frac{g_s^2}{2}\,\frac{\Gamma(1-\epsilon)}{4\pi^{2-\epsilon}}\,,
\end{equation} 
and integrated over the distance scale to obtain an ultraviolet divergence. In doing so we introduced the following integration variables:
\begin{equation}
\label{change_var}
\lambda=\sigma+\tau\qquad \text{and}\qquad x=\frac{\sigma}{\sigma+\tau}\,,
\end{equation}
where the semi-infinite parameter $\lambda$ captures the overall distance of the gluon from the cusp, while $x\in[0,1]$ represents the emission angle, where the two collinear limits correspond to the boundaries $x=0$ and $x=1$.  The $\lambda$ dependence scales out of the propagator and the $\lambda$ integral generates an ultraviolet pole $\Gamma(2\epsilon)\simeq \frac{1}{2\epsilon}$. This leaves only $x$ dependence in the propagator-related function,
\begin{equation}
P(x,\gamma_{ij})\equiv\Big[x^2+(1-x)^2-x(1-x)\gamma_{ij}-{\rm i}\varepsilon\Big]^{\epsilon-1}\,= \Big[1-4x(1-x)\left(\frac12+\frac{\gamma_{ij}}{4}+{\rm i}\varepsilon\right)\Big]^{\epsilon-1}\,.
\label{Pdef}
\end{equation} 

In order to perform the final integration over $x$ in eq.~(\ref{eq:Fijoneloop1}) it is convenient to express the integral in terms of $\alpha_{ij}$, defined by 
\begin{equation}
\label{gamma_alpha_relation}
-\gamma_{ij}=\alpha_{ij}+\frac{1}{\alpha_{ij}}\,.
\end{equation}
Note that this definition inherently introduces an inversion symmetry, $\alpha_{ij}\to {1}/\alpha_{ij}$, which implies that one can either choose $\left|\alpha_{ij}\right|\leq 1$ or $\left|\alpha_{ij}\right|\geq 1$, as both are describing the same kinematics. Throughout this paper we shall assume the former, so all relevant values of $\alpha_{ij}$ are within the unit circle.

Let us now consider the coefficient of the $1/\epsilon$ pole in eq.~(\ref{eq:Fijoneloop1}), which is regularization-independent, in term of $\alpha_{ij}$:
\begin{align}
  \label{eq:F1m1_using_part_frac}
\begin{split}
  \mathcal{F}^{(1,-1)}(\gamma_{ij}) &= -\frac{g_s^2 \gamma_{ij} }{16 \pi^2}
  \int_0^1 {\rm d}x\ P_0(x,\gamma_{ij})\\
&= \frac{g_s^2}{16 \pi^2} \left(\alpha_{ij}+{\displaystyle\frac{1}{\alpha_{ij}}}\right)
  \int_0^1 {\rm d}x\frac{1}{x^2 + (1 - x)^2 + x(1 - x) (\alpha_{ij} + 1/\alpha_{ij})-{\rm i} \varepsilon}
\\
&= -\frac{g_s^2 }{16 \pi^2 }\,\frac{1+\alpha_{ij}^2}{1-\alpha_{ij}^2}
  \int_0^1 {\rm d}x \left(\frac{1}{x - \frac{1}{1 - \alpha_{ij}}-{\rm i} 0} - \frac{1}{x + \frac{\alpha_{ij}}{1 - \alpha_{ij}}+{\rm i}0} \right)
\\
&= -\frac{g_s^2 }{16 \pi^2 }\,2\,\frac{1+\alpha_{ij}^2}{1-\alpha_{ij}^2}\,
 \ln\left({\alpha_{ij}+{\rm i}0}\right)
\end{split}
\end{align}
where $P_0(x,\gamma_{ij})$ is the $\epsilon=0$ limit of $P(x,\gamma_{ij})$ of eq.~(\ref{Pdef}).
In the third line we performed a partial-fraction decomposition of the integrand, obtaining a $d\log$ form. 
The final result in eq.~(\ref{eq:F1m1_using_part_frac}) is the familiar one-loop cusp anomalous dimension (see e.g.~\cite{Korchemsky:1987wg,Ferroglia:2009ep,Kidonakis:2009ev}), often expressed in terms of the Minkowski space cusp angle $\xi_{ij}$:
\begin{align}
  \label{eq:F1m1}
\begin{split}
&\Gamma^{(1)}\alpha_s
=-2w^{(1,-1)} \alpha_s=-2 T_i\cdot T_j \, \mathcal{F}^{(1,-1)}(\gamma_{ij}) \\
&  \mathcal{F}^{(1,-1)}(\gamma_{ij}) 
=
\frac{g_s^2 }{16 \pi^2 }
\left[-2\,\ln(\alpha_{ij})\,\frac{1+\alpha_{ij}^2}{1-\alpha_{ij}^2}\right]
=
\frac{g_s^2 }{16 \pi^2 }
\Big[2\,\xi_{ij}\,\coth(\xi_{ij})\Big]\\
& \text{where}\,\qquad   \xi_{ij}=\cosh^{-1}(-\gamma_{ij}/2)=\ln\alpha_{ij}
\,.
\end{split}
\end{align}
We identify $\alpha_{ij}$, which will be our preferred kinematic variable, as the exponential of the cusp angle between the two lines, $\alpha_{ij}=\exp \xi_{ij}$.
We note that the $\alpha_{ij}\to {1}/\alpha_{ij}$ symmetry mentioned earlier corresponds to $\xi_{ij}\to -\xi_{ij}$. 
As usual, the ${\rm i}\varepsilon$ prescription of the propagator in eq.~(\ref{propdef}) dictates the sign of the imaginary part of $\alpha_{ij}+{\rm i}0$ in eq.~(\ref{eq:F1m1_using_part_frac}), which is important when $\alpha_{ij}$ is negative. From now on we shall not write explicitly the imaginary part; it can always be recovered taking $\alpha_{ij}\to \alpha_{ij}+{\rm i}\varepsilon$ with $\varepsilon>0$.
\begin{figure}[htb]
\begin{center}
\scalebox{0.9}{\includegraphics{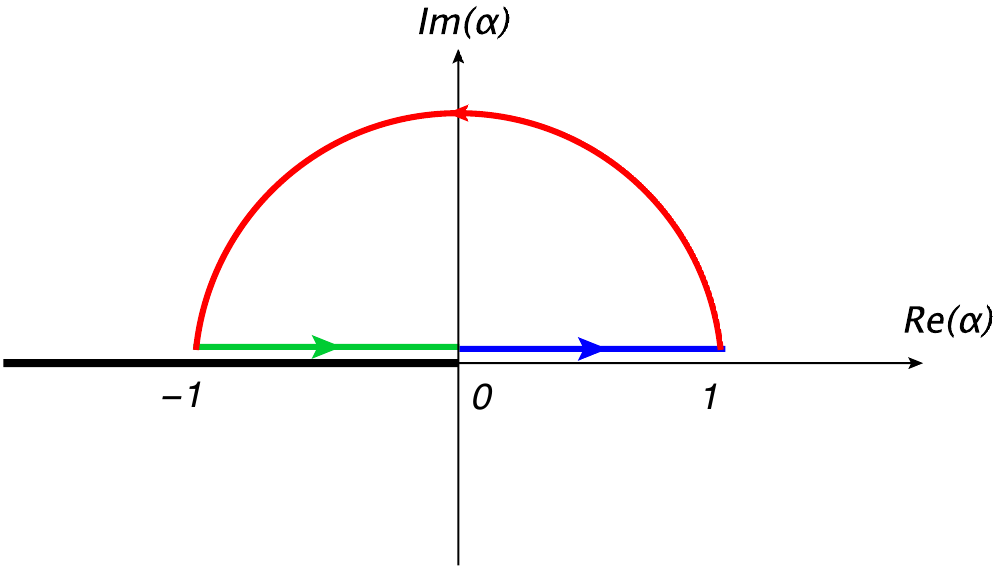}}
\caption{The analytic structure of the one-loop result in the complex $\alpha$ plane -- a logarithmic branch cut along the negative real axis -- shown together with a contour describing the values of $\alpha$ for real values of $\gamma$: the $\alpha\in (0,1)$ region corresponds to space-like kinematics (one incoming and one outgoing partons) where  $\gamma$ varies between $-\infty$ and $-2$;
next, the region of complex $\alpha$ with a positive imaginary part corresponds to the Euclidean region where $-2<\gamma<2$; and finally the region where $\alpha$ is near the branch cut, $\alpha=\alpha_r+{\rm i} \varepsilon$ with $\alpha_r\in (-1,0)$ and $\varepsilon>0$, corresponds to time-like kinematics.  }
\label{alpha_plane}
\end{center}
\end{figure}

figure~\ref{alpha_plane} describes the values\footnote{Here and below we will be using $\alpha$ and $\gamma$ to denote $\alpha_{ij}$ and $\gamma_{ij}$, respectively, whenever specific line indices are not needed.} $\alpha$ takes in the complex plane for real values of $\gamma$. There are two physical regions where $\alpha$ is real: positive $\alpha$ corresponds to space-like kinematics, while negative $\alpha$ to time-like kinematics.
Analytically continuing from one to the other can be done at fixed $|\alpha|$, with $\alpha$ having a positive imaginary part.
As can be verified using eq.~(\ref{gamma_ij}) the limits $\alpha\to \pm 1$ are collinear limits: for $\alpha=-1$ ($\gamma=2$) the two semi-infinite Wilson lines merge into one carrying the sum of the colour charges, while for $\alpha=1$ ($\gamma=-2$)  they join to create a single infinite line.  Physically, $\alpha\to -1$ corresponds to 
heavy-quark production near threshold, a situation where there are Coulomb singularities, while $\alpha\to 1$ corresponds to forward scattering, as occurs in the high-energy limit. Note that in the latter case we do not expect a singularity and indeed the pole of the rational prefactor at $\alpha_{ij}=1$ is compensated in  eq.~(\ref{gamma_alpha_relation}) by the zero of the logarithm.
Finally the limit $\alpha\to 0$ ($|\gamma|\to \infty$) is the lightlike limit.  In this case the logarithmic divergence in eq.~(\ref{eq:F1m1_using_part_frac}) corresponds to the extra collinear singularity characteristics of massless partons.

For the multi-loop analysis we will need the single-gluon-exchange diagram computed to higher orders in the dimensional regularization parameter $\epsilon$. Subleading terms in this expansion enter the expressions for the higher-orders coefficients $\Gamma^{(n)}$ in eq.~(\ref{Gamma_n}). Keeping the full $\epsilon$ dependence in eq.~(\ref{eq:Fijoneloop1}), the kinematic factor takes the form: 
\begin{align}
  \label{eq:Fijoneloop2}
  \begin{split}
  {\cal F}^{(1)}_{ij}(\gamma_{ij},\mu^2/m^2,\epsilon)
  &=
\kappa 
 \,\Gamma(2\epsilon) \,\gamma_{ij}\int_0^1dx \, P(x,\gamma_{ij})\,\, 
  \\
  &=
  \kappa \,\Gamma(2\epsilon) \,\gamma_{ij}\, \,\,
  _2F_1\left([1,1-\epsilon],[3/2],\frac12+\frac{\gamma_{ij}}{4}\right)\,.
  \end{split}
\end{align}
The leading term in the $\epsilon$ expansion of
eq.~\eqref{eq:Fijoneloop2} coincides of course with eq.~(\ref{eq:F1m1_using_part_frac}).
It is straightforward to expand the hypergeometric function in \eqref{eq:Fijoneloop2} to subleading orders in $\epsilon$. To this end we recall the formula from ref. \cite{Kalmykov:2006pu} (eq. (4.24) there continues to ${\cal O}(\epsilon^3)$):
\begin{align}
\label{hyper_exp}
\begin{split}
& _2 F_1\left(\Big[1+a_1\epsilon,1+a_2\epsilon\Big],\Big[\frac32+f\epsilon\Big],z\right)
=\frac{1+2f\epsilon}{2z}\frac{1-\alpha}{1+\alpha}
\Bigg[\ln(\alpha)\\&+\epsilon\bigg(
2(f-a_1-a_2)({\rm Li}_2(-\alpha)+\ln(\alpha)\ln(1+\alpha))\\&
-2f({\rm Li}_2(\alpha)+\ln(\alpha)\ln(1-\alpha))+
\frac{a_1+a_2}2\ln(\alpha)^2+\zeta_2(3f-a_1-a_2)\bigg)+\cdots\Bigg]\,,
\end{split}
\end{align}
where $z=1/2+\gamma/4$ and 
\begin{equation}
\label{alpha_z}
\alpha=\frac{1-\sqrt{z/(z-1)}}{{1+\sqrt{z/(z-1)}}}={\displaystyle \frac{1-\sqrt{\frac{\gamma+2}{\gamma-2}}}{1+\sqrt{\frac{\gamma+2}{\gamma-2}}}}\,.
\end{equation}
The inverse relation between $\gamma$ and $\alpha$ is given in eq.~(\ref{gamma_alpha_relation}) above.
For the expansion of eq.~\eqref{eq:Fijoneloop2} we use eq.~(\ref{hyper_exp}) with $a_2=-1$ and $a_1=f=0$ getting:
\begin{equation}
\label{wf2expnd}
\gamma_{ij}\, _2F_1\left([1,1-\epsilon],[3/2],\frac12+\frac{\gamma_{ij}}{4}\right)
=
2\,r(\alpha_{ij})\,\Big(R_0(\alpha_{ij})+\epsilon R_1(\alpha_{ij}) +\epsilon^2 R_2(\alpha_{ij}) +{\cal O}(\epsilon^2) \Big)
\end{equation}
where the rational factor is
\begin{equation}
\label{r}
r(\alpha_{ij})\equiv \frac{1+\alpha_{ij}^2}{1-\alpha_{ij}^2}\,,
\end{equation} 
and the pure transcendental functions of the first three orders are:
\begin{align}
\label{R12}
\begin{split}
R_0(\alpha)&= \ln(\alpha)\\
R_1(\alpha)&= 2{\rm Li}_2(-\alpha)+2\ln(\alpha)\ln(1+\alpha)-\frac12\ln^2(\alpha)+\zeta_2
\\
R_2(\alpha)&=
2 \text{Li}_3\left(\frac{\alpha}{1+\alpha}\right)
-2 \text{Li}_3\left(\frac{1}{1+\alpha}\right)
-\ln(\alpha)\left[\text{Li}_2\left(\frac{1}{1+\alpha}\right) +\text{Li}_2\left(\frac{\alpha}{1+\alpha}\right) \right]
+\frac{1}{6}\ln^3(\alpha)\,.
\end{split}
\end{align}

We comment that an alternative to writing the general $\epsilon$ integral as a hypergeometric function and then expanding, one may expand under the integral in eq.~(\ref{eq:Fijoneloop1}), defining $R_n(\alpha)$ via\footnote{Throughout the paper  we refrain from expanding the overall coefficient $\kappa$ (as well as the overall $\Gamma(2n\epsilon)$) in order not to clutter the expressions.}
\begin{align}
\label{One_loop_f}
{\cal F}^{(1,n-1)}_{ij}(\gamma_{ij},\mu^2/m^2)
= 
 \,\frac{\kappa}{2} \,\gamma_{ij}\int_0^1dx \, \frac{1}{n!}\,\frac{\ln^n\big(x^2+(1-x)^2-x(1-x)\gamma_{ij}\big)}{x^2+(1-x)^2-x(1-x)\gamma_{ij}}
= \kappa\,r(\alpha_{ij})\,R_{n}(\alpha_{ij}) \,.
\end{align}
Note that the notation here is such that $n$ is the power of the logarithm; the function $R_n$ itself is of weight $n+1$. Note also that upon assigning $\epsilon$ weight $-1$, the entire expansion on the r.h.s. of eq.~(\ref{wf2expnd}) is of uniform weight ($1$). 
These integrals, along with similar integrals which occur in higher-order webs, will be important in determining the anomalous dimension at higher loop orders. For easy reference these definitions are all complied in appendix \ref{sec:summary_of_integrals}.

A final comment is due concerning the choice we made to use $\alpha$ as the default kinematic variable, as opposed to $\gamma$ for example. To this end it is useful to consider the $\alpha\to -1$ limit which corresponds to heavy-quark production near threshold.
In this limit the physics is most transparent when expressed in terms of the velocity of the heavy quark, which is given by $v=\sqrt{1-4m^2/s}$ and tends to zero in this limit. Before making any approximation the relation is $\alpha=-(1-v)/(1+v)$ (see eq.~(\ref{dimensionful})), which indeed tends to $-1$ for $v\to 0$, up to power terms, while the rational factor in eq.~(\ref{eq:F1m1_using_part_frac}) yields $r(\alpha)=(1+v^2)/(2v)$, which is linearly divergent for small $v$, as expected. 
When using the variable $\alpha$ this singularity simply translates into a simple pole at $\alpha\to-1$, however according to eq.~(\ref{alpha_z})  $v=\sqrt\frac{\gamma-2}{\gamma+2}$ implying that in terms of $\gamma$ we obtain a \emph{square-root} singularity at
$\gamma\to2$, rather than a simple pole. 
We see that in terms of $\alpha$ we have a simple analytic structure, which is not the case for $\gamma$. A more complete picture of the analytic structure will be presented in the following sections, after computing the two-loop diagrams. It will transpire that $\alpha$ is a convenient kinematic variable.

\section{Two-loop calculation and the notion of a subtracted web\label{sec:2_loop}}

\begin{figure}[htb]
\begin{center}
\scalebox{1.0}{\includegraphics{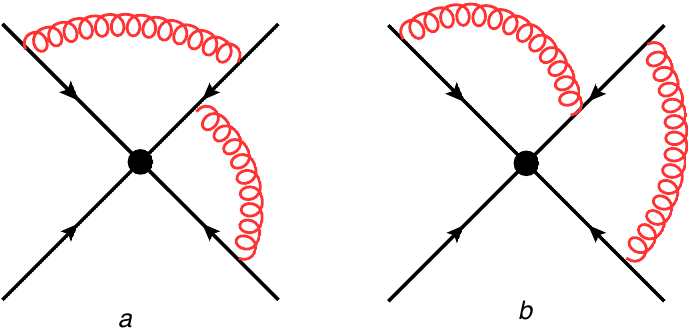}}\\
\scalebox{0.6}{\includegraphics{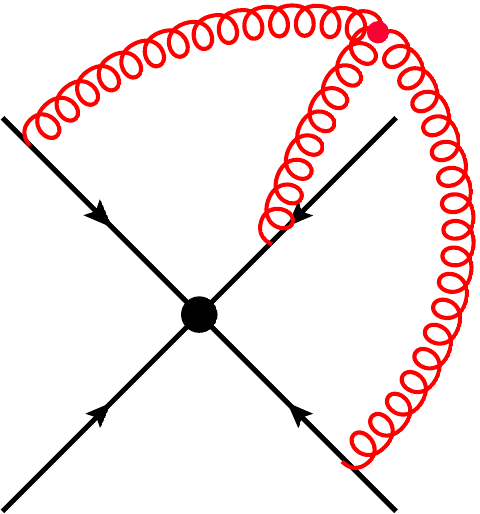}}
\caption{Two-loop graphs connecting three Wilson lines. The two diagrams at the top, (2a) and (2b) respectively, form together a 1-2-1 web, which we denote by $w_{121}^{(2)}$ or alternatively $w_{ijjk}^{(2)}$, where the leg $j$ has two gluon emissions connected respectively to legs $i$ and $k$. The connected diagram at the bottom is a web by itself, denoted by $w_{ijk;\, 3g}^{(2)}$. The two webs have the same colour factor.}
\label{2loopfig}
\end{center}
\end{figure}

In this section we focus on the two-loop calculation of webs connecting three Wilson lines (we will not be concerned with higher-order corrections to the cusp anomalous dimension itself, which was computed to two-loops in refs.~\cite{Korchemsky:1987wg,Kidonakis:2009zc}). The relevant diagrams are shown in figure~\ref{2loopfig}. The figure shows the two relevant types of webs, the  1-2-1 web which comprises two diagrams, each of which has two individual gluon exchanges, and the connected three-gluon-vertex diagram, which is a web on its own. Our focus here is on the former\footnote{The ${\cal O}(\epsilon^{-1})$ term of the latter has been computed analytically in ref.~\cite{Ferroglia:2009ii} in momentum space and numerically, in ref.~\cite{Mitov:2009sv,Mitov:2010xw} in configuration space. An analytic calculation in configuration space will be presented separately~\cite{Almelid:2013tb}.} and we carry out the calculation to ${\cal O}(\epsilon^0)$ as necessary for the computation of the three-loop anomalous dimension. The  ${\cal O}(\epsilon^{-1})$ term of this web, which contributes to the two-loop soft anomalous dimension, has already been computed in refs. \cite{Ferroglia:2009ii,Mitov:2009sv,Mitov:2010xw} and we confirm these results. 

\subsection{Computing the 1-2-1 web}

The 1-2-1 web contribution reads~\cite{Gardi:2010rn}:
\begin{align}
\label{121web}
\begin{split}
W_{(1,2,1)}^{(2)}\,&\,=\,\frac12 \left(C(2a)-C(2b)\right)\left({\cal F}(2a)-{\cal F}(2b)\right)
\\
&\,=\,-\frac12 {\rm i}f^{abc} T_i^aT_j^bT_k^c \left({\cal F}(2a)-{\cal F}(2b)\right)
\end{split}
\end{align}
where the notation $(2a)$ and $(2b)$ refers to the two diagrams in figure~\ref{2loopfig}, respectively. In eq.~(\ref{121web}) we incorporated the relevant mixing matrix and used the colour algebra to write the commutator on line $j$ using the structure constants, exhibiting the connected nature of the colour factor.
Next we need to compute the kinematic factors. Employing the configuration-space Feynman rules detailed in the previous section we have: 
\begin{align}
\label{eq:twoloop_starting_point}  
\begin{split}
    \mathcal{F}^{(2)}(2a)\ &=\  g_s^4\, {\cal N}^2\, (\beta_i\cdot\beta_j) (\beta_j\cdot \beta_k) \int_0^{\infty} ds\, du\, dt_1\, dt_2 \,\theta(t_1>t_2)\\&
\Big(-(s\beta_i-t_1\beta_j)^2\Big)^{\epsilon-1}\Big(-(u\beta_k-t_2\beta_j)^2\Big)^{\epsilon-1}
e^{-{\rm i}m\left(s\sqrt{\beta_i^2-{\rm i}0}+u\sqrt{\beta_k^2-{\rm i}0}+(t_1+t_2)\sqrt{\beta_j^2-{\rm i}0}\right)}\,,
\end{split}
\end{align}
and similarly with $\theta(t_1<t_2)$ for diagram $(2b)$. Rescaling the line-integral parameters such that
$s\sqrt{\beta_i^2-{\rm i}0}=\sigma$ and $t_{1,2}\sqrt{\beta_j^2-{\rm i}0}=\tau_{1,2}$ and 
$u\sqrt{\beta_k^2-{\rm i}0}=\mu$ 
yields
\begin{align}
\label{eq:twoloop_starting_point}  
\begin{split}
    \mathcal{F}^{(2)}(2a)\ &=\frac{ g_s^4 {\cal N}^2 }{4} \gamma_{ij}\gamma_{jk} \int_0^{\infty} d\sigma \,d\mu\, d\tau_1\, d\tau_2 \,\theta(\tau_1>\tau_2)\\&\qquad
\Big(-\sigma^2-\tau_1^2+\gamma_{ij}\sigma\tau_1\Big)^{\epsilon-1}
\,
\Big(-\mu^2-\tau_2^2+\gamma_{jk}\mu\tau_2\Big)^{\epsilon-1}
e^{-{\rm i}(m-{\rm i}0)( \sigma+\tau_1+\tau_2+\mu)}\,.
\end{split}
\end{align}
Repeating now the change of variables of eq.~(\ref{change_var}) for each of the gluons, namely
\begin{align*}
&\lambda_1=\sigma+\tau_1\qquad \text{and}\qquad x=\frac{\tau_1}{\sigma+\tau_1}\,,\\
&\lambda_2=\mu+\tau_2\qquad \text{and}\qquad z=\frac{\tau_2}{\mu+\tau_2}\,,
\end{align*}
followed by 
\[
\lambda=\lambda_1+\lambda_2  \qquad \text{and}\qquad  \omega=\frac{\lambda_1}{\lambda_1+\lambda_2}\,,
\]
we may integrate over the overall distance parameter $\lambda$ obtaining an ultraviolet singularity:
\begin{align}  
\label{eq:twoloop_param}
\begin{split}
    \mathcal{F}^{(2)}(2a)\ &=\  \kappa^2\Gamma(4\epsilon)\, \gamma_{ij}\gamma_{jk} 
\int_0^1dx \int_0^1 dz P(x,\gamma_{ij})\, P(z,\gamma_{jk})  \\&\qquad\int_0^{1} d\omega \Big(\omega(1-\omega)\Big)^{2\epsilon-1} \,\theta\left(\frac{\omega}{1-\omega}>\frac{z}{x}\right)\,,
\end{split}
\end{align}
where $P(x,\gamma_{ij})$ is defined in eq.~(\ref{Pdef}) above. The normalization factor here is written in terms of $\kappa$, defined in eq.~(\ref{kappa1}).
We obtain a similar expression for diagram (2b), where the Heaviside function is $\theta\left(\frac{\omega}{1-\omega}<\frac{z}{x}\right)$. 

Performing the integral over $\omega$ in the second line of eq.~(\ref{eq:twoloop_param}) we get:
\begin{align}  
\begin{split}
I_{\omega}(2a)&=\int_0^{1} d\omega \Big(\omega(1-\omega)\Big)^{2\epsilon-1} \,\theta\left(\frac{\omega}{1-\omega}>\frac{z}{x}\right)=\left(\frac{x}{z}\right)^{2\epsilon} \frac{1}{2\epsilon}\, _2F_1\Big([4\epsilon,2\epsilon],[1+2\epsilon],-\frac{x}{z}\Big)
\\
&=\left(\frac{x}{z}\right)^{2\epsilon} \frac{1}{2\epsilon}\, \Big(1+8\epsilon^2 {\rm Li}_2(-x/z)+{\cal O}(\epsilon^3) \Big)
\\&= \frac12\,\frac{1}{\epsilon}-\ln  \left( {\frac {z}{x}} \right) + \left( 4{\rm Li}_2\left(-{\frac {x}{z}} \right) +\ln^2\left( {\frac 
{z}{x}} \right)  \right) \epsilon+O \left( {\epsilon}^{2} \right) \,,
\end{split}
\end{align}
and similarly for the integral corresponding to diagram $(2b)$ we obtain:
\begin{align}  
\begin{split}
I_{\omega}(2b)&=\int_0^{1} d\omega \Big(\omega(1-\omega)\Big)^{2\epsilon-1} \,\theta\left(\frac{\omega}{1-\omega}<\frac{z}{x}\right)
\\
&={\frac{1}{2}} {\epsilon}^{-1}+\ln  \left( {\frac {z}{x}} \right) - \left( 4
\,{\rm Li}_2\left(-{\frac {x}{z}} \right) + \ln^2\left( 
{\frac {z}{x}} \right)  +4\zeta_2 \right) {\epsilon}+O
 \left( {\epsilon}^{2} \right) \,.
\end{split}
\end{align}
Therefore, for the kinematic factor of the web we obtain:
\begin{align}
\label{121web_kin}  
\begin{split}
    \mathcal{F}^{(2)}(2a)- \mathcal{F}^{(2)}(2b) &=\  2 \kappa^2\Gamma(4\epsilon) \gamma_{ij}\gamma_{jk} 
\int_0^1dx \int_0^1 dz P(x,\gamma_{ij})\, P(z,\gamma_{jk}) \,\phi_2(x,z;\epsilon)\,,
\end{split}
\end{align}
where we denoted the kernel of the integral by
\begin{align}
\label{phi2}  
\begin{split}
\phi_2(x,z;\epsilon)=
\ln\frac{x}{z}+\left(4{\rm Li}_2\left(-\frac{x}{z}\right)+\ln^2\left(\frac{x}{z}\right)+2\zeta_2
\right)\epsilon+{\cal O}(\epsilon^2)\,.
\end{split}
\end{align}
Here the double pole cancels between the two diagrams, as expected\footnote{We recall that the cancellation of the leading poles has been shown to be a general property of webs based on their renormalization properties.} \cite{Gardi:2011yz}. Finally, substituting 
this into eq.~(\ref{121web}) we obtain:
\begin{align}
\label{w121_phi2}
\begin{split}
W_{(1,2,1)}^{(2)}(\alpha_{ij},\alpha_{jk})=- {\rm i}f^{abc} T_i^aT_j^bT_k^c  \,\kappa^2\,\Gamma(4\epsilon) \,\gamma_{ij}\gamma_{jk} &
\int_0^1dx \int_0^1 dz P(x,\gamma_{ij})\, P(z,\gamma_{jk}) \,\phi_2(x,z;\epsilon)\,.
\end{split}
\end{align}

We may now expand the result in $\epsilon$ writing
\[
W_{(1,2,1)}^{(2)}=w_{121}^{(2)}\,\alpha_s^2=\left(
\frac{1}{\epsilon} w_{121}^{(2,-1)}+ w_{121}^{(2,0)} +{\cal O}(\epsilon)\right)\,\alpha_s^2.
\]
To determine the contribution of this web to the two-loop anomalous dimension we only need the coefficient $w_{121}^{(2,-1)}$ of the $1/\epsilon$ pole of (\ref{w121_phi2}).
The required integrals are of the form:
\begin{align}
\label{S1}
\begin{split}
\gamma_{ij} \int_0^1 dx \ln(x) P_0(x,\gamma_{ij})&\equiv r(\alpha_{ij}) \left(-\frac{S_1(\alpha_{ij})}{2}\right)
\\&=
r(\alpha_{ij}) 
\left[-2 {\rm Li}_2(\alpha_{ij})+\frac12 \ln^2(\alpha_{ij})-2 \ln(1-\alpha_{ij}) \ln(\alpha_{ij})+2\zeta_2\right]\,,
\end{split}
\end{align}
with the usual relation (\ref{gamma_alpha_relation}) between $\gamma_{ij}$ and $\alpha_{ij}$, and where the rational function $r(\alpha_{ij})$ is defined in eq.~(\ref{r}), as at one loop.
The fully-integrated result at ${\cal O}(\alpha_s^2\,\epsilon^{-1})$ is:
\begin{align}
\label{w121_result}
\begin{split}
w_{121}^{(2,-1)}(\alpha_{ij},\alpha_{jk})
=&
 -{\rm i}f^{abc} T_i^aT_j^bT_k^c
\, \left(\frac{1}{4\,\pi}\right)^2
r(\alpha_{ij})r(\alpha_{jk})
\Big(\ln(\alpha_{ij})S_1(\alpha_{jk})- \ln(\alpha_{jk})S_1(\alpha_{ij})\Big)\,,
\end{split}
\end{align}
in agreement with refs.~\cite{Ferroglia:2009ii,Mitov:2009sv,Mitov:2010xw}.

\subsection{Two-loop soft anomalous dimension and the subtracted web }

Recall that at one loop we got, according to eq.~(\ref{eq:F1m1}),
\begin{equation}
\label{one_loop_gamma}
\Gamma^{(1)}\alpha_s= T_i\cdot T_j \frac{\alpha_s }{\pi}\,r(\alpha_{ij}) \,\ln(\alpha_{ij})
\,.
\end{equation}
Let us now examine contributions to $\Gamma^{(2)}$ involving the colour indices of three lines, which we denote by~$\Gamma_3^{(2)}$. According to eq.~(\ref{Gamma_n}) there are three sources of such contributions: the 1-2-1 web given by eq.~(\ref{w121_result}), the commutator term and the three-gluon-vertex diagram; all three have the same colour factor $\propto  f^{abc} T_i^aT_j^bT_k^c$.
For the three-gluon-vertex diagram we have~\cite{Ferroglia:2009ii,Mitov:2009sv,Mitov:2010xw,Almelid:2013tb}:
\begin{equation}
\label{3g}
w_{3g}^{(2,-1)}=- {\rm i}f^{abc} T_i^aT_j^bT_k^c\,
2 \left(\frac{1}{4\pi}\right)^2 \,r(\alpha_{ij})
\ln\alpha_{ij}\ln^2\alpha_{jk}\,.
\end{equation}

Let us now evaluate the commutator
$\left[w^{(1,-1)},w^{(1,0)}\right]$
using the results of eq.~(\ref{One_loop_f}).
We have:
\begin{align}
\label{counter_term}
\begin{split}
\left[w^{(1,-1)}\alpha_s,\,w^{(1,0)}\alpha_s\right]&=
\Bigg[{\cal F}_{ij}^{(1,-1)} T_i\cdot T_j 
+{\cal F}_{jk}^{(1,-1)} T_j\cdot T_k,\,
{\cal F}_{ij}^{(1,0)} T_i\cdot T_j 
+{\cal F}_{jk}^{(1,0)} T_j\cdot T_k\Bigg]\\
&\hspace*{-40pt}=T_i^a \left[T_j^a,T_j^b\right] T_k^b
\left({\cal F}_{ij}^{(1,-1)}{\cal F}_{jk}^{(1,0)}-{\cal F}_{jk}^{(1,-1)}{\cal F}_{ij}^{(1,0)}\right)
\\
&\hspace*{-40pt}=-4{\rm i} f^{abc} T_i^a T_j^b T_k^c\, \left(\frac{g_s^2}{16\pi^2}\right)^2
  r(\alpha_{ij}) \, r(\alpha_{jk})
\,\Big(\ln(\alpha_{ij}) R_1(\alpha_{jk})-\ln(\alpha_{jk}) R_1(\alpha_{ij})\Big)
\end{split}
\end{align}
where $R_1(\alpha)$ is given by eq.~(\ref{R12}) above.

Given the similar structure of the commutator and 1-2-1 web contributions, both having the same rational factor $r(\alpha_{ij}) \, r(\alpha_{jk})$, it is natural to combine them, writing the anomalous dimension as
\begin{equation}
\label{Subtracted_2-loop_web}
\Gamma_3^{(2)}=-4w^{(2,-1)}_{3g}\underbrace{-4w_{121}^{(2,-1)}-2\left[w^{(1,-1)},w^{(1,0)}\right]}_{\displaystyle -4\overline{w}_{121}^{(2,-1)}}
\end{equation}
and then:
\begin{align}
\label{two_loop_sum}
\begin{split}
\overline{w}_{121}^{(2,-1)}=&\,\,
w_{121}^{(2,-1)}+\frac12\left[w^{(1,-1)},w^{(1,0)}\right]\\
=&\,\,- {\rm i}f^{abc} T_i^aT_j^bT_k^c 
\, \left(\frac{1}{4\pi}\right)^2
\,r(\alpha_{ij})\,r(\alpha_{jk})\Big(\ln(\alpha_{ij})U_1(\alpha_{jk})- \ln(\alpha_{jk})U_1(\alpha_{ij})\Big)\,,
\end{split}
\end{align}
where we defined the following transcendental function:
\begin{equation}
\label{U1_SR}
U_1(\alpha)=S_1(\alpha)+2R_1(\alpha)\,.
\end{equation}
We shall refer to $\overline{w}_{121}^{(2,-1)}$ as the 1-2-1 \emph{subtracted web}. More generally a subtracted web will be defined as the contributions to the anomalous dimension from the web and all the commutators terms comprised of its subdiagrams.
Using the previous results for $R_1$ and $S_1$, defined respectively in eqs.~(\ref{R12}) and (\ref{S1}), together with the identity:
\[
{\rm Li}_2(\alpha)+{\rm Li}_2(-\alpha)=\frac12 {\rm Li}_2(\alpha^2)\,,
\] 
we get
\begin{align}
\label{U_SR_calculation}
\begin{split}
U_1(\alpha)&=S_1(\alpha)+2R_1(\alpha)=2{\rm Li}_2(\alpha^2)+4\ln(\alpha)\ln(1-\alpha^2)-2 \ln^2(\alpha)-2\zeta_2\,.
\end{split}
\end{align}
Eq.~(\ref{two_loop_sum}) with (\ref{U_SR_calculation}) agrees with both the momentum-space calculation of eq.~(23) in ref.~\cite{Ferroglia:2009ii} and the configuration space one in ref.~\cite{Mitov:2009sv,Mitov:2010xw}.

Note that given the similar structure of the integrals for the web (\ref{w121_phi2}) and commutator (\ref{counter_term}) (see eqs.~(\ref{Ri_def})): 
\begin{align}
\label{counter_term_2}
\begin{split}
\left[w^{(1,-1)},w^{(1,0)}\right]
=-{\rm i} f^{abc}T_i^a T_j^b T_k^c \left(\frac{1}{4\pi}\right)^2
\int_0^1 &dx dz \, p_0(x,\alpha_{ij})\, p_0(z,\alpha_{jk})
\\&\qquad \times 
 \Big( \ln q(z,\alpha_{jk})-\ln q(x,\alpha_{ij}\Big)
\end{split}
\end{align}
where we defined 
\begin{align}
\label{p0def_}
\begin{split}
&p(x,\alpha)\equiv \gamma P(x,\gamma)= \frac{\gamma }{(q(x,\alpha))^{1-\epsilon}}\\
&\text{where} \qquad \quad q(x,\alpha)=x^2 + (1 - x)^2 - x(1 - x)\, \gamma;\qquad -\gamma=\alpha+\frac{1}{\alpha}
\end{split}
\end{align}
and $p_0(x,\alpha)=\lim_{\epsilon\to 0} p(x,\alpha)$, we can also form the subtracted web combination at the level of the integrand, namely
\begin{align}
\label{overline_w121_phi2}
\begin{split}
\overline{w}_{121}^{(2,-1)}(\alpha_{ij},\alpha_{jk})=- {\rm i}f^{abc} T_i^aT_j^bT_k^c  \left(\frac{1}{4\pi}\right)^2 &
\int_0^1dx \, dz \,p_0(x,\alpha_{ij})\, p_0(z,\alpha_{jk}) \,\overline{\phi}_2^{(0)}(x,z)\,,
\end{split}
\end{align}
where
\begin{align}
\label{overline_phi2}  
\begin{split}
\overline{\phi}_2^{(0)}(x,z)=
\ln\frac{x}{z} +\frac12 \Big( \ln q(z,\alpha_{jk})-\ln q(x,\alpha_{ij}\Big)\,
\end{split}
\end{align}
is the subtracted web kernel. Formulating subtracted webs as integrals over $p_0(x,\alpha)$ factors would be the starting point for our general analysis in the next section. We will see that it is useful to postpone the $x$-type integrations until after having formed the subtracted webs combination.

\subsection{Inversion symmetry and the forward limit}

Let us now return to the inversion symmetry mentioned above. We will also comment on the connection of this property with the behaviour of the result in the forward limit. 

The one-loop (cusp) anomalous dimension, of eq.~(\ref{one_loop_gamma})
is symmetric under inversion, $\alpha_{ij}\to 1/\alpha_{ij}$, as must be the case owing to the definition of $\alpha_{ij}$ in  eq.~(\ref{gamma_alpha_relation}). This symmetry is realised through the fact that both the rational function $r(\alpha_{ij}) =\frac{1+\alpha_{ij}^2}{1-\alpha_{ij}^2}$ and the transcendental one, $\ln(\alpha_{ij})$, are separately odd under inversion. 

Consider now the two-loop result presented above. Using the inversion formula:
\begin{equation}
\label{Li2_inversion}
{\rm Li}_2(1/x)=-{\rm Li}_2(x)-\frac12 \ln^2(-x)-\zeta_2
\end{equation}
in eqs.~(\ref{S1}), (\ref{R12}) and (\ref{U_SR_calculation}) 
we find:
\begin{equation}
\label{odd_under_inversion}
S_1(1/\alpha)=-S_1(\alpha)\,,\qquad R_1(1/\alpha)=-R_1(\alpha)\,,\qquad U_1(1/\alpha)=-U_1(\alpha)\,.
\end{equation}
We see that the situation here is similar that at one loop: in particular, in $\overline{w}_{121}^{(2,-1)}(\alpha_{ij},\alpha_{jk})$ of eq.~(\ref{two_loop_sum}) both the rational function and the transcendental function are odd under inversion of each of the $\alpha$ variables, making the final result symmetric, as expected. 
Note that the inversion symmetry is realised differently for the three-gluon vertex contribution $w_{3g}^{(2,-1)}$ of eq.~(\ref{3g}). This function has just one rational factor $r(\alpha_{ij})$: there is none associated with $\alpha_{jk}$. Here the transcendental function $\ln^2\alpha_{jk}$ is by itself even under inversion.

Consider now the the forward-scattering (or straight-line) limit where $\alpha_{ij}\to 1$. In this limit the rational factor $r(\alpha_{ij})$ is singular, and since physically there should not be any singularity there, one expects that the transcendental function should vanish. We now observe that when the function is odd under inversion -- as occurs 
at one loop in eq.~(\ref{one_loop_gamma}) or for two-loops functions in eq.~(\ref{odd_under_inversion}) -- vanishing at $\alpha_{ij}\to 1$ follows automatically.
We can also use the relation
\[
{\rm Li}_2(1-x)=-{\rm Li}_2(x)-\ln(1-x)\ln(x) +\zeta_2
\]
to express the dilogarithmic function $U_1(\alpha)$ 
in eq.~(\ref{U_SR_calculation}) as
\begin{align}
\label{U1_result}
\begin{split}
U_1(\alpha)&=-2{\rm Li}_2(1-\alpha^2)-2 \ln^2(\alpha)\,,
\end{split}
\end{align}
where each of the terms vanishes for $\alpha\to 1$. In the next section we will see how these properties generalise to the entire class of multiple-gluon-exchange webs.

\subsection{Integrating to order $\epsilon^0$}

We may now proceed to perform the integrals required for the ${\cal O}(\epsilon^0)$ term of $w_{121}^{(2)}$, which will be needed for the three-loop anomalous dimension. According to eq. (\ref{w121_phi2}) with $\phi_2$ of eq.~(\ref{phi2}), these include for example the integral $\int_0^1 dx \ln^2(x) p_0(x,\alpha)$, where $p_0(x,\alpha)$ is defined following eq.~(\ref{p0def_}); this integral is given by eq.~(\ref{S2_def}). In addition the ${\cal O}(\epsilon^0)$ calculation requires integrals such as eq.~(\ref{V2_def}) involving $\ln(q(x,\alpha)$ from the expansion of 
$p(x,\alpha)$ to higher order in $\epsilon$. These integrals are all summarised in appendix~\ref{sec:summary_of_integrals}.

On the face of it, examining eq.~(\ref{w121_phi2}) with (\ref{phi2}) one would conclude that also the following integral is required: 
\begin{align}
\label{dilog_integral}
\int_0^1 dx \int_0^1 dz \,{\rm Li}_2\left(-\frac{x}{z}\right)  p_0(x,\alpha_{ij}) p_0(z,\alpha_{jk})\,.
\end{align}
Here, in contrast to the other integrals we encountered, the dilogarithm appearing in the kernel couples the two gluons and prevents the factorization of the result into polylogarithms of $\alpha_{ij}$ and those of $\alpha_{jk}$. Indeed, performing this integral we obtain a rather complicated function, which can be expressed in terms of Goncharov multiple  polylogarithm~\cite{Goncharov:arXiv0908.2238,Goncharov:2010jf}, some of which depend on both cusp angles, rather than a sum of products of polylogarithm of a single cusp angle.
Interestingly, the integral of eq.~(\ref{dilog_integral}) is actually not needed for the three-loop webs we are considering: as we shall see below all dilogarithms cancel out in the integrand in combinations of webs and commutators of their subdiagrams, so if we postpone the integration until after subtracted web combinations are formed, multiple polylogarithms never arise.  As we discuss in the following section this cancellation is not accidental, but rather points to a general structure of multi-gluon-exchange webs. The simplification achieved by these cancellations is a significant incentive for arranging the calculation in terms of subtracted webs.

\section{Properties of functions appearing in multi-gluon-exchange webs  \label{sec:functions}}

The functions we have encountered through two loops have a rather simple analytic structure and several symmetries which call for interpretation. The purpose of this section is to analyse these properties in order to gauge how general they are, gain some physical understanding and prepare the grounds for the three-loop calculation that follows. 

\subsection{The structure of multiple-gluon-exchange integrals}

Let us examine the kinematic dependence of multiple-gluon-exchange webs. We will analyse the general form of the corresponding integrals, their analytic structure and their symmetries. The most obvious among these is the inversion symmetry, $\alpha\to 1/\alpha$, which is an immediate consequence of the definition of $\alpha$ in eq.~(\ref{gamma_alpha_relation}).
We have already seen that the transcendental functions associated with gluon-exchange diagrams at one and two loops are odd under inversion, compensating the fact that the rational factor associated with each exchange, $r(\alpha)$ of eq.~(\ref{r}), is itself odd, and making the anomalous dimension even, as it must be.  We will see that for the class of webs we are concerned with, namely those involving multiple-gluon-exchange diagrams (without any three of four gloun vertices) this property generalises to any order. 

The first observation is that the rational functions in these webs can be determined without doing any integrals, and they simply amount to a factor of $r(\alpha_{ij})$ for each gluon exchange between lines $i$ and $j$. 
Recall that the Feynman rules for each gluon exchange involve a propagator and two parameter integrals corresponding to the positions at which the gluon attaches to the two Wilson lines. Upon using the parametrization of eq.~(\ref{change_var}) we see that each gluon exchange gives rise to precisely one factor of $p(x,\alpha)$, defined in
eq.~(\ref{p0def_}), along with a corresponding parametric integral over $x$, representing the gluon emission angle. The overall distance of this gluon from the vertex, which has been scaled out of the propagator, can be integrated over along with similar variables associated with the other gluons, where the limits of integration for any particular web diagram are determined by the order of attachments of the gluons to a given Wilson line. 
The general structure of these integrals is:
\begin{align}
\label{general_form_of_int}
\begin{split}
{\cal F}&\sim g_s^{2n} 
\int_0^{\infty}d\lambda_1 \, d\lambda_2\,\ldots\,d\lambda_n
\prod_{k=1}^{n}\lambda_k^{2\epsilon-1}
 \\&\times \int_0^{1} dx_1\,dx_2\,\ldots\, dx_n \Theta\left(\left\{x_i/x_j,\lambda_i/\lambda_j\right\}\right)\,
{\rm e}^{-{\rm i}(m-{\rm i}0)\sum_{k=1}^n\lambda_k}
\prod_{k=1}^{n} p(x_k,\alpha_k)
\end{split}
\end{align}
where the function $\Theta\left(\left\{x_i/x_j,\lambda_i/\lambda_j\right\}\right)$ denotes a product of Heaviside functions of the form \hbox{$\theta(x_i\lambda_i>x_j\lambda_j)$}, which represent the order of gluon emissions along each of the Wilson lines. 
The integral over the overall distance scale of the $n$ gluon exchanges, 
$\lambda\equiv \sum_{k=1}^n\lambda_k$,
yields an ultraviolet divergence of the form $\Gamma(2n\epsilon)$ and the remaining $(n-1)$ integrals involving $\Theta$ yield a pure polylogarithmic function\footnote{To see why such integrals yield a polylogarithmic function at any order in $\epsilon$, note that in 4-dimensions the $\lambda_k$ integrals in eq.~(\ref{general_form_of_int}) are of the $d\log$ form, $d\lambda_k/\lambda_k$. In $d$-dimensions the integrand has extra logarithms, which raise the weight of the polylogarithm by one per power of $\epsilon$. The limits of integration, which depend on the order of attachments, determine the specific polylogarithmic function that is obtained. An example was given in eq.~(\ref{eq:twoloop_param}) at two-loops and we will see further examples in appendix \ref{sec:3_loop_calc}.} of weight $(n-1)$. The latter function, which we call the \emph{kernel} and denote by $\phi_{n-1}$, depends solely on the $x$-type angular variables of which there are $n$. To be precise eq.~(\ref{general_form_of_int}) implies that the kernel may only involve polylogarithms of ratios such as $x_i/x_j$ and Heaviside functions of the form $\theta(x_i>x_j)$.
In webs where $n$ gluons connect $(n+1)$ Wilson lines no Heaviside functions remain after the $\{\lambda_i\}$ integrations. In contrast, in more entangled webs, where $n$ gluons connect fewer than $(n+1)$ Wilson lines, some Heaviside functions may appear in the kernel. 

The final integration for a given $n$ gluon exchange diagram takes the form:
\begin{align}
\label{mge_kin}
\begin{split}
 {\cal F}^{(n)}&\sim\kappa^n
\Gamma(2n\epsilon)\int dx_1 dx_2\ldots dx_n\, \phi_{n-1}(x_1,x_2,\ldots, x_n;\epsilon)\, \prod_{k=1}^n p(x_k,\alpha_k) 
\\
&= \kappa^n \Gamma(2n\epsilon) \left(\prod_{k=1}^{n} \, r(\alpha_k) \right) \, s_n(\{\alpha_k\};\epsilon)
\end{split}
\end{align}
where $k$ runs over the $n$ gluons. To perform the $x_k$ integrals we may expand the integrand in powers of $\epsilon$ and integrate term by term. Contributions to this expansion arise in general from the kernel $\phi_{n-1}$ and from the powers of the propagators, the latter yielding 
\begin{equation}
\label{expand_p}
p(x,\alpha)= p_0(x,\alpha) \sum_{j=0}^{\infty}\frac{\epsilon^j}{j!} \ln^j q(x,\alpha)\,,
\end{equation}
where $q(x,\alpha)$ is defined in eq.~(\ref{p0def_}).
Having made this expansion we readily see how the final expression in eq.~(\ref{mge_kin}) is obtained: the rational functions $r(\alpha_k)$ simply emerge upon partial fractioning each $p_0(x_k,\alpha_k)$ as in eq.~(\ref{eq:F1m1_using_part_frac}), while the resulting integral $s_n$ is a pure transcendental function of weight $(2n-1)$ depending on all the kinematic variables. Note that beyond that overall $\Gamma(2n\epsilon)$ singularity which we factored out, this integral may have up to $(n-1)$ extra powers of $1/\epsilon$; such a maximal power is attained for maximally reducible diagrams, where each gluon can be independently contracted towards the hard-interaction vertex. As before, $\epsilon^{-1}$ is assigned a weight $1$. At any order in the $\epsilon$ expansion $s_n$ is again a polylogarithmic function of uniform weight.

We observe that integrals of the form of eq.~(\ref{mge_kin}) with the same product of $p(x_k,\alpha_k)$, albeit with different kernel functions, are associated to any individual diagram in a web. 
The contribution of the web as a whole to the anomalous dimension, ${w}^{(n,-1)}$, is simply a linear combination of the integrals of the various diagrams expanded to ${\cal O}(\epsilon^{-1})$, with numerical coefficients dictated by the corresponding web mixing matrix.  
Furthermore, the very same integrals appear in commutator contributions to the anomalous dimension at order $n$ which are formed by webs defined by subdiagrams of the web under consideration. Thus the subtracted web $\overline{w}^{(n,-1)}$ also has the form of eq.~(\ref{mge_kin}) with the same rational factor.

Given the common structure, it is most natural to combine the web diagrams at the level of the \emph{integrand} of eq.~(\ref{mge_kin}), yielding a \emph{web kernel}, and similarly, include the relevant commutator terms, defining a \emph{subtracted web kernel}.
In the 1-2-1 case the combined integral is given by eq.~(\ref{121web_kin}) and the corresponding web kernel is given in eq.~(\ref{phi2}); the subtracted web is given by eq.~(\ref{overline_w121_phi2}).  We now see that the general structure, summarised by eq.~(\ref{mge_kin}), is obtained for any multiple-gluon-exchange web. We will see further examples of this at three loops in what follows (see appendix \ref{sec:3_loop_calc}).

The conclusions from this analysis can be summarised as follows: 
\begin{itemize}
\item{} A given multiple-gluon-exchange diagram, and likewise the web and the corresponding subtracted web, has a kinematic dependence of the form of eq.~(\ref{mge_kin}) where the rational function is simply a product of $n$ factors of $r(\alpha)$ of eq.~(\ref{r}) with the relevant $\alpha=\alpha_{ij}$. 
\item{} The latter rational factor is multiplied by a transcendental function, $s_n(\{\alpha_k\};\epsilon)$ of weight $(2n-1)$, potentially having extra poles ${\cal O}(\epsilon^{-k})$ with $k\leq (n-1)$. The coefficient of $\epsilon^{-k}$ in this function is a polylogarithmic function of weight $(2n-1-k)$; the contribution to the anomalous dimension, $\overline{w}^{(n,-1)}$, corresponding to $k=0$, has weight $(2n-1)$.  This class of functions is amenable  
to the symbol analysis~\cite{Goncharov:arXiv0908.2238,Goncharov:2010jf,Duhr:2011zq,Duhr:2012fh}, which we will use below.
\item{} Given the rational factor, the symmetry of the anomalous dimension under inversion, $\alpha\to 1/\alpha$, dictates how $s_n(\{\alpha_k\};\epsilon)$ transforms. This in turn relates to the forward limit $\alpha_k\to 1$ where $s_n(\{\alpha_k\};\epsilon)$  is expected to vanish. 
\end{itemize}

\subsection{Crossing symmetry and the symbol alphabet}

Let us now turn to discuss another symmetry property, which is far less obvious. 
We note that $U_1(\alpha)$, in eq.~(\ref{U_SR_calculation}), or equivalently in eq.~(\ref{U1_result}), appears to be a function of $\alpha^2$. That is, assuming at this stage that $\alpha>0$ (see below!) one may replace $2\ln \alpha$ by $\ln \alpha^2$ to obtain a function of $\alpha^2$. It is evident that neither the 1-2-1-web function $S_1(\alpha)$ nor the commutator function, $R_1(\alpha)$, have such a property, while their combination entering the anomalous dimension, the subtracted web, does:
\begin{equation}
\label{square_relation}
U_1(\alpha)=S_1(\alpha)+2R_1(\alpha)=\frac12 S_1(\alpha^2)\,,\qquad \quad \forall \alpha>0\,.
\end{equation} 

The fact that the subtracted web appears to be a function of~$\alpha^2$, rather than just a function of~$\alpha$ can be understood on general grounds, as we shall see below. 
Before turning to the explanation, it should be pointed out that a related observation was recently made in refs.~\cite{Henn:2012qz,Henn:2012ia,Henn:2013wfa}. It was found there that the angle-dependent cusp anomalous dimension in ${\cal N}=4$ supersymmertic Yang-Mills can be expressed as a function of $\alpha^2$ through three loops, and at least for multiple-gluon-exchange diagrams this persists through six loops.  Our observation above generalises this to the three-leg soft anomalous dimension at two-loops.

The fact that these functions appear to be functions of $\alpha_{ij}^2$ calls for considering the symmetry under, 
\begin{equation}
\label{alpha_minus_alpha} 
\alpha_{ij}\to -\alpha_{ij}\,. 
\end{equation}
Physically such a transformation is indeed interesting since it is associated with crossing symmetry: it may be realised by reversing the 4-velocity of one of the partons, e.g. $\beta_i\to -\beta_i$ keeping the other one ($\beta_j$) unchanged (it follows from eq.~(\ref{gamma_ij}) that this amounts to $\gamma_{ij}\to - \gamma_{ij}$, and therefore to
$\alpha_{ij}\to -\alpha_{ij}$). Thus, it corresponds to a relation between spacelike kinematics with $\alpha_{ij}>0$, where the two Wilson lines $i$ and $j$ correspond to two partons one of which belonging to the initial state and one to the final state, and timelike kinematics with $\alpha_{ij}< 0$, where the two partons are both in the initial state or both in the final state. For example, the amplitude where $i$ is an incoming quark and $j$ is an outgoing quark, as in deep-inelastic scattering, may be related by eq.~(\ref{alpha_minus_alpha}) to one where $i$ is an outgoing antiquark (while $j$ is still an outgoing quark) as in quark anti-quark production. 

Recall that the observation that $U_1$ is a function of $\alpha^2$ in eq.~(\ref{square_relation}) relied on $\alpha$ being positive. Specifically it relies on replacing $2\ln \alpha$ by $\ln \alpha^2$ which does not hold for negative~$\alpha$.  The correct analytical continuation of 
$U_1(\alpha)$ to complex $\alpha$, and to the timelike axis in particular, is given by eq.~(\ref{U_SR_calculation}), where $\alpha$ is in the upper half of the complex plane (see figure \ref{alpha_plane}). For $\alpha<0$: 
\begin{equation}
\label{analytic_cont}
\ln(\alpha)=\ln(\alpha+{\rm i}0)=\ln (|\alpha|\,{\rm e}^{{\rm i}\pi})=\ln|\alpha|+{\rm i}\pi 
\end{equation}
which is different from $\frac12 S_1(\alpha^2)$.
Indeed physically we expect such an imaginary part to be generated for $\alpha< 0$ where the two partons are both in the final (or both in the initial) state. A similar logarithm, generating an ${\rm i}\pi$ term, appears already at one-loop in eq.~(\ref{one_loop_gamma}).
We are then led to conclude that, strictly speaking, there is no $\alpha_{ij}\to -\alpha_{ij}$ symmetry: this symmetry is broken by ${\rm i}\pi$ terms from the analytical continuation. 

The precise statement is that the crossing symmetry of eq.~(\ref{alpha_minus_alpha}) is realised \emph{at the level of the symbol}. The symbol~\cite{Goncharov:arXiv0908.2238,Goncharov:2010jf,Duhr:2011zq,Duhr:2012fh,Gaiotto:2011dt} represent the branch point structure of the function, but, in contrast to the function itself, it is not sensitive to the kinematic region (or to the way the cuts are routed) and specifically it eliminates ${\rm i}\pi$ terms:
\begin{equation}
\label{noIpi}
{\cal S}\left[\ln(-\alpha)\right]=\otimes \alpha ={\cal S}\left[\ln(\alpha)\right].
\end{equation}
Thus in the one-loop case the crossing symmetry is trivially realised at the symbol level. 
In the two-loop case we have, 
\begin{subequations}
\begin{align}
{\cal S}\left[S_1(\alpha)\right] &=4 \alpha\otimes (1-\alpha)-2 \alpha\otimes \alpha
\\
{\cal S}\left[R_1(\alpha)\right] &=2 \alpha\otimes (1+\alpha)-\alpha\otimes \alpha
\end{align}
\end{subequations}
neither of which admits the symmetry, while the combination (\ref{U1_SR}) that appears in the anomalous dimension, does:
\begin{align}
\label{SYMB_U1_sym}
{\cal S}\left[U_1(\alpha)\right]&=4 \Big[\alpha\otimes (1-\alpha)+ \alpha\otimes (1+\alpha)- \alpha\otimes \alpha\Big]=4\left[\alpha\otimes (1-\alpha^2) -\alpha\otimes\alpha\right]\,,
\end{align}
where the last expression is written explicitly in terms of the alphabet $\alpha$ and $1-\alpha^2$.
Note that the symbol may also be expressed using $\alpha^2$ and $1-\alpha^2$, as $2{\cal S}[\ln(\alpha)]=\otimes \alpha^2$. So at the symbol level we have
\begin{align}
\label{URS_symbol}
{\cal S}\left[U_1(\alpha)\right]
=\left. \,{\cal S}\left[R_1(\alpha)\right]\right\vert_{\alpha\to-\alpha^2}
=\left.\frac12\, {\cal S}\left[S_1(\alpha)\right]\right\vert_{\alpha\to\alpha^2}\,.
\end{align}
In conclusion we saw that at the level of the symbol, contributions to the anomalous dimensions are invariant under $\alpha\to-\alpha$. As expected by crossing symmetry, the result for timelike kinematics only differs from that for spacelike kinematics by the terms generated through analytic continuation. Furthermore, we learnt that the crossing symmetry is realised (at least at two loops) only after combining the web with the corresponding commutators of its subdiagrams appearing in eqs.~(\ref{Gamma_n}) to form the subtracted web. 

To fully understand these observations it is useful to recall relation between $\alpha_{ij}$ and the dimensionful kinematic invariants using eqs.~(\ref{alpha_z}) and (\ref{gamma_ij}):
\begin{equation}
\label{dimensionful}
\alpha_{ij}=\frac{\sqrt{1-\frac{\sqrt{m_i^2m_j^2}}{p_i\cdot p_j}}-\sqrt{1+\frac{\sqrt{m_i^2m_j^2}}{p_i\cdot p_j}}}{\sqrt{1-\frac{\sqrt{m_i^2m_j^2}}{p_i\cdot p_j}}+\sqrt{1+\frac{\sqrt{m_i^2m_j^2}}{p_i\cdot p_j}}}
\end{equation}
and consider the expansion for small masses, $m_i^2,\, m_j^2\to 0$, corresponding to the lightlike limit\footnote{I would like to thank Lance Dixon for illuminating discussions on this subject.}. We observe that in this limit
\begin{equation}
\label{small_alpha}
\alpha_{ij}\to \frac{\sqrt{m_i^2m_j^2}}{-2p_i\cdot p_j}\,\left[1+{\cal O}\left(\frac{m_i^2m_j^2}{(2p_i\cdot p_j)^2}\right)\right]
\end{equation}
where the square brackets is a Taylor expansion in powers of ${m_i^2m_j^2}/{(2p_i\cdot p_j)^2}$. 
Given what we know about the functions contributing to the anomalous dimension for this class of webs (eq.~(\ref{mge_kin}) above) we expect logarithmic and polylogarithmic functions depending on $\alpha_{ij}$.

A logarithm of $\alpha_{ij}$ corresponds in the lightlike limit a logarithm of $\sqrt{m_i^2m_j^2}/({-2p_i\cdot p_j})$, plus subleading dependence taking the form of integer powers of ${m_i^2m_j^2}/{(2p_i\cdot p_j)^2}$, consistently with the expected analytic properties. 
Logarithms of $\alpha_{ij}$ are sensitive to the sign of $p_i\cdot p_j$, namely they depend on whether the partons are incoming or outgoing, and they will generate ${\rm i}\pi$ terms when analytically continued to timelike kinematics, $\alpha_{ij}<0$, but importantly, the powers are insensitive to this. 
 
Consider now a logarithm of $(1-\alpha_{ij})$. Upon expanding near the lightlike limit, $m_i^2,\, m_j^2\to 0$, such a term would generate power terms proportional to ${\sqrt{m_i^2m_j^2}}/({-2p_i\cdot p_j})$ having square-root branch points for small masses, which should never occur in a scattering amplitude. 
A similar behaviour would arise in more complicated functions containing $\otimes (1-\alpha_{ij})$ in their symbol, as occurs in $S_1(\alpha_{ij})$. This is avoided of course if there is also a corresponding term with  $\otimes (1+\alpha_{ij})$ building up dependence on $\otimes (1-\alpha_{ij}^2)$ in the symbol, as we do indeed observe in the example considered.

On this basis we formulate the following general conjecture:
\\
\\
{\bf Alphabet conjecture:}\,\,\emph{The alphabet of the symbol of all multiple-gluon-exchange subtracted webs is restricted to $\otimes \alpha_{ij}$ and $\otimes (1-\alpha_{ij}^2)$.}
\\
\\
This guarantees, in particular, an exact crossing symmetry at the level of the symbol at any loop order.
This analytic structure is consistent with all we have observed through three loops as well as with what authors of refs.~\cite{Henn:2012qz,Henn:2012ia} have found through six loops in the two-Wilson-line case.
While we do not have a formal proof of the above conjecture, it looks unavoidable given the following physical considerations:
\begin{itemize}
\item{} The physical region extends throughout the range 
$0<|\alpha_{ij}|<1$.
\item{} One expects a branch cut on the negative real $\alpha_{ij}$ axis corresponding to timelike kinematics, and no other cuts in the physical region.
\item{} Given this and the inversion symmetry $\alpha_{ij}\to 1/\alpha_{ij}$, one  expects branch points to appear at boundaries, $\alpha_{ij}\to 0$,\, $\alpha_{ij}\to\pm 1$ and at $\alpha_{ij}\to \infty$, but nowhere else.
\item{} Considering the small-mass limit discussed above, analyticity of power terms in $m_i^2$ requires dependence through $\otimes (1-\alpha_{ij}^2)$ rather than separately on $\otimes (1-\alpha_{ij})$ and on $\otimes (1+\alpha_{ij})$.
\end{itemize}

\subsection{Subtracted webs}

We observed that the crossing symmetry is realised only at the level of the subtracted web, namely only once we form the relevant combination of a web with the corresponding commutators of 
lower-order webs corresponding to its subdiagrams. At two loops we have seen that neither the symbol of the 1-2-1 web in eq.~(\ref{w121_result}) nor that of the commutator of 
eq.~(\ref{counter_term}) admit the $\alpha_{ij}\to -\alpha_{ij}$ symmetry, while the symbol of their combinations which enters the anomalous dimension, the subtracted web of eq.~(\ref{SYMB_U1_sym}), does.  Our next question is then why is $\alpha\to-\alpha$ symmetry violated before forming the subtracted-web combination. Equivalently, we want to understand why $\otimes (1-\alpha_{ij})$ appears in the symbol of non-subtracted webs without a corresponding $\otimes (1+\alpha_{ij})$, while in subtracted webs only $\otimes (1-\alpha_{ij}^2)$ is allowed.

To address this question recall that the $1/\epsilon$ pole whose coefficient we are extracting is not the leading pole of individual diagrams. For example the two-loop diagrams of the 1-2-1 web are separately of ${\cal O}(1/\epsilon^2)$, implying that the ${\cal O}(1/\epsilon)$ terms are regularization dependent. When forming the web the leading poles cancel, but a residual regularization dependence remains at ${\cal O}(1/\epsilon)$. 

We conclude that in general the kinematic dependence of the web, before forming the subtracted-web combination, is regularization dependent. Indeed our infrared regularization in eq.~(\ref{FRmod}) \emph{introduces}\footnote{Recall that without any regulator the result vanishes as a scaleless integral. The suppression factor is introduced in a manner that preserves the rescaling symmetry. This requirement along with the chosen linear dependence of the exponential on $\lambda$ implies dependence on the square root $\sqrt{\beta_i^2-{\rm i}0}$.} dependence on the kinematic invariants through the exponential suppression factor $\exp\left\{-{\rm i} m\lambda \sqrt{\beta_i^2-{\rm i}0}\right\}$, involving the square-root of the Wilson-line mass squared.
The particular form of the cutoff dictates the specific functional dependence of $w^{(n)}$ on~$\alpha$, and there is no surprise that  square roots appear there. However the subtracted web $\overline{w}^{(n,-1)}$, which contributes directly to the anomalous dimension $\Gamma^{(n,-1)}$, must be independent of the details of the regulator, and no square roots can survive.

The fact that regularization-independence along with the crossing symmetry must be there for subtracted webs while not for non-subtracted ones, strongly suggests one should organise the calculation in terms of subtracted webs. Indeed, we have already seen that the structure of the integrals for products of subdiagrams of a web according to eq.~(\ref{Gamma_n}) admits a form similar to the web itself, all sharing the same product of $p_0(x_k,\alpha_k)$ functions, so it is straightforward to form the subtracted-web combinations at the level of the integrand. 

Expanding the kinematic functions of the form of eq.~(\ref{mge_kin}) in $\epsilon$ and forming the subtracted web combination we obtain at ${\cal O}(\alpha_s^n\,\epsilon^{-1})$: 
\begin{align}
\label{subtracted_web_mge_kin}
\begin{split}
 \overline{w}^{(n,-1)}= \left(\frac{1}{4\pi}\right)^n \, C_{i_1,i_2,\ldots i_{n+1}}\,&
 \int dx_1 dx_2\ldots dx_n\, \times\,\prod_{k=1}^n p_0(x_k,\alpha_k) \,\times \,\\&{\cal G}_{n-1}\Big(x_1,x_2,\ldots, x_n;
q(x_1,\alpha_1), q(x_2,\alpha_2), \ldots q(x_n,\alpha_n)
\Big)\, ,
\end{split}
\end{align}
where $C_{i_1,i_2,\ldots i_{n+1}}$ is a (connected) colour factor involving the generators of up to $(n+1)$ Wilson lines and ${\cal G}_{n-1}$ is a polylogarithmic function of uniform transcendental weight $(n-1)$.
The contributions to the anomalous dimension need to be summed over all subtracted webs of order $n$, $\Gamma^{(n)}=-2n \,\sum_i \overline{w}^{(n,-1)}_i$ (where we discarded running coupling terms).
It should be noted that a given web may in general contribute to several different colour factors (these are enumerated by the eigenvectors of the mixing matrix corresponding to unit eigenvalue~\cite{Gardi:2010rn,Gardi:2013ita}). Also note that different webs may contribute to the same colour factor. Thus webs, or for that matter subtracted webs, are not by themselves gauge invariant, while the sum of all subtracted webs contributing to a given colour factor is.

\subsection{Basis of integrals for subtracted webs}

Given that the integral in eq.~(\ref{subtracted_web_mge_kin}) is expected to admit the analytic properties established above, namely its symbol alphabet should be restricted to $\otimes \alpha$ and $\otimes (1-\alpha^2)$, it is natural to express the result in terms of a basis of functions which themselves have these analytic properties.
Our final task in this section is to explicitly construct this basis of functions. To this end it is useful to recall the structure of the integrand on the r.h.s of eq.~(\ref{mge_kin}) along with the constraints on the transcendental weight of the kernel.
For $n$ gluon exchanges $\phi_{n-1}$ has weight $(n-1)$, thus at one loop it is a constant, at two loops a logarithm and at three loops a product of two logarithms or a dilogarithm, and so on. Similarly in ${\cal G}_{n-1}$, we expect to get products of logarithms and polylogarithms of ratios\footnote{It must be ratios of $x_k$ because $\Theta$ in eq.~(\ref{general_form_of_int}) only depends on such ratios.} of $x_k$, as well as logarithms\footnote{Note that the logarithms of $q(x_k,\alpha_k)$ all originate from expanding $p(x,\alpha)$ in $\epsilon$ as in eq.~(\ref{expand_p}); thus polylogarithms of $q(x_k,\alpha_k)$ do not ever arise in ${\cal G}_{n-1}$.} of $q(x_k,\alpha_k)$. Thus we conclude that ${\cal G}_0$ is a constant, ${\cal G}_{1}$ may be a linear combination of $\ln x_k$ and $\ln q(x_k,\alpha_k)$ terms, and ${\cal G}_{2}$ may contain products of the above including $\ln^2 x_k$, $\ln x_k \ln q(x_k,\alpha_k)$, and $\ln^2 q(x_k,\alpha_k)$. It may a priori contain also a dilogarithm ${\rm Li}_2(-x_k/x_j)$, which we encountered in eq.~(\ref{dilog_integral}), and as we already noted, it does not appear in subtracted webs; this issue will be further discussed below.

Another ingredient which can appear in ${\cal G}_{n-1}$ is a Heaviside function such as $\theta(x_k>x_j)$. As already mentioned, a Heaviside function may survive the $\{\lambda_i\}$ integrations in eq.~(\ref{general_form_of_int}) and appear in the web kernel in webs where the $n$ gluons connect less than the maximal number of $(n+1)$ Wilson lines. 
Here we will not consider this possibility, which will be studied in detail in a forthcoming publication \cite{Falcioni:2013tb} where we consider three-loop webs connecting three Wilson lines.

A compilation of the integrals which may appear in non-subtracted webs through three loops appears in table \ref{tab:4leg3loop_transc_func_symb}.
Importantly the symbol of such integrals would not in general admit the $\alpha_k\to -\alpha_k$ symmetry. Thus, imposing the requirement that the symbol alphabet would be restricted to $\otimes \alpha$ and $\otimes (1-\alpha^2)$ places a severe restriction on the basis of required functions for subtracted webs.  
It is straightforward to construct those linear combinations of the integrals in table~\ref{tab:4leg3loop_transc_func_symb} which, at the symbol level, admit the $\alpha_k\to -\alpha_k$ symmetry. Examining the entries in table~\ref{tab:4leg3loop_transc_func_symb} we note that an extra power of $\ln q(x_k,\alpha_k)$ in the integrand results in an extra $\otimes (1+\alpha)$ in the symbol of the integral, while an extra power of $\ln x_k$ translates into an extra 
$\otimes (1-\alpha)$. Thus to impose the crossing symmetry we need to balance factors of $\ln q(x_k,\alpha_k)$ with factors of $\ln x_k$, selecting very few basis functions.  Note that this is sufficient to realise the symmetry since, owing to eq.~(\ref{noIpi}), entries with the letter $\alpha$ are insensitive to sign reversal. 

At one and two loops there is just one basis function at each order:
\begin{subequations}
\begin{align}
R_0(\alpha)&=\frac{1}{2r}\int_0^1 dx  p_0(x,\alpha)
\label{R0_sym} \\
U_1(\alpha)&=S_1(\alpha)+2R_1(\alpha)=\frac{1}{r}\int_0^1 dx  p_0(x,\alpha) \ln \left(\frac{q(x,\alpha)}{x^2}\right)\,,
\label{U1_sym}
\end{align}
\end{subequations}
where the symbol of the latter is given in eq.~(\ref{SYMB_U1_sym}).
Also beyond two loops this symmetry provides a highly non-trivial constraint, yet there are more functions that arise. At three loops we get the following set of functions, all having a symbol which is symmetric under $\alpha\to-\alpha$:
\begin{subequations}
\label{symfunc}
\begin{align}
U_2(\alpha)&=R_2(\alpha)-S_2(\alpha)-V_2(\alpha)=\frac{1}{r}\int_0^1 dx\,  p_0(x,\alpha) \frac14 \ln^2 \left(\frac{q(x,\alpha)}{x^2}\right)
\label{U2_sym}
\\
\Sigma_2(\alpha)&=\widetilde{S}_2(\alpha)-S_2(\alpha)=\frac{1}{r}\int_0^1 dx \, p_0(x,\alpha) \frac12 \ln^2 \left(\frac{x}{1-x}\right)
\label{Sig2_sym}
\\
V_2(\alpha)&=\frac{1}{r}\int_0^1 dx \,p_0(x,\alpha) \ln(x)\ln(q(x,\alpha)) 
\label{V2_sym}
\end{align}
\end{subequations}
with the corresponding symbols:
\begin{subequations}
\label{SYMB_symfunc}
\begin{align}
\begin{split}
{\cal S}\left[U_2(\alpha)\right]&= 
4\alpha\otimes \Big[\alpha\otimes\alpha-\alpha\otimes (1-\alpha^2)-(1-\alpha^2)\otimes\alpha+(1-\alpha^2)\otimes(1-\alpha^2)\Big]
\label{SYMB_U2_sym}
\end{split}
\\
{\cal S}\left[\Sigma_2(\alpha)\right]&= 2 \alpha\otimes \alpha\otimes \alpha
\label{SYMB_Sig2_sym}
\\
\begin{split}
{\cal S}\left[V_2(\alpha)\right]&=
2 \alpha\otimes (1-\alpha)\otimes \alpha
-4 \alpha\otimes (1-\alpha)\otimes (1+\alpha)
+2 \alpha\otimes \alpha\otimes (1-\alpha)
\\&-\alpha\otimes \alpha\otimes \alpha
+2 \alpha\otimes \alpha\otimes (1+\alpha)
-4 \alpha\otimes (1+\alpha)\otimes (1-\alpha)
+2 \alpha\otimes (1+\alpha)\otimes \alpha\,.
\label{SYMB_V2_sym}
\end{split}
\end{align}
\end{subequations} 
The definitions and the symbols of the non-symmetric functions $R_2$, $S_2$ and $\widetilde{S}_2$ can be found in table~\ref{tab:4leg3loop_transc_func_symb}.
It is interesting to observe, in analogy with eq.~(\ref{URS_symbol}), that the symbol for $U_2$ is related to those of the functions $R_2$ and $S_2$ as follows:
\begin{align}
\label{URS2_symbol}
{\cal S}\left[U_2(\alpha)\right]
=\left.\frac12 \,{\cal S}\left[R_2(\alpha)\right]\right\vert_{\alpha\to-\alpha^2}
=\left.-\frac12\, {\cal S}\left[S_2(\alpha)\right]\right\vert_{\alpha\to\alpha^2}
-4\alpha\otimes\alpha\otimes\alpha\,.
\end{align}
The latter relation is useful for deriving a compact expression for the integral (\ref{U2_sym}) in terms of polylogs, eq.~(\ref{U2_func}) below.

We note that there is an important difference between the first two functions, $U_2$ and $\Sigma_2$ on the one hand, and $V_2$ on the other: while the symbol of the former may be written solely in terms of the letters $\alpha$ and $1-\alpha^2$, the symbol of the latter requires using $1-\alpha$ and $1+\alpha$ (and yet it admits the $\alpha\to -\alpha$ symmetry). 
The difference between these two types of functions can also be seen upon taking an expansion about the lightlike limit, where $\alpha\to 0$. For $V_2(\alpha)$ this expansion yields:
\begin{equation}
\label{V2_lightlike}
V_2(\alpha)=-\frac16\ln^3(\alpha) -\frac{\pi^2}{6}\ln(\alpha) -3\zeta_3\,+\,
\pi^2\alpha\,+\,{\cal O}(\alpha^2)\,,
\end{equation}
which evidently includes an ${\cal O}(\alpha^1)$ term corresponding, according to eq.~(\ref{small_alpha}), to a square root of the Wilson-line mass squared, which violates the expected analytic properties. Such odd powers cannot appear in functions whose symbols consist exclusively of $\otimes \alpha$ and $\otimes (1-\alpha^2)$. This can be checked explicitly for the functions $U_2$ and $\Sigma_2$, whose expansions will be given in eq.~(\ref{lightlike_limit}) below. 
We will see that the contributions of three-loop subtracted webs considered in the next section can indeed be written in terms of $U_2$ and $\Sigma_2$ (as well as the lower-order functions $U_1$ and $R_0$).
 
Upon integration eqs.~(\ref{symfunc}) yield:
\begin{align}
\label{U2_func}
\begin{split}
U_2(\alpha)&=\zeta_3
-\text{Li}_3\left(\alpha^2\right)
-2 \text{Li}_3\left(1-\alpha^2\right)
+2 \text{Li}_2\left(\alpha^2\right)\ln \left(1-\alpha^2\right)
+2 \text{Li}_2\left(1-\alpha^2\right) \ln\left(1-\alpha^2\right) 
\\
&-2 \ln\left(1-\alpha^2\right)  \ln^2\left(\alpha\right)
+4 \ln (\alpha ) \ln^2\left(1-\alpha^2\right)
+\frac{2}{3} \ln^3\left(\alpha \right)
-\frac{\pi^2}{3}  \ln\left(1-\alpha^2\right) 
+\frac{\pi^2}{3}  \ln(\alpha)
\end{split}
\end{align}
and 
\begin{equation}
\label{Sigma_func}
\Sigma_2(\alpha)=\frac{1}{3} \ln (\alpha) \left(\ln ^2(\alpha)+\pi ^2\right)\,.
\end{equation}
We note that the functions, much like their symbols, can be conveniently written in terms of $\alpha$ and $1-\alpha^2$. We emphasize that the functions of eqs.~(\ref{U2_func}) and (\ref{Sigma_func}) correctly represent the defining integrals of eqs.~(\ref{U2_sym}) and (\ref{Sig2_sym}), respectively, for complex values of $\alpha$, and in particular, making the analytic continuation through the upper half plane, they are valid near the timelike axis.
These are the two weight-three functions which will be needed for the three-loop subtracted webs 1-2-2-1 and 1-1-1-3 in the next section.
The third function, $V_2$, which as explained above should not appear, is given in the appendix (eq.~(\ref{V2})).

A comment is due concerning the fact that the integral of 
eq.~(\ref{dilog_integral}), where the kernel involves a dilogarithm, does not occur in subtracted webs. 
Indeed this integral\footnote{We do not present the result for the integral nor its symbol since these are rather lengthy, and will not be needed.} does not fulfill our expectation that the letters in the symbol would be drawn from the set $1-\alpha^2$ and $\alpha$; instead these are drawn from a much longer list:
\[
\Big\{\alpha_1,1+\alpha_1,\alpha_2,1-\alpha_2,2+\alpha_1-\alpha_2,1+2\alpha_1,1-2\alpha_2,1-2\alpha_2-\alpha_1\alpha_2, 1+2\alpha_1-\alpha_1\alpha_2,\alpha_1-\alpha_2-2\alpha_1\alpha_2\Big\}\,,
\]
where we denoted the two $\alpha$ variables by $\alpha_1$ and $\alpha_2$.
From this list it is already clear that this integral violates the symbol-level symmetries  $\alpha\to -\alpha$ for the two $\alpha$ variables. 
We cannot rigorously exclude at this point the possibility that at some higher-loop order polylogarithms of ratios of $x_k$ variables would appear, but we consider this unlikely: because ${\cal G}_{n-1}$ can only depends on ratios $x_i/x_j$, such integrals are bound to contain a richer alphabet and violate the $\alpha\to -\alpha$ symmetries.   

Given our conclusion that polylogarithms do not appear in ${\cal G}_{n-1}$, the basis of functions appearing in subtracted webs (of the multiple-gluon-exchange type) is very limited and can be constructed to all orders by considering powers of logarithms. This is a rather remarkable simplification: due to the basic property of the logarithm, this would imply that the following conjecture must hold. 
\\
\\
{\bf Factorization conjecture:}\,\,\emph{all multiple-gluon-exchange subtracted webs can be written as sums of products of polylogarithms, each depending on a single $\alpha_{ij}$ variable.} 
\\
\\
In other words eq.~(\ref{subtracted_web_mge_kin}) does not entangle two or more kinematic variables through multiple polylogarithms. 
There is one potential caveat though: recall that beyond logarithms ${\cal G}_{n-1}$ may also contain some Heaviside functions. These should occur in more entangled webs than the ones computed in this paper, where $n$ gluons connect $n$ or fewer Wilson lines. While it is quite clear that for such webs we will need to extend the class of functions by allowing for a Heaviside function in the integrand, we expect that these would not violate the factorization of the final integrals into sums of products of polylogarithms. The reason for this is that these Heaviside functions are related to \emph{non-singular configurations} which occur in diagrams that are not maximally reducible, such crossed gluons (as in the 1-2-3 web) or the Esher straircase configuration of the 2-2-2 web (figures can be found in section A.2 or ref.~\cite{Gardi:2013ita}). 
This needs to be examined in detail, and it will be done as part of a forthcoming study of webs connecting three Wilson lines at three loops~\cite{Falcioni:2013tb}.
  
As another consistency check of the basis of functions we presented, consider the forward limit where $\alpha\to 1$. Recall that each rational factor $r(\alpha)=\frac{1+\alpha^2}{1-\alpha^2}$ has a simple pole in this limit, and since physically we do not expect a singularity for $\alpha=1$, the transcendental function must vanish there to compensate for this pole. We have already seen this at one and two loops in eqs.~(\ref{gamma_alpha_relation}) and (\ref{U1_result}), respectively. Taking the $\alpha\to 1$ limit in eqs. (\ref{Sigma_func}) and (\ref{U2_func}) we do indeed find that the two functions
$U_2(\alpha)$ and $\Sigma_2(\alpha)$ vanish.  

Finally, consider expansion around the lightlike limit, $\alpha\to 0$. The first observation is that in this limit the rational factor of any multiple-gluon-exchange web tends to one, since $r(\alpha)=1+{\cal O}(\alpha^2)$.
As discussed above, the transcendental functions are expected to give rise to a series of logarithms of plus, potentially, ${\cal O}(\alpha^2)$ power suppressed terms. This is indeed what we get for our basis functions: 
\begin{subequations}
\label{lightlike_limit}
\begin{align}
R_0(\alpha)&=\ln\alpha\,,\\
U_1(\alpha)&= -2 \ln^2(\alpha)-\frac{\pi^2}{3} \,+\, {\cal O}(\alpha^2)\,,\\
U_2(\alpha)&= \frac{2}{3} \ln^3(\alpha)+\frac{\pi^2}{3} \ln(\alpha)-\zeta_3\, +\, {\cal O}(\alpha^2)\,,\\
\Sigma_2(\alpha)&=\frac{1}{3} \ln^3 (\alpha) +\frac{\pi^2}{3} \ln(\alpha)\,,
\end{align}
\end{subequations}
where the absence of odd powers of $\alpha$ guaranties the correct analytic structure in the Wilson-line mass squared.

\section{Three-loop results \label{sec:3_loop}}

In the previous section we analysed the symmetries and analytic properties characterizing multiple-gluon-exchange subtracted webs which contribute directly to the soft anomalous dimension and deduced a basis of functions in terms of which three-loop subtracted webs may be expressed. 
We now return to perform explicit calculations and illustrate that the results meet our expectations. We focus here on the two three-loop webs that span four Wilson lines: the 1-2-2-1 and the 1-1-1-3 webs. These are the most interesting gluon-exchange webs in what concerns the lightlike limit of the soft anomalous dimension. These webs are also less entangled than ones that span fewer Wilson lines at the same loop order; this restricts the basis of relevant integrals to the ones we explicitly constructed above, those which do not contain any Heaviside functions. The extension to webs connecting three Wilson lines at three loops is under way~\cite{Falcioni:2013tb}. 

Given that the calculation is lengthy we relegate most of the details to the appendices. These involve many important issues with regards to the integration over the Wilson lines\footnote{A particularly subtle point dealt with in appendix~\ref{sec:3_loop_calc} is the $\epsilon$ expansion in the presence of end-point singularities.}
 and the simplifications achieved by forming the web at the first stage (appendix~\ref{sec:3_loop_calc}) and the subtracted web (appendix \ref{sec:combining}) at the final stage of the calculation. These steps will be useful for future calculations of other three-loop and higher-order webs~\cite{Falcioni:2013tb}. In this section we will only summarize the main results. We start with the 1-2-2-1 web and then consider the 1-1-1-3 one.

\subsection{The 1-2-2-1 subtracted web}

\begin{figure}[htb]
\begin{center}
\scalebox{0.8}{\includegraphics{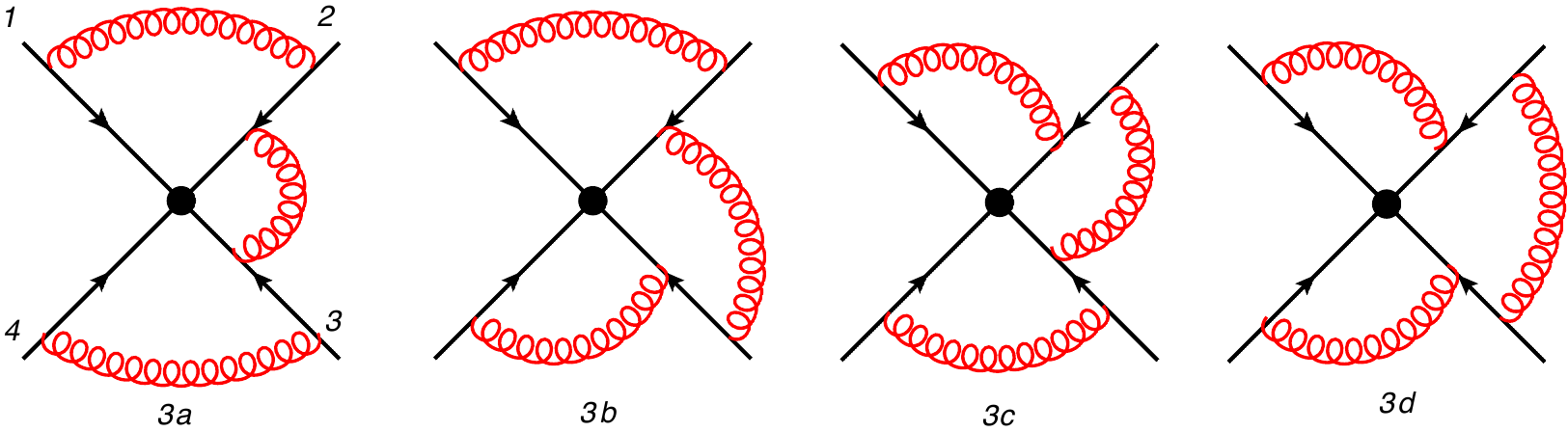}}
\caption{The four 3-loop diagrams forming the 1-2-2-1 web in which four eikonal lines are linked by three gluon exchanges.}
\label{3lfour}
\end{center}
\end{figure}

The web $W_{(1,2,2,1)}^{(3)}$ is composed of four diagrams, depicted in figure~\ref{3lfour}. Following refs.~\cite{Gardi:2010rn} and~\cite{Gardi:2011wa} (section 3.3) we know that this web contribute with a unique colour factor as follows:
\begin{eqnarray}
\label{3lfourmat}
\begin{split}
W_{(1,2,2,1)}^{(3)}&=\left(\begin{array}{c}{\cal F}(3a)\\{\cal F}(3b)\\{\cal F}(3c)
\\{\cal F}(3d)\end{array}\right)^T\frac{1}{6}\left(\begin{array}{rrrr}
1&-1&-1&1\\-2&2&2&-2\\-2&2&2&-2\\1&-1&-1&1
\end{array}\right)\left(\begin{array}{c}C(3a)\\C(3b)\\C(3c)\\C(3d)
\end{array}\right)\\
&\,\\
&=\frac16 \Big( C(3a)-C(3b)-C(3c)+C(3d) \Big) \Big(
    \mathcal{F}(3a)-2\mathcal{F}(3b) - 2\mathcal{F}(3c) + \mathcal{F}(3d)
  \Big) \,.
\end{split}
\end{eqnarray}
In appendix~\ref{sec:3_loop_calc} we compute the four diagrams and express them all as integrals over the three parameters $x$, $y$ and $z$ associated respectively with emission angles of the three gluons. 
Combining the four according to 
eq.~(\ref{3lfourmat}) and evaluating the nested commutator generated by the combination of the colour factors we obtain:
\begin{align}
\label{3lfourmat_}
\begin{split}
W_{(1,2,2,1)}^{(3)}
&=-f^{dce} f^{abe} T_1^aT_2^bT_3^cT_4^d\, \kappa^3\,
     \frac{3\Gamma(6\epsilon)}{4\epsilon}\int_0^1 {\rm 
      d}x{\rm d} y {\rm d}z\ p(x,\alpha_{12}) p(y,\alpha_{23}) p(z,\alpha_{34})\times\\&
 \Bigg\{ \left(\ln\left(\frac{y}{x}\right) +\ln\left(\frac{y}{z}\right) \right)+\epsilon
\Bigg[
12\Bigg({\rm{Li}}_2\left(-\frac{y}{x}\right)+{\rm{Li}}_2\left(-\frac{1-y}{z}\right) +\zeta_2\Bigg)
\\&\hspace*{80pt}+\ln^2\left(\frac{1-y}{z}\right) + 8\ln\left(\frac{1-y}{z}\right) \ln\left(\frac{y}{x}\right) +
\ln^2\left(\frac{y}{x}\right) \Bigg] +{\cal O}(\epsilon^2)\Bigg\}\,,
\end{split}
\end{align}  
where we used the definition of $\kappa$ from eq.~(\ref{kappa1}) and $p(x,\alpha)$ from eq.~(\ref{p0def_}).  The result is consistent with the general form of multi-gluon-exchange webs described by eq.~(\ref{mge_kin}). The function of $x$, $y$ and $z$ in the curly brackets is the 1-2-2-1 web kernel.
 Note that the web, as it stands, has a double pole, which exactly matches the commutator term involving its subdiagrams, as shown in ref.~\cite{Gardi:2011yz}. The contribution to the anomalous dimension emerges from the ${\cal O}(\epsilon^{-1})$ term which is displayed explicitly in eq.~(\ref{3lfourmat_2_expanded}); note that besides the contribution of the ${\cal O}(\epsilon)$ term in the curly brackets of (\ref{3lfourmat_}), this includes logarithms such as $\ln q(x,\alpha_{12})$ from the expansion of the $d$-dimensional propagators which multiply the ${\cal O}(\epsilon^0)$ term in the curly brackets in (\ref{3lfourmat_}).

The next stage of the calculation, which is presented in detail in appendix \ref{sec:combining}, is to form the subtracted web, combining the web of eq.~(\ref{3lfourmat_}), or eq.~(\ref{3lfourmat_2_expanded}), with commutators of lower-order webs according to 
\begin{align}
\label{sub_web_3l_}
\begin{split}
\overline{w}^{(3,-1)}&= w^{(3,-1)}
-\frac12\left[w^{(1,0)},w^{(2,-1)}\right]
-\frac12\left[w^{(2,0)},w^{(1,-1)}\right]
\\&
-\frac16\left[w^{(1,0)},\left[w^{(1,-1)},w^{(1,0)}\right]\right]
-\frac16\left[w^{(1,-1)},\left[w^{(1,1)},w^{(1,-1)}\right]\right]\,,
\end{split}
\end{align}
which is the combination entering the anomalous dimension coefficient ${\Gamma_4^{(3)}}$. It should be understood that by ${w}^{(2,k)}$ and ${w}^{(1,k)}$ in eq.~(\ref{sub_web_3l_}) we refer to a sum over all relevant webs corresponding to the subdiagrams of the three-loop web under consideration. In the case of the 1-2-2-1 considered here, this sum is spelled out explicitly in eq.~(\ref{w3_122331}). Importantly, all the terms give rise to the same colour factor, and the same general form of kinematic integrals, as they all involve the same three gluon exchanges: between the pairs of lines (1,2), (2,3) and (3,4). This facilitates combining 
the terms under the integral in eq.~(\ref{122334_factorization}) which is consistent with the general form anticipated in eq.~(\ref{subtracted_web_mge_kin}).
Importantly, in eq.~(\ref{122334_factorization}) we observe the cancellation of the dilogarithmic term present in the web kernel of eq.~(\ref{3lfourmat_}), a cancellation which is due to the interplay between the web and the commutator terms. As discussed in the previous section this cancellation is necessary for the subtracted web to be consistent with the predicted analytic structure.

All the relevant expressions for $w^{(n,k)}$ are summarised in appendix \ref{sec:summary_of_integrals}. Substituting these into eq.~(\ref{sub_web_3l_}) one arrives, after some algebra which we relegate to 
appendix \ref{sec:combining}, to the final result for the 1-2-2-1 subtracted web, which reads:
\begin{align}
\begin{split}
\overline{w}^{(3,-1)}_{(122334)}&=-\frac16 f^{abe}f^{cde}T_1^aT_2^bT_3^cT_4^d\,\,
\left(\frac{1}{4\pi}\right)^3\,
G(\alpha_{12},\alpha_{23}, \alpha_{34})\,
\end{split}
\end{align}
with
\begin{align}
\label{G1221_sym}
\begin{split}
&G(\alpha_{12},\alpha_{23}, \alpha_{34})=r(\alpha_{12})\,r(\alpha_{23})\,r(\alpha_{34})\, \Bigg[-8 U_2(\alpha_{12})\, \ln \alpha_{23}\, \ln \alpha_{34}
-8 U_2(\alpha_{34}) \,\ln \alpha_{12}\, \ln \alpha_{23}
\\&\hspace*{30pt}+16\Big(U_2(\alpha_{23})-2\Sigma_2(\alpha_{23})\Big) \,\ln \alpha_{12}\, \ln \alpha_{34}
\\&\hspace*{30pt}-2\ln \alpha_{12} \, U_1(\alpha_{23}) \, U_1(\alpha_{34})
-2\ln \alpha_{34}\, U_1(\alpha_{12}) \, U_1(\alpha_{23})
+4\ln \alpha_{23}\, U_1(\alpha_{12})\, U_1(\alpha_{34})\Bigg]\,,
\end{split}
\end{align}
where the rational factor is a product of three $r(\alpha)$ factors of eq.~(\ref{r}), each of them associated with one of the gluons, while the transcendental part of  $G(\alpha_{12},\alpha_{23},\alpha_{34})$, which is a pure function of weight 5, is expressed in terms of the basis of functions constructed in the previous section, with $U_1(\alpha)$ defined as an integral in eq.~(\ref{U1_sym}) and given explicitly in eq.~(\ref{U_SR_calculation}) and  $U_2(\alpha)$ and $\Sigma_2(\alpha)$ are defined in eqs.~(\ref{U2_sym}) and~(\ref{Sig2_sym}), respectively, and given explicitly in eqs.~(\ref{U2_func}) and~(\ref{Sigma_func}), respectively.
We emphasise that the fact that the final result may be written in terms of a sum of products of  polylogarithms, each depending on a single cusp angle (a single $\alpha_{ij}$) is a highly non-trivial property of subtracted webs, which does not hold for individual diagrams, nor for the non-subtracted web in this case, owing to the presence of the dilogarithm ${\rm{Li}}_2\left(-\frac{y}{x}\right)$ in eq.~(\ref{3lfourmat_}).
As explained in the previous section this remarkably simple form of the subtracted web is associated with the purely logarithmic nature of the function $\cal G$ in eq.~(\ref{subtracted_web_mge_kin}), and it is ultimately related to analyticity and crossing symmetry.

\subsection{The 1-1-1-3 subtracted web}

\begin{figure}[htb]
\begin{center}
\scalebox{0.8}{\includegraphics{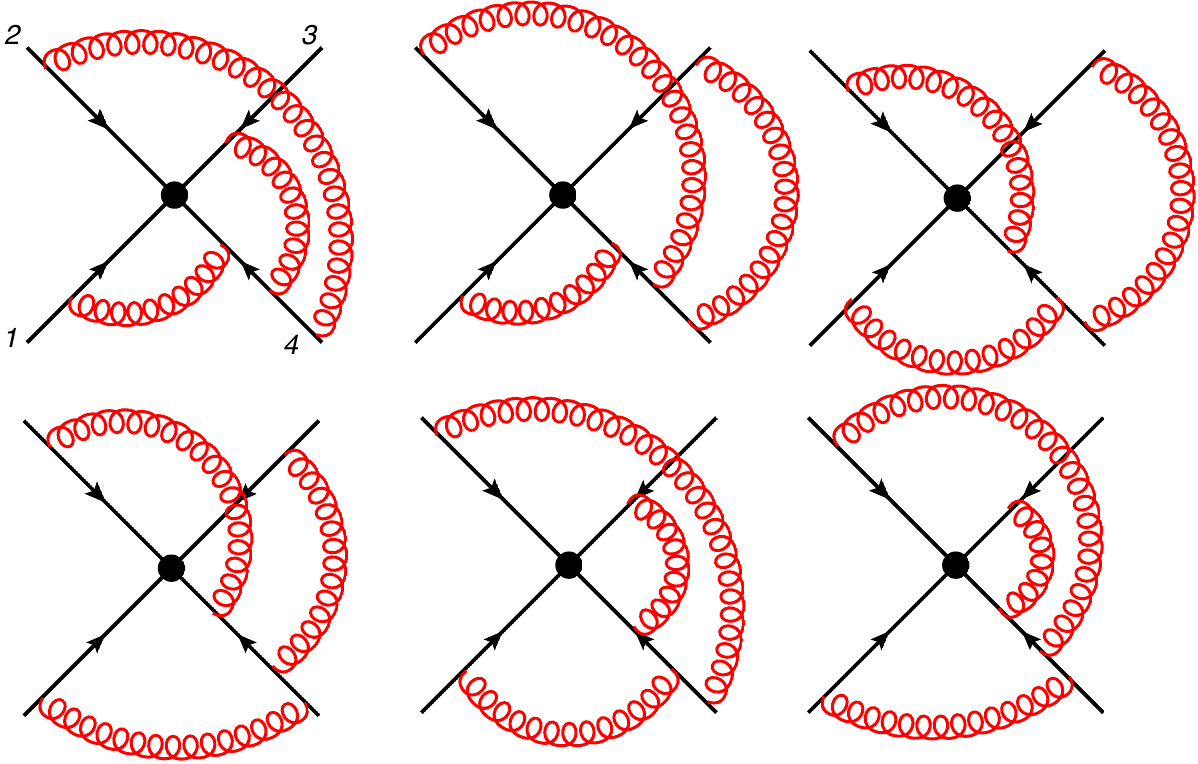}}
\caption{The six 3-loop diagrams forming the 1-1-1-3 web in which four eikonal lines are linked by three gluon exchanges. We label the diagrams in the first row from left to right by A, B and C, and the ones in the second row by D, E and F, respectively.}
\label{1113}
\end{center}
\end{figure}

Let us now turn to consider the 1-1-1-3 web in which the three gluons connect to line 4. The six (3!) diagrams forming this web are shown in figure~\ref{1113}; the six are denoted respectively by $A$ through $F$.
The mixing matrix for this web was presented in 
ref.~\cite{Gardi:2013ita} (see appendix A.1.2 there). In contrast to the 1-2-2-1 case, this matrix has rank two, thus giving rise to two independent (connected) colour factors, each of which involves a different linear combination of kinematic integrals. The result reads:
\begin{align}
\label{W1113_P_result}
\begin{split}
W_{(1,1,1,3)}(\alpha_{14},\alpha_{24},\alpha_{34})  =T_1^{a}T_2^{b}T_3^{c}T_4^d\, \Bigg[&
f^{bce}f^{ade} {\cal F}_{(1,1,1,3); 1}(\alpha_{14},\alpha_{24},\alpha_{34}) 
\\&+
f^{ace}f^{bde} {\cal F}_{(1,1,1,3); 2}(\alpha_{14},\alpha_{24},\alpha_{34}) \Bigg]\,,
\end{split}
\end{align}
with
\begin{subequations}
\label{F1113_12}
\begin{align}
{\cal F}_{(1,1,1,3); 1}(\alpha_{14},\alpha_{24},\alpha_{34}) =& \,\frac{1}{6}\,\Big(-{\cal F}_A+2{\cal F}_B-{\cal F}_C-{\cal F}_D-{\cal F}_E+2{\cal F}_F\Big)\,,\\
{\cal F}_{(1,1,1,3); 2}(\alpha_{14},\alpha_{24},\alpha_{34}) =& \, \frac{1}{6}\,\Big(-{\cal F}_A-{\cal F}_B+2{\cal F}_C-{\cal F}_D+2{\cal F}_E-{\cal F}_F\Big)\,.
\end{align}
\end{subequations}
The calculation of the integrals for the six diagrams is presented in detail in appendix~\ref{sec:3_loop_calc}.
As we have seen in previous examples, we combine the integrands under the triple integral over $x$, $y$ and $z$, which correspond respectively to the angles of the three gluon exchanges between the lines (1,4), (2,4) and (3,4).  The resulting combinations of kinematic integrals in eq.~(\ref{F1113_12}) are summarised in appendix~\ref{sec:summary_of_integrals} in eqs.~(\ref{ijklll}) and (\ref{phi3}).  Note that similarly to the 1-2-2-1 case, the web as a whole has a double pole in $\epsilon$, while only the single pole would contribute to the anomalous dimension.  Also here the ${\cal O}(1/\epsilon)$ term of the non-subtracted web contains a dilogarithm, which upon integration over the $x$, $y$ and $z$ variables would generate multiple polylogarithms; also here these terms will cancel upon forming the subtracted web combination.

As in the previous example, the calculation proceed by using eq.~(\ref{sub_web_3l_}) to form the subtracted web, eq.~(\ref{sub_web_1113}).
The details of the calculation are again relegated to 
appendix~\ref{sec:combining} and the final result reads:
\begin{align}
\begin{split}
\overline{w}^{(3,-1)}_{(123444)}=-\frac16
\left(\frac{1}{4\pi}\right)^3\,
T_1^aT_2^bT_3^cT_4^d \bigg[
&f^{ade}f^{bce} F(\alpha_{14},\alpha_{24},\alpha_{34})
+f^{ace}f^{bde}
 \, F(\alpha_{24},\alpha_{14},\alpha_{34})
\bigg]
\end{split}
\end{align}
with
\begin{align}
\label{F1113_sym}
\begin{split}
F&(\alpha_{14},\alpha_{24},\alpha_{34})=
r(\alpha_{14})\,r(\alpha_{24})\,r(\alpha_{34})
\,\,\times \\&\bigg[
8 U_2(\alpha_{14})\, \ln \alpha_{24}\, \ln \alpha_{34}
+8 U_2(\alpha_{34}) \,\ln \alpha_{14}\, \ln \alpha_{24}
-16 U_2(\alpha_{24})\,\ln \alpha_{14}\, \ln \alpha_{34}
\\&+2\ln \alpha_{14} \, U_1(\alpha_{24}) \, U_1(\alpha_{34})
+2\ln \alpha_{34}\, U_1(\alpha_{14}) \, U_1(\alpha_{24})
-4\ln \alpha_{24}\, U_1(\alpha_{14})\, U_1(\alpha_{34})\bigg]\,.
\end{split}
\end{align}
Again, as anticipated, there are three rational factors $r(\alpha)$ each corresponding to a single gluon exchange, multiplying a transcendental function of weight 5 which is written as a sum of products of functions of single kinematic variables using basis of section \ref{sec:functions}, where the symbol-level crossing symmetry is manifest. 
Note that $F(\alpha_{14},\alpha_{24},\alpha_{34})$ is symmetric under swapping the third and first arguments, but not the second. Also note the similarity between this result for the 1-1-1-3 web and that of the 1-2-2-1 web in eq.~(\ref{G1221_sym}): apart from the necessarily different set of arguments (here all gluon connect to line 4) and an overall sign, the only difference is the $\Sigma_2$ term which appears in the 1-2-2-1 case but not here. 

\section{Conclusions\label{sec:conclusions}}

In this paper we took a step towards calculating the soft anomalous dimension for multi-leg scattering at three-loop order. The state of the art, as of a few years ago, is two loops~\cite{Aybat:2006wq,Aybat:2006mz,Ferroglia:2009ii,Mitov:2009sv,Mitov:2010xw}, where up to three  Wilson lines are connected. 
Here we computed explicitly the three-loop contributions to the soft anomalous dimension of a product of non-lightlike Wilson lines from a specific class of webs consisting of multiple gluon exchanges which connect four lines, the 1-2-2-1 and the 1-1-1-3 webs. These are the first three-loop results to become available beyond the case of the angle-dependent cusp anomalous dimension which was studied recently in the context of the ${\cal N}=4$ theory~\cite{Korchemsky:1987wg,Kidonakis:2009zc,Correa:2012nk,Henn:2012qz,Henn:2012ia}.
Our final results for these webs, presented in section~\ref{sec:3_loop}, display a remarkably simple structure involving a pure function of weight 5, which is a sum of products of specific polylogarithmic functions, each depending on a single cusp angle.

In order to complete the computation of the angle-dependent soft anomalous dimension several other webs need to be evaluated; the complete list can be found in ref.~\cite{Gardi:2013ita}. Specifically, the remaining webs that connect four Wilson lines, thus having the same colour factors as the webs we computed here, fall into two categories: those of figure~\ref{Connected_four}, which are comprised of a single connected graph with either one four-gluon vertex or two three-gluon vertices, and the 1-1-1-2 web of figure~\ref{two_pieces} where each diagram consists of two connected subdiagrams, one of which involving a three-gluon vertex. Each of these types of webs requires different techniques, and their calculation is in progress.

Besides computing specific diagrams we developed in this paper a general strategy for computing webs that consist of multiple gluon exchanges between any number of non-lightlike Wilson lines, and elucidated the analytic structure of the contributions of such webs to the anomalous dimension. The calculation itself is formulated entirely in configuration space, where loop-momentum integrals are replaced by integrals along the Wilson lines. By choosing appropriate integration variables the latter are done in two steps, as summarised in eqs.~(\ref{general_form_of_int}) and~(\ref{mge_kin}). In the first step all integrals associated with the distances of the gluons to the hard-interaction vertex are performed, making use of an infrared regulator. The result is a polylogarithmic kernel multiplying a product of propagator-related functions, $p(x,\alpha)$, each depending on the emission angle $x$ of a given gluon, and the exponent of the corresponding cusp angle between the respective Wilson lines, $\alpha$. In the second step the integrals over the gluon emission angles are performed; in practice this is only done after combining diagrams into webs, and further combining webs with commutators of their subdiagrams to form subtracted webs. It is at this stage that regularization independence is recovered, along with the analytic properties associated with crossing symmetry.

Organising the calculation in terms of webs, and then subtracted webs, has been absolutely essential for the present work. Recall that the notion of a web as a set of diagrams which are interrelated by permutations emerged not long ago through the formulation of a diagrammatic approach to exponentiation in refs.~\cite{Gardi:2010rn,Mitov:2010rp}.
Progress was achieved there owing to the replica trick of statistical physics which led to a general algorithm for determining web mixing matrices. The study of these~\cite{Gardi:2011wa,Gardi:2011yz,Gardi:2013ita,Dukes:2013wa,Dukes:2013gea} proceeded using both combinatorial methods and the connection with the renormalizability of the Wilson-line vertex. The most striking feature of webs is that -- despite the fact that they contain disconnected, often reducible diagrams -- their colour factors always correspond to connected graphs~\cite{Gardi:2013ita}. More subtle is the fact that webs renormalize independently. This implies that the kinematic integrals in a web conspire to cancel certain subdiveregnces, and all remaining multiple poles match precisely the commutators of lower-order webs~\cite{Gardi:2011yz}. The contributions to the anomalous dimension, summarised by eq.~(\ref{Gamma_n}), are associated with the single $1/\epsilon$ pole.  These properties have been essential for the present calculation.

Doing the calculation we have seen that combining diagrams into webs at the integrand level has numerous advantages. The first, rather obvious advantage is that there are fewer integrals to evaluate; this is because the rank of a mixing matrix is always smaller than its dimension. More important is the fact that the web combinations one computes are projections onto particular connected colour factors. Ref.~\cite{Gardi:2013ita} provided a systematic way of determining a basis for these colour factors. This is important because a gauge invariant result is only obtained upon summing all the webs that contribute with a given colour factor. A further advantage of organising the calculation in terms of webs is the possibility to predict the precise structure of all multiple poles, $1/\epsilon^k$ for $k\geq2$, in webs through commutators of lower-order webs~\cite{Gardi:2011yz}. This is a highly constrained structure, facilitating non-trivial consistency checks.

All of these points were fully appreciated when setting up the present calculation. What was less obvious a priori were the advantages of the further step of combining webs with commutators to form subtracted webs. Such a combination is natural, since all combined terms have the same set of subdiagrams, hence the common colour factors. Similarly to the previous step this combination can be done before integrating over the gluon emission angles, owing to the common set of propagators. The key point is that it is only at this stage that regularization invariance is restored, and this is linked with the restoration of crossing symmetry. This amounts to a major simplification whereby all polylogarithmic terms in the kernel cancel. As a consequence the final result is free of multiple polylogarithm that couple the kinematic variables associated with different cusp angles. Furthermore, the fact crossing symmetry is violated before forming the subtracted web combinations turns out to be very useful in practice, since it provides stringent checks of the new
${\cal O}(\epsilon^{-1})$ terms through their correspondence with commutators of lower-order webs.

Analysing the structure of multiple-gluon-exchange integrals we have shown in section~\ref{sec:functions} that each web is associated with a unique rational factor which is common to all the diagrams in a web, and all the contributions to  the corresponding subtracted web. This factor is simply the product of~$r(\alpha_{ij})$ of eq.~(\ref{r}) for each gluon exchange.
These rational factors are particularly important for timelike kienamtics near absolute threshold where $\alpha_{ij}\to -1$, since they encapsulate the coulomb singularity. They also have a special role for space-like kinematics near the forward limit, $\alpha\to 1$, where the pole cancels by a zero of the corresponding transcendental function.
We comment that for webs that do not consist exclusively of gluon exchanges, and have some three- or four-gluon vertices off the Wilson lines, the number of factors of $r(\alpha_{ij})$ is lower than the loop order, and it seems to be related instead to the number of connected subdiagrams\footnote{We emphasise that similarly to other diagrammatic observations this comment concerns the result in the Feynman gauge only.}. An example is provided by the two-loop three-leg web of eq.~(\ref{3g}) which has just one factor of $r(\alpha_{ij})$.
It is likely that the rational functions of general multi-loop webs would be rather more complex than seen in gluon-exchange diagrams, but this  deserves further study. 

Some important comments are due concerning the lightlike limit, where $\alpha_{ij}\to 0$. The first observation is that in this limit the rational factors tend to one (up to power suppressed terms) and thus become irrelevant. 
Consequently the lightlike soft anomalous dimension receives, on equal footing, contributions from terms which have different rational factors away from this limit. As has already been noted in refs.~\cite{Ferroglia:2009ii,Mitov:2009sv,Mitov:2010xw} the terms originating in the two webs of figure~\ref{2loopfig} are separately logarithmically divergent in this limit, but they conspire to cancel, consistently with the calculation of ref.~\cite{Aybat:2006wq,Aybat:2006mz} and the dipole 
formula~\cite{Becher:2009cu,Gardi:2009qi,Becher:2009qa,Dixon:2009gx,Dixon:2009ur}. 
The implication for three loops is that all webs that span four Wilson lines need to be computed in order to deduce whether there are corrections to the dipole formula. The three-loop webs computed here, eqs. (\ref{G1221_sym}) and (\ref{F1113_sym}), are also logarithmically divergent in the lightlike limit; in fact, all the terms which do not vanish in this limit are logarithmically divergent. Similarly to the two-loop case such terms are expected to appear in other webs and conspire to cancel in the sum. In contrast to the two-loop case, non-trivial functions of confomally-invariant cross ratios $\frac{\alpha_{ij}\alpha_{kl} }{\alpha_{ik}\alpha_{jl}}$ may survive in this limit~\cite{Gardi:2009qi,Becher:2009qa,Dixon:2009ur,Gardi:2009zv,Gehrmann:2010ue,Bret:2011xm,DelDuca:2011ae,Ahrens:2012qz,Naculich:2013xa,Caron-Huot:2013fea}. To determine them we need to complete the calculation of the diagrams in figures~\ref{Connected_four} and~\ref{two_pieces}.

The analysis of multiple-gluon-exchange integrals in section~\ref{sec:functions} led, aside from a determination of the rational factors, to a detailed understanding of the transcendental functions appearing in the corresponding subtracted webs. Considering the analytic structure of these functions in terms of $\alpha_{ij}$, which is depicted in figure~\ref{alpha_plane}, along with the crossing symmetry which relates spacelike and timelike kinematics, we were led to conjecture that the alphabet of the symbol of subtracted webs is restricted to $\otimes \alpha_{ij}$ and $\otimes (1-\alpha_{ij}^2)$. This in turn severely restricts the space of functions which may appear in the final results for the contributions of this class webs to the anomalous dimension: these, we conjectured, can always be written as sums of products of specific polylogarithmic functions, each depending on a single $\alpha_{ij}$ variable. The basis of these functions can be systematically constructed at any loop order by considering the range of functions of the appropriate weight that may appear in the relevant subtracted web kernel of eq.~(\ref{subtracted_web_mge_kin}). This, we argued, consists exclusively of powers of logarithms and Heaviside functions. Specifically, for $n$ gluon exchange webs that span $(n+1)$ Wilson lines, Heaviside functions cannot appear, restricting the basis of functions further. At three loops this construction yields just two basis functions $U_2(\alpha_{ij})$ and $\Sigma_2(\alpha_{ij})$, which together with the functions already encountered at one and two loops,
were indeed sufficient to express the three-loop webs we computed. We now have the means to construct a similar basis for other gluon-exchange webs at three loops and beyond. Specifically, work on the extension of the basis for webs connecting three Wilson lines at three loops, where Heaviside functions are allowed, is under way~\cite{Falcioni:2013tb}. 

Beyond its inherent significance, the study of infrared singularities is also a natural testing grounds for techniques to explore the structure of gauge-theory scattering amplitudes in general. This work significantly benefited from the recent progress in computing iterated polylogarithmic integrals using the symbol technique. This has proven important for understanding the analytic structure of infrared-singular corrections to amplitudes. We believe that our results, notably with respect to the realisation of crossing symmetry, will feed into the study of non-singular corrections as well.

In conclusion, we have taken a step towards computing the soft anomalous dimension for multi-leg scattering at the three-loop order, and developed methods that would facilitate more calculations of this kind. 
The progress we made here on evaluating web integrals was facilitated by, and is highly complementary to, the recent advances in soft gluon exponentiation and the generalization of the non-Abelian exponentiation theorem to the multi-leg case. We acquired some understanding of the kinematic dependence and the analytic structure of webs, and we believe that much further progress in this direction is now achievable.

\vspace*{38pt}
\noindent {\bf Acknowledgements}
\vspace*{10pt}

\noindent
I would like to thank Adrian Bodnarescu for verifying many of the results of appendix~\ref{sec:3_loop_calc}, and to  Samuel Abreu, $\O$yvind Almelid, Ruth Britto, Lance Dixon, Claude Duhr, Mark Dukes, Mark Harley, Johannes Henn, Gregory Korchemsky, Lorenzo Magnea, Mairi McKay, Jenni Smillie and Chris White, for stimulating discussions and collaboration on related projects. 
This research is supported by the STFC grant ``Particle Physics at the Tait Institute''.

\newpage
\appendix


\section{Summary of integrals through three loops\label{sec:summary_of_integrals}}

Let us now summarize the results of the webs computed in this paper in a form that will be useful for forming the subtracted web combination in which each web is combined with the commutators of the webs corresponding to its subdiagrams. To this end we keep the Wilson-line indices general, for example, we denote the 1-2-2-1 web by $W^{(3)}(ijjkkl)$ meaning that line $i$ has a single attachment, line $j$ has two, line $k$ has two, and line $l$ one.  
We start with the one-loop summary, continue with two loops and end with the thee-loop webs.

\subsection{One-loop integral summary}

Starting with the one-loop result of eq.~(\ref{eq:Fijoneloop1}) we have:
\begin{align}
  \label{eq:Fijoneloop_appendx}
  \begin{split}
&W^{(1)}(ij)=T_i\cdot T_j {\cal F}^{(1)}_{ij}(\alpha_{ij},\mu^2/m^2,\epsilon)
\\
 & {\cal F}^{(1)}_{ij}(\alpha_{ij},\mu^2/m^2,\epsilon)
  =
   \kappa
  \,\Gamma(2\epsilon) \,\int_0^1dx \, p(x,\alpha_{ij})\,\, 
  \end{split}
\end{align}
where $\kappa$ is defined in eq.~(\ref{kappa1})
and
\begin{equation}
p(x,\alpha_{ij})\equiv \gamma_{ij} P(x,\gamma_{ij})\,,
\label{pdef1}
\end{equation}
with
\begin{equation}
\label{gamma_alpha_relation_appendix}
P(x,\gamma_{ij})\equiv\Big[q(y,\alpha_{ij})\Big]^{\epsilon-1}\,;\qquad
q(x,\alpha_{ij})=x^2+(1-x)^2-x(1-x)\gamma_{ij}\,
;\qquad
-\gamma_{ij}=\alpha_{ij}+\frac{1}{\alpha_{ij}}
\end{equation}
For later convenience we also define the $\epsilon\to 0$ limit of $p$ as
\begin{equation}
p_0(x,\alpha_{ij})\equiv\frac{\gamma_{ij} }{q(x,\alpha_{ij})}\,,
\label{pdef}
\end{equation}
Upon integration, eq.~(\ref{eq:Fijoneloop_appendx}) yields a hypergeometric function as follows:
\begin{align}
  \label{eq:Fijoneloop_int}
  \begin{split}
{\cal F}^{(1)}_{ij}(\alpha_{ij},\mu^2/m^2,\epsilon)
  =
   \kappa
  \,\Gamma(2\epsilon) \,\int_0^1dx \, p(x,\alpha_{ij})\,\, 
  &=\kappa
  \,\Gamma(2\epsilon) \,\gamma_{ij}\, \,\,
  _2F_1\left([1,1-\epsilon],[3/2],\frac12+\frac{\gamma_{ij}}{4}\right)\,,
  \end{split}
\end{align}

Upon taking the $\epsilon$ expansion of the one-loop result we get
\begin{subequations}
\label{Ri_def}
\begin{align}
{\cal F}^{(1,-1)}_{ij}(\alpha_{ij})&= \frac{\kappa}{2}\, \int_0^1dx \, p_0(x,\alpha_{ij}) =\,\kappa\, r(\alpha_{ij})\, \ln \alpha_{ij}\\
{\cal F}^{(1,0)}_{ij}(\alpha_{ij}) &=\frac{\kappa}{2}\,\int_0^1dx \, p_0(x,\alpha_{ij})\, \ln q(x,\alpha_{ij}) = \,\kappa\, r(\alpha_{ij})\, R_1(\alpha_{ij})
\\
{\cal F}^{(1,1)}_{ij}(\alpha_{ij}) &= 
\frac{\kappa}{2}\, \int_0^1dx \, p_0(x,\alpha_{ij})\, \frac12 \ln^2 q(x,\alpha_{ij}) =\,\kappa\, r(\alpha_{ij})\, R_2(\alpha_{ij})
\end{align}
\end{subequations}
where we denoted the rational factor by eq.~(\ref{r}) and $R_i$ in the first three orders are given in eq.~(\ref{R12}). Note that here we have not expanded the overall factor $\kappa$ of eq.~(\ref{kappa1}) and we replaced $\Gamma(2\epsilon)$ by $1/(2\epsilon)$; a similar replacement is systematically applied to the overall $\Gamma$ factors for higher order webs and the resulting constants ($\gamma_E$, $\zeta_2$) cancel in the final results.

\subsection{Two-loop integral summary}

At two loops, starting with eq.~(\ref{w121_phi2}), we have:
\begin{align}
\begin{split}
W^{(2)}(ijjk)&=\frac12 \left(C(2a)-C(2b)\right)\left({\cal F}(2a)-{\cal F}(2b)\right)= {\rm i}f^{abc} T_i^aT_j^cT_k^b \, {\cal F}^{(2)}_{ijjk}(\alpha_{ij},\alpha_{jk},\mu^2/m^2,\epsilon)
\end{split}
\end{align}
with
\begin{align}
\label{two_loop_kin}
\begin{split}
 {\cal F}^{(2)}_{ijjk}(\alpha_{ij},\alpha_{jk},\mu^2/m^2,\epsilon)
&=\kappa^2
\,\Gamma(4\epsilon)\, 
\int_0^1dx \int_0^1 dz \,p(x,\alpha_{ij})\, p(z,\alpha_{jk})\\&\qquad\times  \left[
\phi_{2}^{(0)}(x,z)+\phi_{2}^{(1)}(x,z)\epsilon+{\cal O}(\epsilon^2)
\right]\\&
=\kappa^2
\,\Gamma(4\epsilon)\, 
\int_0^1dx \int_0^1 dz \,p(x,\alpha_{ij})\, p(z,\alpha_{jk})\\&\qquad\times \left[
\ln\frac{x}{z}+\left(4{\rm Li}_2\left(-\frac{x}{z}\right)+\ln^2\left(\frac{x}{z}\right)+2\zeta_2
\right)\epsilon+{\cal O}(\epsilon^2)
\right]
\end{split}
\end{align}
The $\epsilon$ expansion of this result is:
\begin{align}
\label{two_loop_kin_expanded}
\begin{split}
 {\cal F}^{(2,-1)}_{ijjk}(\alpha_{ij},\alpha_{jk})&=
\left(\frac{\kappa}{2}\right)^2
\, 
\int_0^1dx \int_0^1 dz \,p_0(x,\alpha_{ij})\, p_0(z,\alpha_{jk})\, \phi_2^{(0)}(x,z)
\\
{\cal F}^{(2,0)}_{ijjk}(\alpha_{ij},\alpha_{jk})&=\left(\frac{\kappa}{2}\right)^2
\int_0^1dx \int_0^1 dz \,p_0(x,\alpha_{ij})\, p_0(z,\alpha_{jk})\\&\qquad \times \left[ \phi_2^{(1)}(x,z) +\phi_2^{(0)}(x,z) \left(\ln q(x,\alpha_{ij})+\ln q(z,\alpha_{jk})\right)\right]
\end{split}
\end{align}
where 
\begin{subequations}
\begin{align}
\begin{split}
\phi_2^{(0)}(x,z)&=\ln\frac{x}{z}\\
\phi_2^{(1)}(x,z)&=4{\rm Li}_2\left(-\frac{x}{z}\right)+\ln^2\left(\frac{x}{z}\right)+2\zeta_2
\end{split}
\end{align}
\end{subequations}

Note that by Bose symmetry ${\cal F}^{(2)}_{ijjk}$ is antisymmetric under interchanging its arguments $\alpha_{ij}$ and $\alpha_{jk}$, similarly to its colour factor. As a consequence $\phi_{2}^{(n)}(x,z)$ are all antisymmetric with respect to $x$ and $z$, as can be explicitly verified for the expressions given above (note that the function $f(\alpha)=4{\rm Li}_2(-\alpha)+\ln^2(\alpha)+2\zeta_2$ admits $f(-1/\alpha)=-f(\alpha)$).

The integrals arising in eq.~(\ref{two_loop_kin}) are distinguished from the ones of the one-loop case only though the appearance of logarithmic and polylogarithmic functions on $x$ and $z$ in the integrand. 
Logarithmic ones factorize in such a way that each integral involves a single cusp angle: they never couple the two cusp angles. Polylogarithms do couple the cusp angles, but these do not appear in subtracted web kernels, so we will not need to consider them.  

The integrals we shall need are the following:
\begin{align}
\int_0^1 dx \ln(x) p_0(x,\alpha)=-\frac{1}{2}\frac{1+\alpha^2}{1-\alpha^2}\,S_1(\alpha)
\end{align}
and
\begin{align}
\label{S2_def}
\begin{split}
\int_0^1 dx \frac12 \ln^2(x) p_0(x,\alpha)
&=-\frac12 \frac{1+\alpha^2}{1-\alpha^2} S_2(\alpha)
\end{split}
\end{align}
where
\begin{subequations}
\label{S12}
\begin{align}
S_1(\alpha)&=-2\left[-2 {\rm Li}_2(\alpha)+\frac12 \ln^2(\alpha)-2 \ln(1-\alpha) \ln(\alpha)+2\zeta_2\right]
\\
\begin{split}
S_2(\alpha)&=-\bigg[-4 \text{Li}_3(1-\alpha)-2 \text{Li}_3(\alpha)+4 \text{Li}_2(1-\alpha) \ln (1-\alpha)\\&\qquad +4 \text{Li}_2(\alpha) \ln (1-\alpha)+\frac{\ln ^3(\alpha)}{3}-\ln (1-\alpha) \ln ^2(\alpha)\\&\qquad
+4 \ln ^2(1-\alpha) \ln (\alpha)+\frac{1}{3} \pi ^2 \ln (\alpha)-\frac{2}{3} \pi ^2 \ln (1-\alpha)+2 \zeta_3\bigg]
\end{split}
\end{align}
\end{subequations}
A similar integral, which will be important at three loops is the following:
\begin{align}
\label{S1_append}
\begin{split}
\int_0^1 dx  \ln(x)\ln(1-x) p_0(x,\alpha)&= - \frac{1+\alpha^2}{1-\alpha^2}\, \widetilde{S}_2(\alpha)
\end{split}
\end{align}
where
\begin{align}
\label{widetildeS2}
\begin{split}
\widetilde{S}_2(\alpha)&=
4 \text{Li}_3(1-\alpha)
+2 \text{Li}_3(\alpha)
-4 \text{Li}_2(1-\alpha) \ln (1-\alpha)
-4 \text{Li}_2(\alpha) \ln (1-\alpha)
\\&\qquad
-4 \ln (\alpha) \ln ^2(1-\alpha)
+ \ln ^2(\alpha) \ln (1-\alpha)
+\frac{2}{3} \pi ^2 \ln (1-\alpha)
-2 \zeta_3
\end{split}
\end{align}
Finally, we shall also encounter the integral 
\begin{align}
\label{V2_def}
\begin{split}
\int_0^1 dx  \ln(x)\ln(q(x,\alpha)) p_0(x,\alpha)&=  \frac{1+\alpha^2}{1-\alpha^2}\, V_2(\alpha)
\end{split}
\end{align}
where
\begin{align}
\label{V2}
\begin{split}
V_2(\alpha)&=
\text{Li}_3\left(\alpha^2\right)
+2 \text{Li}_3\left(1-\alpha^2\right)
-4 \text{Li}_3(1-\alpha)-2 \text{Li}_3(\alpha)
+2 \text{Li}_3\left(\frac{\alpha}{1+\alpha}\right)
-2 \text{Li}_3\left(\frac{1}{1+\alpha}\right)
\\
&\hspace*{-30pt}
+\zeta_3
-\left(\text{Li}_2\left(\frac{1}{1+\alpha}\right)
+\text{Li}_2\left(\frac{\alpha}{1+\alpha}\right)\right) \ln (\alpha) -\frac16 \ln (\alpha)
+ ( 2 \ln(1+\alpha) + \ln (1-\alpha) ) \ln^2(\alpha)
\end{split}
\end{align}

\subsection{Three-loop integral summary}

The results for the 1-2-2-1 web are:
\begin{align}
\label{3lfourmat_1}
\begin{split}
W^{(3)}(ijjkkl)&= \Big( C(3a)-C(3b)-C(3c)+C(3d) \Big) {\cal F}^{(3)}_{ijjkkl}(\alpha_{ij},\alpha_{jk},\alpha_{kl},\mu^2/m^2,\epsilon)
\\
&=-f^{dce} f^{abe} T_i^aT_j^bT_k^cT_l^d\,  {\cal F}^{(3)}_{ijjkkl}(\alpha_{ij},\alpha_{jk},\alpha_{kl},\mu^2/m^2,\epsilon)
\end{split}
\end{align}
with
\begin{align}
\label{3lfourmat_2}
\begin{split}
&{\cal F}_{ijjkkl}(\alpha_{ij},\alpha_{jk},\alpha_{kl},\mu^2/m^2,\epsilon)= 
\frac16\left(\mathcal{F}(3a)-2\mathcal{F}(3b) - 2\mathcal{F}(3c) + \mathcal{F}(3d)\right)\\
&=\kappa^3
     \frac{\Gamma(6\epsilon)}{8\epsilon}\int_0^1 {\rm 
      d}x{\rm d} y {\rm d}z\, p(x,\alpha_{ij}) p(y,\alpha_{jk}) p(z,\alpha_{kl})
 \Bigg[ \psi_{3}^{(0)}(x,y,z) +\psi_{3}^{(1)}(x,y,z) \epsilon +{\cal O}(\epsilon^2)
 \Bigg]\,,
\end{split}
\end{align}  
where
\begin{subequations}
\begin{align}
\label{psi}
\begin{split}
\!\psi_{3}^{(0)}(x,y,z) &=\ln\left(\frac{y}{x}\right) +\ln\left(\frac{y}{z}\right) 
\\
\!\psi_{3}^{(1)}(x,y,z) &=
12\left[{\rm{Li}}_2\left(-\frac{y}{x}\right)+{\rm{Li}}_2\left(-\frac{y}{z}\right) +\zeta_2\right]
+\ln^2\left(\frac{y}{z}\right) + 8\ln\left(\frac{1-y}{z}\right) \ln\left(\frac{y}{x}\right) +
\ln^2\left(\frac{y}{x}\right) .
\end{split}
\end{align}
\end{subequations}
Thus, taking the $\epsilon$ expansion we obtain:
\begin{align}
\label{3lfourmat_2_expanded}
\begin{split}
{\cal F}_{ijjkkl}^{(3,-2)}(\alpha_{ij},\alpha_{jk},\alpha_{kl})
&=\frac16 \left(\frac{\kappa}{2}\right)^3
\int_0^1 {\rm 
      d}x{\rm d} y {\rm d}z\, p_0(x,\alpha_{ij}) p_0(y,\alpha_{jk}) p_0(z,\alpha_{kl}) \psi_{3}^{(0)}(x,y,z)
\\
{\cal F}_{ijjkkl}^{(3,-1)}(\alpha_{ij},\alpha_{jk},\alpha_{kl})
&=\frac16 \left(\frac{\kappa}{2}\right)^3
\int_0^1 {\rm 
      d}x{\rm d} y {\rm d}z\, p_0(x,\alpha_{ij}) p_0(y,\alpha_{jk}) p_0(z,\alpha_{kl})\\&\,\, \hspace*{-30pt}\times 
\left[
\psi_{3}^{(1)}(x,y,z) +\psi_3^{(0)}(x,y,z) \left(\ln q(x,\alpha_{ij})+\ln q(y,\alpha_{jk})+\ln q(z,\alpha_{kl})\right)
 \right]\,.
\end{split}
\end{align}  

For the 1-1-1-3 web the results are as follows:
\begin{align}
\label{Wijklll}
W^{(3)}(ijklll)=T_i^{a}T_j^{b}T_k^{c}T_l^d \left[f^{bce}f^{ade} {\cal F}_{ijklll\,;1}(\alpha_{il},\alpha_{jl},\alpha_{kl})+
f^{ace}f^{bde} {\cal F}_{ijklll\,;2}(\alpha_{il},\alpha_{jl},\alpha_{kl})\right]\,,
\end{align}
with
\begin{align}
\label{ijklll}
\begin{split}
{\cal F}_{ijklll\,;n}(\alpha_{il},\alpha_{jl},\alpha_{kl})&=
\frac{\Gamma(6\epsilon)}{8\epsilon}
\kappa^3\,
\int_0^{1}dx dy dz \,p(x,\alpha_{il})\, p(y,\alpha_{jl})\, p(z,\alpha_{kl}) \, \phi_{3,n}(x,y,z;\epsilon)
\end{split}
\end{align}
where
\begin{align}
\label{phi3}
\begin{split}
\phi_{3,1}(x,y,z;\epsilon) 
&=
 \left(-2\ln(y)+\ln(x)+\ln(z)\right)
+\bigg(-8\ln(x)\ln(z)-2\ln(y)\ln(x)+10\ln(z)\ln(y)\\&+12{\rm Li}_2(-x/y)-\ln(z)^2+5\ln(x)^2-4\ln(y)^2-12{\rm Li}_2( -y/z)\bigg)\epsilon+{\cal O}(\epsilon^2)
\\
\phi_{3,2}(x,y,z;\epsilon) 
&=
 \left(\ln(y)+\ln(z)-2\ln(x)\right)
+\bigg(-2\ln(x)\ln(z)+10\ln(y)\ln(x)-8\ln(z)\ln(y)\\&-4\ln(x)^2-\ln(y)^2-12{\rm Li}_2( -x/y)+12 {\rm Li}_2(-z/x)+5\ln(z)^2\Bigg)\epsilon+{\cal O}(\epsilon^2)\,.
\end{split}
\end{align}
Thus, in an expanded form we have, for each $n=1,2$:
\begin{align}
\label{ijklll_expanded}
\begin{split}
{\cal F}_{ijklll\,;n}^{(3,-2)}(\alpha_{il},\alpha_{jl},\alpha_{kl})
&=\frac16 \left(\frac{\kappa}{2}\right)^3
\int_0^1 {\rm 
      d}x{\rm d} y {\rm d}z\, p_0(x,\alpha_{il}) p_0(y,\alpha_{jl}) p_0(z,\alpha_{kl}) \phi_{3,n}^{(0)}(x,y,z)
\\
{\cal F}_{ijklll\,;n}^{(3,-1)}(\alpha_{il},\alpha_{jl},\alpha_{kl})
&=\frac16 \left(\frac{\kappa}{2}\right)^3
\int_0^1 {\rm 
      d}x{\rm d} y {\rm d}z\, p_0(x,\alpha_{il}) p_0(y,\alpha_{jl}) p_0(z,\alpha_{kl})\\&\,\,\hspace*{-40pt} \times 
\left[
\phi_{3,n}^{(1)}(x,y,z) +\phi_{3,n}^{(0)}(x,y,z) \left(\ln q(x,\alpha_{il})+\ln q(y,\alpha_{jl})+\ln q(z,\alpha_{kl})\right)
 \right]\,.
\end{split}
\end{align}  
where the definition of the functions $\phi_{3,n}^{(0)}$ and $\phi_{3,n}^{(1)}$ is implied by comparing eq.~(\ref{ijklll_expanded}) to eqs. (\ref{ijklll}) and (\ref{phi3}).

\begin{table}[htb]
  \centering
  \begin{tabular}{|c|c|c|}
    \hline
    Definition & Name & symbol  \\ \hline
&&\\
${\displaystyle \frac{1}{2\, r} \int_0^1dx \, p_0(x,\alpha)}\,=\ln \alpha\,$& $ R_0(\alpha)$&$\otimes \alpha$\\ 
&&\\
\hline
&&\\
${\displaystyle \frac{1}{2\, r} \int_0^1dx \, p_0(x,\alpha)\, \ln q(x,\alpha)}$ &$R_1(\alpha)$& 
$2 \alpha\otimes (\alpha+1)-\alpha\otimes \alpha$
\\
&&\\
\hline
&&\\
${\displaystyle \frac{1}{2\, r} \int_0^1dx \, p_0(x,\alpha)\, \frac12 \ln^2 q(x,\alpha)}$ & $R_2(\alpha)$&
$\alpha\otimes \alpha\otimes \alpha-2 \alpha\otimes \alpha\otimes (\alpha+1)$
\\&
&$-2 \alpha\otimes (\alpha+1)\otimes \alpha+4 \alpha\otimes (\alpha+1)\otimes (\alpha+1)$
\\
&&\\
\hline
&&\\
${\displaystyle -\frac{2}{r}\int_0^1 dx \ln(x) p_0(x,\alpha)}$
&
$S_1(\alpha)$
& $4 \alpha\otimes (1-\alpha)-2 \alpha\otimes \alpha$\\
&&\\
\hline
&&\\
${\displaystyle -\frac{1}{r}\int_0^1 dx \ln^2(x) p_0(x,\alpha)}$
&
$S_2(\alpha)$
& $-4 \alpha\otimes (1-\alpha)\otimes (1-\alpha)+2 \alpha\otimes (1-\alpha)\otimes \alpha$
\\&&$+2 \alpha\otimes \alpha\otimes (1-\alpha)-2 \alpha\otimes \alpha\otimes \alpha$\\
&&\\
\hline
&&\\
${\displaystyle  -\frac{1}{r} \int_0^1 dx  \ln(x)\ln(1-x) p_0(x,\alpha)}$
&
$\widetilde{S}_2(\alpha)$
& $-4 \alpha\otimes (1-\alpha)\otimes (1-\alpha)+2 \alpha\otimes (1-\alpha)\otimes \alpha$
\\&&$+2 \alpha\otimes \alpha\otimes (1-\alpha)$\\
&&\\
\hline
&&\\
${\displaystyle \frac{1}{r}\int_0^1 dx  \ln(x)\ln(q(x,\alpha)) p_0(x,\alpha)}$
&
$V_2(\alpha)$
& $2 \alpha\otimes (1-\alpha)\otimes \alpha-4 \alpha\otimes (1-\alpha)\otimes (\alpha+1)$
\\&&$+2 \alpha\otimes \alpha\otimes (1-\alpha)+2 \alpha\otimes \alpha\otimes (\alpha+1)$
\\&&$-\alpha\otimes \alpha\otimes \alpha$
\\&&$-4 \alpha\otimes (\alpha+1)\otimes (1-\alpha)+2 \alpha\otimes (\alpha+1)\otimes \alpha$\\
&&\\
\hline
  \end{tabular}
  \caption{The symbols of the transcendental functions entering the three-loop expression for the anomalous dimension for multiple-gluon exchange webs (the 1-1-1-3 and 1-2-2-1 webs). The subscript next to the name of each function indicates its transcendental weight.   The normalization in all cases involves the rational function $r\equiv r(\alpha)=\frac{1+\alpha^2}{1-\alpha^2}$.}
  \label{tab:4leg3loop_transc_func_symb}
\end{table}
The symbols of all the transcendental functions entering the three-loop webs we are considering are summarized in table~\ref{tab:4leg3loop_transc_func_symb}. We observe that the first entry of all is $\alpha$, each power of $\ln q(x,\alpha)$ or $\ln(x)$ (or $\ln(1-x)$) in the integrand increases the transcendental weight by one unit, adding either $\alpha$ or $(1+\alpha)$ entry to the symbol in the case of  $\ln q(x,\alpha)$, and either $\alpha$ or $(1-\alpha)$ entry in the case of $\ln(x)$ (or $\ln(1-x)$).

\section{Calculation of diagrams connecting four Wilson lines\label{sec:3_loop_calc}}

In this appendix we compute the multiple-gluon-exchange three-loop webs connecting four Wilson lines. There are two such webs, 1-2-2-1 will be discussed in section~\ref{3lwprime}, and 1-1-1-3 in section~\ref{3lw}. Starting from the configuration-space Feynman rules, our aim here is to integrate over the distance parameters of the three gluons, leaving the angular integration undone, thus obtaining a parameter representation for the web in the general form of eq.~(\ref{mge_kin}). 

\subsection{The four-leg three-loop web $W_{(1,2,2,1)}^{(3)}$~\label{3lwprime}}

The web $W_{(1,2,2,1)}^{(3)}$ is composed of four diagrams as shown in figure~\ref{3lfour}. According to eq.~(\ref{3lfourmat}) we need to compute a single linear combination of the kinematic integrals corresponding to $\mathcal{F}(3a)-2\mathcal{F}(3b) - 2\mathcal{F}(3c) + \mathcal{F}(3d)$. Let us begin by considering diagram $(3a)$.
Using the configuration-space Feynman rules it evaluates to 
\begin{align}
\begin{split}
 \mathcal{F}(3a)&=\, g_s^6 {\cal N}^3
(\beta_1\cdot \beta_2) (\beta_2\cdot \beta_3) (\beta_3\cdot \beta_4)
\int_0^{\infty} ds   dt_1  dt_2 du_1 du_2 dv\, \theta(t_1<t_2) \theta(u_1<u_2)
\\&\times \Big(-(s\beta_1-t_2\beta_2)^2\Big)^{\epsilon-1}
\Big(-(t_1\beta_2-u_1\beta_3)^2\Big)^{\epsilon-1}
\Big(-(v\beta_4-u_2\beta_3)^2\Big)^{\epsilon-1}
\\
& \times \exp \left\{-{\rm i}m\left(s\sqrt{\beta_1^2-{\rm i}0}+(t_1+t_2)\sqrt{\beta_2^2-{\rm i}0}
+(u_1+u_2)\sqrt{\beta_3^2-{\rm i}0}+v\sqrt{\beta_4^2-{\rm i}0}\right)\right\}\,.
\end{split} 
\end{align}
Rescaling the line parameters by the respective Wilson-line masses  we get
\begin{align}
\begin{split}
 \mathcal{F}(3a)=&\frac{ g_s^6 {\cal N}^3}{8}
\gamma_{12}\,\gamma_{23}\, \gamma_{34}\,
\int_0^{\infty} d\sigma   d\tau_1  d\tau_2 d\mu_1 d\mu_2 d\nu\, \theta(\tau_1<\tau_2) \theta(\mu_1<\mu_2)
\\&\hspace*{-30pt}\times \Big(-\sigma^2-\tau_2^2+\gamma_{12}\sigma\tau_2\Big)^{\epsilon-1}
\Big(-\tau_1^2-\mu_1^2+\gamma_{23}\tau_1\mu_1\Big)^{\epsilon-1}
\Big(-\nu^2-\mu_2^2+\gamma_{34}\nu\mu_2\Big)^{\epsilon-1}
\\&\hspace*{-30pt}\times \exp \left\{-{\rm i}(m-{\rm i}0)(\sigma+\tau_1+\tau_2+\mu_1+\mu_2+\nu)\right\}\,.
\end{split} 
\end{align}
At this point we introduce our usual change of variables for each gluon, as in 
eq.~(\ref{change_var}), defining:
\begin{align*}
\sigma+\tau_2&=\lambda_1\qquad& \tau_1+\mu_1&=\lambda_2\qquad&
\nu+\mu_2&=\lambda_3\\
\frac{\tau_2}{\sigma+\tau_2}&=x\qquad&  \frac{\tau_1}{\tau_1+\mu_1}&=y\qquad&
\frac{\mu_2}{\nu+\mu_2}&=z
\end{align*}
and further defining 
\begin{align}
\begin{split}
\lambda_1&=\lambda\omega\\
\lambda_2&=\lambda(1-\omega)\zeta\\
\lambda_3&=\lambda(1-\omega)(1-\zeta)
\end{split}
\end{align}
to extract the overall ultraviolet divergence and bring the remaining integrals into a canonical parametric form. Repeating the same procedure for the remaining three diagrams we obtain the following expressions:
\begin{subequations}
  \label{eq:threeloopsecondweb}
\begin{align}
\allowdisplaybreaks
  \begin{split}
    \mathcal{F}(3a) &=\kappa^3\,\gamma_{12}\gamma_{23}\gamma_{34}\, \Gamma(6\epsilon)\int_0^1 {\rm
      d}x{\rm d} y {\rm d}z\ P(x,\gamma_{12}) P(y,\gamma_{23})
    P(z,\gamma_{34}) 
\, \int_0^1 d\omega (1-\omega)^{4\epsilon-1}\omega^{2\epsilon-1}\\
  & \hspace{1.5cm}   \,\int_0^1 d\zeta (1-\zeta)^{2\epsilon-1}\zeta^{2\epsilon-1}
 \,\,\,
\theta(\omega x>(1-\omega)\zeta y) \,\,\theta((1-\zeta)z>\zeta(1-y))
  \end{split}
\\
  \begin{split}
     \mathcal{F}(3b) &=\kappa^3\,\gamma_{12}\gamma_{23}\gamma_{34}\,\Gamma(6\epsilon)\int_0^1 {\rm
      d}x{\rm d} y {\rm d}z\ P(x,\gamma_{12}) P(y,\gamma_{23}) P(z,\gamma_{34})\
    \int_0^1 d\omega (1-\omega)^{4\epsilon-1}\omega^{2\epsilon-1}\\
  & \hspace{1.5cm}   \,\int_0^1 d\zeta (1-\zeta)^{2\epsilon-1}\zeta^{2\epsilon-1}
 \,\,\,
\theta(\omega x>(1-\omega)\zeta y) \,\,\theta((1-\zeta)z<\zeta(1-y))
\end{split}
\\
  \begin{split}
     \mathcal{F}(3c) &=\kappa^3\,\gamma_{12}\gamma_{23}\gamma_{34}\, \Gamma(6\epsilon)\int_0^1 {\rm
      d}x{\rm d} y {\rm d}z\ P(x,\gamma_{12}) P(y,\gamma_{23}) P(z,\gamma_{34})\
    \int_0^1 d\omega (1-\omega)^{4\epsilon-1}\omega^{2\epsilon-1}\\
  & \hspace{1.5cm}   \,\int_0^1 d\zeta (1-\zeta)^{2\epsilon-1}\zeta^{2\epsilon-1}
 \,\,\,
\theta(\omega x<(1-\omega)\zeta y) \,\,\theta((1-\zeta)z>\zeta(1-y))
\end{split}
\\
  \begin{split}
     \mathcal{F}(3d) &=\kappa^3\,\gamma_{12}\gamma_{23}\gamma_{34}\, \Gamma(6\epsilon)\int_0^1 {\rm
      d}x{\rm d} y {\rm d}z\ P(x,\gamma_{12}) P(y,\gamma_{23}) P(z,\gamma_{34})\
    \int_0^1 d\omega (1-\omega)^{4\epsilon-1}\omega^{2\epsilon-1}\\
  & \hspace{1.5cm}   \,\int_0^1 d\zeta (1-\zeta)^{2\epsilon-1}\zeta^{2\epsilon-1}
 \,\,\,
\theta(\omega x<(1-\omega)\zeta y) \,\,\theta((1-\zeta)z<\zeta(1-y))
  \end{split}
\end{align}
\end{subequations}
where $\kappa$ is defined in eq.~(\ref{kappa1}).
Note that the only difference between the different diagrams in (\ref{eq:threeloopsecondweb}) is in the Heaviside functions defining the range of integration over $\omega$ and $\zeta$.
The first integral we perform is over $\omega$ and it yields:
\begin{align}
\label{I_expand_ab}
\begin{split}
\text{\,Diagrams a and b:} \qquad I^{\text{a,b}}_\omega(A) &=  \int_0^1 d\omega (1-\omega)^{4\epsilon-1}\omega^{2\epsilon-1} \theta(\omega/(1-\omega)>A)
\\&= \int_0^\infty d\alpha (1+\alpha)^{-6\epsilon} \,\alpha^{2\epsilon-1} \,\theta(\alpha>A)
\\&=\int_0^\infty d\alpha (1+\alpha)^{-6\epsilon} \,\alpha^{2\epsilon-1}
\Big(1-\theta(\alpha<A)\Big)
\\&
=\frac{\Gamma(4\epsilon)\,\Gamma(2\epsilon)}{\Gamma(6\epsilon)}-\frac{A^{2\epsilon}}{2\epsilon}\left[1+12\epsilon^2\,{\rm Li}_2(-A)+{\cal O}(\epsilon^3)\right]\,
\\&
=\frac{1}{2\epsilon}\left(\frac32-12\zeta_2\epsilon^2+\ldots\right)-\frac{A^{2\epsilon}}{2\epsilon}\left[1+12\epsilon^2\,{\rm Li}_2(-A)+{\cal O}(\epsilon^3)\right]\,
\\&
=\frac{1}{2\epsilon}\left[\frac32-A^{2\epsilon}-12\epsilon^2\left({\rm Li}_2(-A)+\zeta_2\right)+{\cal O}(\epsilon^3)\right]\,,
\end{split}
\end{align}
and
\begin{align}
\label{I_expand_cd}
\begin{split}
\text{\,Diagrams c and d:} \qquad I^{\text{c,d}}_\omega(A) &=  \int_0^1 d\omega (1-\omega)^{4\epsilon-1}\omega^{2\epsilon-1} \theta(\omega/(1-\omega)<A)
\\&= \int_0^\infty d\alpha (1+\alpha)^{-6\epsilon} \,\alpha^{2\epsilon-1} \,\theta(\alpha<A)
\\&=\frac{A^{2\epsilon}}{2\epsilon}\left[1+12\epsilon^2\,{\rm Li}_2(-A)+{\cal O}(\epsilon^3)\right]\,,
\end{split}
\end{align}
where $A=\zeta y/x$. 

This relies on the general results~\cite{Kalmykov:2006pu}:
\begin{align}
\label{Hyper_expan}
\begin{split}
\int_0^\infty d\alpha (1+\alpha)^{-n\epsilon} \,\alpha^{m\epsilon-1} \,\theta(\alpha>A)
&=\frac{A^{(m-n)\epsilon}}{(n-m)\epsilon}\,_2F_1
\left([n\epsilon, (n-m)\epsilon],[1+(n-m)\epsilon],-1/A\right)
\\
&=\frac{A^{(m-n)\epsilon}}{(n-m)\epsilon}\left(1-\sum_{i=2}^{\infty}\epsilon^i\sum_{k=1}^{i-1} n^k (m-n)^{i-k} S_{i-k,k}(-1/A)
\right)
\\
\int_0^\infty d\alpha (1+\alpha)^{-n\epsilon} \,\alpha^{m\epsilon-1} \,\theta(\alpha<A)
&=\frac{A^{m\epsilon}}{m\epsilon}\,_2F_1\left([n\epsilon, m\epsilon],[1+m\epsilon],-A\right)
\\&=\frac{A^{m\epsilon}}{m\epsilon}\left(1-\sum_{i=2}^{\infty}\epsilon^i\sum_{k=1}^{i-1} n^k (-m)^{i-k} S_{i-k,k}(-A)\right)
\end{split}
\end{align}
where $S_{a,b}(z)$ is the Nielsen Generalized Polylogarithm
\[
S_{a,b}(z)=\frac{(-1)^{a+b-1}}{(a-1)!b!}\,\int_0^1\frac{dx}{x} \ln^{a-1}(x)\ln^b(1-xz)
\]
and $S_{1,1}(z)={\rm Li}_2(z)$.
We note that while both expansions in eq.~(\ref{Hyper_expan}) are valid in general for any $A$, only the second can be used in our application. To see this note that below we are going to take the integral over $\zeta$ from zero to 1; thus $A$ eventually varies between zero and a finite limit $y/x$ (the integrals over $x$ and $y$ converge independently of these factors). The $A\to 0$ limit would be problematic if we were to use the first expansion in eq.~(\ref{Hyper_expan}) as individual terms in the expansion diverge. In contract, using the second expansion in  eq.~(\ref{Hyper_expan}) the limit $A\to 0$ is smooth. This explains the way we chose to expand the integrals in eq.~(\ref{I_expand_ab}).

So up to constant terms we get:
\begin{subequations}
  \label{eq:threeloopsecondweb_expansion}
\begin{align}
\allowdisplaybreaks
\begin{split}
    \mathcal{F}(3a) &=\kappa^3\,\gamma_{12}\gamma_{23}\gamma_{34}\, \frac{\Gamma(6\epsilon)}{2\epsilon}\int_0^1 {\rm
      d}x{\rm d} y {\rm d}z\ P(x,\gamma_{12}) P(y,\gamma_{23})
    P(z,\gamma_{34}) 
\, \\
  & \hspace{1.5cm}   \,\int_0^1 d\zeta (1-\zeta)^{2\epsilon-1}\zeta^{2\epsilon-1}
 \,\,\,
 \,\,\theta\left(\frac{1-\zeta}{\zeta}>\frac{1-y}{z}\right) 
\\&\hspace{1.5cm} 
\left[\frac32-\left(\zeta\frac{y}{x}\right)^{2\epsilon}-12\epsilon^2\left({\rm Li}_2\left(-\left(\zeta\frac{y}{x}\right)\right)+\zeta_2\right)+{\cal O}(\epsilon^3)\right]
  \end{split}
\\
  \begin{split}
     \mathcal{F}(3b) &=\kappa^3\,\gamma_{12}\gamma_{23}\gamma_{34}\,\frac{\Gamma(6\epsilon)}{2\epsilon}\int_0^1 {\rm
      d}x{\rm d} y {\rm d}z\ P(x,\gamma_{12}) P(y,\gamma_{23}) P(z,\gamma_{34})\
    \\
  & \hspace{1.5cm}   \,\int_0^1 d\zeta (1-\zeta)^{2\epsilon-1}\zeta^{2\epsilon-1}
 \,\,\,
 \,\,\theta\left(\frac{1-\zeta}{\zeta}<\frac{1-y}{z}\right)
\\&\hspace{1.5cm} 
\left[\frac32-\left(\zeta\frac{y}{x}\right)^{2\epsilon}-12\epsilon^2\left({\rm Li}_2\left(-\left(\zeta\frac{y}{x}\right)\right)+\zeta_2\right)+{\cal O}(\epsilon^3)\right]
\end{split}
\\
  \begin{split}
     \mathcal{F}(3c) &=\kappa^3\,\gamma_{12}\gamma_{23}\gamma_{34}\, \frac{\Gamma(6\epsilon)}{2\epsilon}\int_0^1 {\rm
      d}x{\rm d} y {\rm d}z\ P(x,\gamma_{12}) P(y,\gamma_{23}) P(z,\gamma_{34})\
    \\
  & \hspace{1.5cm}  
\left(\frac{y}{x}\right)^{2\epsilon}
 \,\int_0^1 d\zeta (1-\zeta)^{2\epsilon-1}\zeta^{4\epsilon-1}
 \,\,\,
 \,\,\theta\left(\frac{1-\zeta}{\zeta}>\frac{1-y}{z}\right)
\\& \hspace{1.5cm}
\left[1+12\epsilon^2\,{\rm Li}_2\left(-\zeta\frac{y}{x}\right)+{\cal O}(\epsilon^3)\right]
\end{split}
\\
  \begin{split}
     \mathcal{F}(3d) &=\kappa^3\,\gamma_{12}\gamma_{23}\gamma_{34}\, \frac{\Gamma(6\epsilon)}{2\epsilon}\int_0^1 {\rm
      d}x{\rm d} y {\rm d}z\ P(x,\gamma_{12}) P(y,\gamma_{23}) P(z,\gamma_{34})\
    \\
   & \hspace{1.5cm}  
\left(\frac{y}{x}\right)^{2\epsilon}
 \,\int_0^1 d\zeta (1-\zeta)^{2\epsilon-1}\zeta^{4\epsilon-1}
 \,\,\,
 \,\,\theta\left(\frac{1-\zeta}{\zeta}<\frac{1-y}{z}\right)
\\& \hspace{1.5cm}
\left[1+12\epsilon^2\,{\rm Li}_2\left(-\zeta\frac{y}{x}\right)+{\cal O}(\epsilon^3)\right]
  \end{split}
\end{align}
\end{subequations}

To compute all the singular terms, down to the single pole, let us split each integral into two, ${\cal F}={\cal F}_1+{\cal F}_2$, where $\mathcal{F}_1$ corresponds to the leading term, $1$, in the square brackets above (the leading term in the expansion of the hypergeometric function), and $\mathcal{F}_2$ corresponds to the dilogarithm at ${\cal O}(\epsilon^2)$.
We get:
\begin{subequations}
  \label{eq:threeloopsecondweb_expansion_F1}
\begin{align}
\allowdisplaybreaks
\begin{split}
    \mathcal{F}_1(3a) &=\kappa^3\,\gamma_{12}\gamma_{23}\gamma_{34}\, \frac{\Gamma(6\epsilon)}{2\epsilon}\int_0^1 {\rm
      d}x{\rm d} y {\rm d}z\ P(x,\gamma_{12}) P(y,\gamma_{23})
    P(z,\gamma_{34}) 
\, \\
  & \hspace{1.5cm}   \,\int_0^1 d\zeta (1-\zeta)^{2\epsilon-1}\zeta^{2\epsilon-1}
 \,\,\,
 \,\,\theta\left(\frac{1-\zeta}{\zeta}>\frac{1-y}{z}\right) 
\,
\left[\frac32-\left(\zeta\frac{y}{x}\right)^{2\epsilon}\right]
\\
&=\kappa^3\,\gamma_{12}\gamma_{23}\gamma_{34}\, \frac{\Gamma(6\epsilon)}{2\epsilon}\int_0^1 {\rm
      d}x{\rm d} y {\rm d}z\ P(x,\gamma_{12}) P(y,\gamma_{23})
    P(z,\gamma_{34}) 
\, \\
  & \hspace{1.5cm}   
 \,\,\,
\Bigg[\frac32 \,\int_0^1 d\zeta (1-\zeta)^{2\epsilon-1}\zeta^{2\epsilon-1}\,\,\theta\left(\frac{1-\zeta}{\zeta}>\frac{1-y}{z}\right) 
\,
\\&\hspace{1.5cm}  -\left(\frac{y}{x}\right)^{2\epsilon}\,\int_0^1 d\zeta (1-\zeta)^{2\epsilon-1}\zeta^{4\epsilon-1}
\theta\left(\frac{1-\zeta}{\zeta}>\frac{1-y}{z}\right) \Bigg]
  \end{split}
\\
  \begin{split}
     \mathcal{F}_1(3b) &=\kappa^3\,\gamma_{12}\gamma_{23}\gamma_{34}\,\frac{\Gamma(6\epsilon)}{2\epsilon}\int_0^1 {\rm
      d}x{\rm d} y {\rm d}z\ P(x,\gamma_{12}) P(y,\gamma_{23}) P(z,\gamma_{34})\
    \\
  & \hspace{1.5cm}   \,\int_0^1 d\zeta (1-\zeta)^{2\epsilon-1}\zeta^{2\epsilon-1}
 \,\,\,
 \,\,\theta\left(\frac{1-\zeta}{\zeta}<\frac{1-y}{z}\right)
\, 
\left[\frac32-\left(\zeta\frac{y}{x}\right)^{2\epsilon}\right]
\\ &=\kappa^3\,\gamma_{12}\gamma_{23}\gamma_{34}\,\frac{\Gamma(6\epsilon)}{2\epsilon}\int_0^1 {\rm
      d}x{\rm d} y {\rm d}z\ P(x,\gamma_{12}) P(y,\gamma_{23}) P(z,\gamma_{34})\
    \\
  & \hspace{1.5cm}   \Bigg[\frac32\,\int_0^1 d\zeta (1-\zeta)^{2\epsilon-1}\zeta^{2\epsilon-1}
 \,\,\,
 \,\,\theta\left(\frac{1-\zeta}{\zeta}<\frac{1-y}{z}\right)
\, 
\\& \hspace{1.5cm} -\left(\frac{y}{x}\right)^{2\epsilon}
\int_0^1 d\zeta (1-\zeta)^{2\epsilon-1}\zeta^{4\epsilon-1}
 \,\,\,
 \,\,\theta\left(\frac{1-\zeta}{\zeta}<\frac{1-y}{z}\right)
\Bigg]
\end{split}
\\
  \begin{split}
     \mathcal{F}_1(3c) &=\kappa^3\,\gamma_{12}\gamma_{23}\gamma_{34}\, \frac{\Gamma(6\epsilon)}{2\epsilon}\int_0^1 {\rm
      d}x{\rm d} y {\rm d}z\ P(x,\gamma_{12}) P(y,\gamma_{23}) P(z,\gamma_{34})\
    \\
  & \hspace{1.5cm}  
\left(\frac{y}{x}\right)^{2\epsilon}
 \,\int_0^1 d\zeta (1-\zeta)^{2\epsilon-1}\zeta^{4\epsilon-1}
 \,\,\,
 \,\,\theta\left(\frac{1-\zeta}{\zeta}>\frac{1-y}{z}\right)
\end{split}
\\
  \begin{split}
     \mathcal{F}_1(3d) &=\kappa^3\,\gamma_{12}\gamma_{23}\gamma_{34}\, \frac{\Gamma(6\epsilon)}{2\epsilon}\int_0^1 {\rm
      d}x{\rm d} y {\rm d}z\ P(x,\gamma_{12}) P(y,\gamma_{23}) P(z,\gamma_{34})\
    \\
   & \hspace{1.5cm}  
\left(\frac{y}{x}\right)^{2\epsilon}
 \,\int_0^1 d\zeta (1-\zeta)^{2\epsilon-1}\zeta^{4\epsilon-1}
 \,\,\,
 \,\,\theta\left(\frac{1-\zeta}{\zeta}<\frac{1-y}{z}\right)
  \end{split}
\end{align}
\end{subequations}
and
\begin{subequations}
  \label{eq:threeloopsecondweb_expansion_F2}
\begin{align}
\allowdisplaybreaks
\begin{split}
    \mathcal{F}_2(3a) &=-\kappa^3\,\gamma_{12}\gamma_{23}\gamma_{34}\,\int_0^1 {\rm
      d}x{\rm d} y {\rm d}z\ P(x,\gamma_{12}) P(y,\gamma_{23})
    P(z,\gamma_{34}) 
\, \\
  & \hspace{1.5cm}   \,\int_0^1 d\zeta (1-\zeta)^{2\epsilon-1}\zeta^{2\epsilon-1}
 \,\,\,
 \,\,\theta\left(\frac{1-\zeta}{\zeta}>\frac{1-y}{z}\right) 
\left({\rm Li}_2\left(-\left(\zeta\frac{y}{x}\right)\right)+\zeta_2\right)
  \end{split}
\\
  \begin{split}
     \mathcal{F}_2(3b) &=-\kappa^3\,\gamma_{12}\gamma_{23}\gamma_{34}\,\int_0^1 {\rm
      d}x{\rm d} y {\rm d}z\ P(x,\gamma_{12}) P(y,\gamma_{23}) P(z,\gamma_{34})\
    \\
  & \hspace{1.5cm}   \,\int_0^1 d\zeta (1-\zeta)^{2\epsilon-1}\zeta^{2\epsilon-1}
 \,\,\,
 \,\,\theta\left(\frac{1-\zeta}{\zeta}<\frac{1-y}{z}\right)
\left({\rm Li}_2\left(-\left(\zeta\frac{y}{x}\right)\right)+\zeta_2\right)
\end{split}
\\
  \begin{split}
     \mathcal{F}_2(3c) &=\kappa^3\,\gamma_{12}\gamma_{23}\gamma_{34}\, \int_0^1 {\rm
      d}x{\rm d} y {\rm d}z\ P(x,\gamma_{12}) P(y,\gamma_{23}) P(z,\gamma_{34})\
    \\
  & \hspace{1.5cm}  
 \,\int_0^1 d\zeta (1-\zeta)^{2\epsilon-1}\zeta^{4\epsilon-1}
 \,\,\,
 \,\,\theta\left(\frac{1-\zeta}{\zeta}>\frac{1-y}{z}\right)
\,{\rm Li}_2\left(-\zeta\frac{y}{x}\right)
\end{split}
\\
  \begin{split}
     \mathcal{F}_2(3d) &=\kappa^3\,\gamma_{12}\gamma_{23}\gamma_{34}\, \int_0^1 {\rm
      d}x{\rm d} y {\rm d}z\ P(x,\gamma_{12}) P(y,\gamma_{23}) P(z,\gamma_{34})\
    \\
   & \hspace{1.5cm}  
 \,\int_0^1 d\zeta (1-\zeta)^{2\epsilon-1}\zeta^{4\epsilon-1}
 \,\,\,
 \,\,\theta\left(\frac{1-\zeta}{\zeta}<\frac{1-y}{z}\right)
\,{\rm Li}_2\left(-\zeta\frac{y}{x}\right)
\end{split}
\end{align}
\end{subequations}

To perform the $\zeta$ integrals in ${\cal F}_1$ of eq.~(\ref{eq:threeloopsecondweb_expansion_F1}) we use the following results:
\begin{align}
\begin{split}
I^>_{\zeta}(p,q;B)&=\int_0^1(1-\zeta)^{p\epsilon-1}\zeta^{q\epsilon-1}\theta\left(\frac{1-\zeta}{\zeta}>B\right) \\
&= \int_0^{\infty}d\beta \beta^{p\epsilon-1}(1+\beta)^{-(p+q)\epsilon}\theta(\beta>B)
\\&=\frac{B^{-\epsilon q}}{\epsilon q} \, _2F_1([(q+p)\epsilon,q\epsilon],[1+q\epsilon],-1/B)
\\&=\frac{B^{-\epsilon q}}{\epsilon q} \left(1-\sum_{i=2}^{\infty}\epsilon^i\sum_{k=1}^{i-1} (p+q)^k (-q)^{i-k} S_{i-k,k}(-1/B)\right)
\end{split}
\end{align}
and 
\begin{align}
\begin{split}
I^<_{\zeta}(p,q;B)&=\int_0^1(1-\zeta)^{p\epsilon-1}\zeta^{q\epsilon-1}\theta\left(\frac{1-\zeta}{\zeta}<B\right) \\
&= \int_0^{\infty}d\beta \beta^{p\epsilon-1}(1+\beta)^{-(p+q)\epsilon}\theta(\beta<B)
\\&=\frac{B^{\epsilon p}}{\epsilon p} \, _2F_1([(q+p)\epsilon,p\epsilon],[1+p\epsilon],-B)
\\&=\frac{B^{\epsilon p}}{\epsilon p} \left(1-\sum_{i=2}^{\infty}\epsilon^i\sum_{k=1}^{i-1} (p+q)^k (-p)^{i-k} S_{i-k,k}(-B)\right)
\end{split}
\end{align}
This yields:
\begin{subequations}
  \label{eq:threeloopsecondweb_expansion}
\begin{align}
\allowdisplaybreaks
\begin{split}
    \mathcal{F}_1(3a)
&=\kappa^3\,\gamma_{12}\gamma_{23}\gamma_{34}\, \frac{\Gamma(6\epsilon)}{8\epsilon^2}\int_0^1 {\rm
      d}x{\rm d} y {\rm d}z\ P(x,\gamma_{12}) P(y,\gamma_{23})
    P(z,\gamma_{34}) 
\, \\
  & \hspace{1.5cm}   
 \,\,\,
\Bigg[3\left(\frac{1-y}{z}\right)^{-2\epsilon}\left(1+8\epsilon^2\,{\rm Li}_2(-z/(1-y))+\ldots\right)
\\&\hspace{1.5cm}- \left(\frac{y}{x}\right)^{2\epsilon} \left(\frac{1-y}{z}\right)^{-4\epsilon}\left(1+24\epsilon^2\,{\rm Li}_2(-z/(1-y))+\ldots\right)
\Bigg]
  \end{split}
\\
  \begin{split}
     \mathcal{F}_1(3b)  &=\kappa^3\,\gamma_{12}\gamma_{23}\gamma_{34}\,\frac{\Gamma(6\epsilon)}{8\epsilon^2}\int_0^1 {\rm
      d}x{\rm d} y {\rm d}z\ P(x,\gamma_{12}) P(y,\gamma_{23}) P(z,\gamma_{34})\
    \\
  & \hspace{1.5cm}   \Bigg[3\left(\frac{1-y}{z}\right)^{2\epsilon}\,\left(1+8\epsilon^2\,{\rm Li}_2(-(1-y)/z)+\ldots\right)
\\& \hspace{1.5cm} -2\left(\frac{y}{x}\right)^{2\epsilon}\left(\frac{1-y}{z}\right)^{2\epsilon}
\,\left(1+12\epsilon^2\,{\rm Li}_2(-(1-y)/z)+\ldots\right)
\Bigg]
\end{split}
\\
  \begin{split}
     \mathcal{F}_1(3c) &=\kappa^3\,\gamma_{12}\gamma_{23}\gamma_{34}\, \frac{\Gamma(6\epsilon)}{8\epsilon^2}\int_0^1 {\rm
      d}x{\rm d} y {\rm d}z\ P(x,\gamma_{12}) P(y,\gamma_{23}) P(z,\gamma_{34})\
    \\
  & \hspace{1.5cm}  
\left(\frac{y}{x}\right)^{2\epsilon}\left(\frac{1-y}{z}\right)^{-4\epsilon}
 \,\left(1+24\epsilon^2\,{\rm Li}_2(-z/(1-y))+\ldots\right)
\end{split}
\\
  \begin{split}
     \mathcal{F}_1(3d) &=\kappa^3\,\gamma_{12}\gamma_{23}\gamma_{34}\, \frac{\Gamma(6\epsilon)}{8\epsilon^2}\int_0^1 {\rm
      d}x{\rm d} y {\rm d}z\ P(x,\gamma_{12}) P(y,\gamma_{23}) P(z,\gamma_{34})\
    \\
   & \hspace{1.5cm}  
2\left(\frac{y}{x}\right)^{2\epsilon}
 \left(\frac{1-y}{z}\right)^{2\epsilon}
 \,\left(1+12\epsilon^2\,{\rm Li}_2(-(1-y)/z)+\ldots\right)
  \end{split}
\end{align}
\end{subequations}

To perform the $\zeta$ integration in the ${\cal F}_2$ terms (\ref{eq:threeloopsecondweb_expansion_F2}) we use the following results. First for the constants:
\begin{align}
\text{for a\qquad}\int_0^1 d\zeta (1-\zeta)^{p\epsilon-1} \zeta^{q\epsilon-1} \,\,\theta\left(\zeta<\frac{1}{1+B}\right)
&=\frac{1}{q\epsilon}\,(1+{\cal O}(\epsilon))
\\
\text{for b\qquad}\int_0^1 d\zeta (1-\zeta)^{p\epsilon-1} \zeta^{q\epsilon-1}\,\, \theta\left(\zeta>\frac{1}{1+B}\right)
&=\frac{1}{p\epsilon}\,(1+{\cal O}(\epsilon))
\end{align}
For the dilogarithmic functions we get:
\begin{align}
\text{for a\qquad}\int_0^1 d\zeta (1-\zeta)^{p\epsilon-1} \zeta^{q\epsilon-1} \,{\rm Li}_2(-r\zeta) \,\,\theta\left(\zeta<\frac{1}{1+B}\right)
&={\cal O}(\epsilon)
\\
\text{for b\qquad}\int_0^1 d\zeta (1-\zeta)^{p\epsilon-1} \zeta^{q\epsilon-1}\,{\rm Li}_2(-r\zeta) \,\,\theta\left(\zeta>\frac{1}{1+B}\right)
&=\frac{\,{\rm Li}_2(-r)}{p\epsilon}\,(1+{\cal O}(\epsilon))
\end{align}
These results can be proven by expanding the 
${\rm Li}_2(-r\zeta)=\sum_{n=1}^{\infty} (-t\zeta)^n/n^2$ under the integrals.

We then get:
\begin{subequations}
  \label{eq:threeloopsecondweb_expansion_F2_results}
\begin{align}
\allowdisplaybreaks
\begin{split}
    \mathcal{F}_2(3a) &=-\kappa^3\,\gamma_{12}\gamma_{23}\gamma_{34}\, \frac{\zeta_2}{2\epsilon}\int_0^1 {\rm
      d}x{\rm d} y {\rm d}z\ P(x,\gamma_{12}) P(y,\gamma_{23})
    P(z,\gamma_{34}) 
  \end{split}
\\
  \begin{split}
     \mathcal{F}_2(3b) &=-\kappa^3\,\gamma_{12}\gamma_{23}\gamma_{34}\,\frac{1}{2\epsilon} \int_0^1 {\rm
      d}x{\rm d} y {\rm d}z\ P(x,\gamma_{12}) P(y,\gamma_{23}) P(z,\gamma_{34}) 
\left({\rm Li}_2\left(-\frac{y}{x}\right)+\zeta_2\right)
\end{split}
\\
  \begin{split}
     \mathcal{F}_2(3c) &={\cal O}(\epsilon^0)
\end{split}
\\
  \begin{split}
     \mathcal{F}_2(3d) &=\kappa^3\,\gamma_{12}\gamma_{23}\gamma_{34}\, \frac{1}{2\epsilon}  \int_0^1 {\rm
      d}x{\rm d} y {\rm d}z\ P(x,\gamma_{12}) P(y,\gamma_{23}) P(z,\gamma_{34})\
\,{\rm Li}_2\left(-\frac{y}{x}\right)\,.
\end{split}
\end{align}
\end{subequations}

Finally, combining ${\cal F}_1$ and ${\cal F}_2$ we obtain the following  parametric integrals for the four diagrams, down to single-pole terms:
\begin{subequations}
\begin{align}
\begin{split}\label{eq:slp2_a}
     \mathcal{F}(3a) &=\kappa^3\,\gamma_{12}\gamma_{23}\gamma_{34}\,
     \frac{\Gamma(6\epsilon)}{4\epsilon}\int_0^1 {\rm 
      d}x{\rm d} y {\rm d}z\ P(x,\gamma_{12}) P(y,\gamma_{23}) P(z,\gamma_{34})\\
    &  \times
\Bigg\{\left(\ln\left(\frac{x}{y}\right) + \ln\left(\frac{z}{y}\right) \right) \\&+\epsilon \Bigg[-\ln^2\left(\frac{1-y}{z}\right)-\ln^2\left(\frac{y}{x}\right)+4\ln\left(\frac{y}{x}\right)\ln\left(\frac{1-y}{z}\right)-12\zeta_2\Bigg] \Bigg\}
\end{split} \\
\begin{split}\label{eq:slp2_b}
     \mathcal{F}(3b) &=\kappa^3\,\gamma_{12}\gamma_{23}\gamma_{34}\,
     \frac{\Gamma(6\epsilon)}{4\epsilon}\int_0^1 {\rm 
      d}x{\rm d} y {\rm d}z\ P(x,\gamma_{12}) P(y,\gamma_{23}) P(z,\gamma_{34})\\
    & \times \Bigg\{
\left(2\ln\left(\frac{x}{y}\right)
+\ln\left(\frac{y}{z}\right)\right)\\&+\epsilon
\Bigg[\ln^2\left(\frac{1-y}{z}\right)-2\ln^2\left(\frac{y}{x}\right)-4\ln\left(\frac{y}{x}\right)\ln\left(\frac{1-y}{z}\right)-12 \left({\rm Li}_2\left(-\frac{y}{x}\right)+\zeta_2\right)\Bigg]
\Bigg\}
\end{split} \\
\begin{split}\label{eq:slp2_c}
     \mathcal{F}(3c) &=\kappa^3\,\gamma_{12}\gamma_{23}\gamma_{34}\,
     \frac{\Gamma(6\epsilon)}{4\epsilon}\int_0^1 {\rm 
      d}x{\rm d} y {\rm d}z\ P(x,\gamma_{12}) P(y,\gamma_{23}) P(z,\gamma_{34})\\
    &  \times \Bigg\{
\left(2 \ln\left(\frac{z}{y}\right) +\ln\left(\frac{y}{x}\right)\right) \\&+\epsilon
\Bigg[  4\ln^2\left(\frac{1-y}{z}\right)+\ln^2\left(\frac{y}{x}\right)-4\ln\left(\frac{y}{x}\right)\ln\left(\frac{1-y}{z}\right)
\,+12\,{\rm Li}_2\left(-\frac{z}{1-y}\right) \Bigg]
 \Bigg\}
\end{split} \\
\begin{split}\label{eq:slp2_d}
     \mathcal{F}(3d) &=\kappa^3\,\gamma_{12}\gamma_{23}\gamma_{34}\,
     \frac{\Gamma(6\epsilon)}{4\epsilon}\int_0^1 {\rm 
      d}x{\rm d} y {\rm d}z\ P(x,\gamma_{12}) P(y,\gamma_{23}) P(z,\gamma_{34})\\
    &  \times \Bigg\{\left(2\ln\left(\frac{y}{x}\right)  +2\ln\left(\frac{y}{z}\right)\right)+\epsilon
\Bigg[2\ln^2\left(\frac{1-y}{z}\right)+2\ln^2\left(\frac{y}{x}\right)
+4\ln\left(\frac{y}{x}\right)\ln\left(\frac{1-y}{z}\right)
\\&\,+12\,{\rm Li}_2\left(-\frac{1-y}{z}\right)+12 {\rm Li}_2\left(-\frac{y}{x}\right)\Bigg] \Bigg\}
  \end{split}
\end{align}
\end{subequations}
The four diagrams contribute to the web through the combination in eq.~(\ref{3lfourmat}). Combining the corresponding integrands yields the final expression in eq.~(\ref{3lfourmat_}).

\subsection{The four-leg three-loop web $W_{(1,1,1,3)}^{(3)}$~\label{3lw}}

Let us now compute the diagrams of the 1-1-1-3 web shown in figure~\ref{1113}. One immediately observes that there is a permutation symmetry allowing to deduce the result for the integral for all these diagrams from any one in the set. Let us compute diagram B.
\begin{align}
\begin{split}
{\cal F}_{(1,1,1,3)}^B&=g_s^6 {\cal N}^3
(\beta_1\cdot \beta_4)(\beta_2\cdot\beta_4)(\beta_3\cdot \beta_4)
\int_0^{\infty}ds_1ds_2ds_3 \\&\int_0^{\infty}dsdtdu \,\theta(t>s)\theta(u>t){\rm e}^{-{\rm i} m\left(s_1\sqrt{\beta_1^2-{\rm i}0 }+s_2\sqrt{\beta_2^2-{\rm i}0 }+s_3\sqrt{\beta_3^2-{\rm i}0 }
+(s+t+u)\sqrt{\beta_4^2-{\rm i}0 }\right)}\\&
\left(-(\beta_4s-\beta_1s_1)^2\right)^{\epsilon-1}
\left(-(\beta_4t-\beta_2s_2)^2\right)^{\epsilon-1}
\left(-(\beta_4u-\beta_3s_3)^2\right)^{\epsilon-1}
\end{split}
\end{align} 
Defining $s_i\sqrt{\beta_i^2-{\rm i}0}=\sigma_i$ and $s\sqrt{\beta_4^2-{\rm i}0}=\sigma$, $t\sqrt{\beta_4^2-{\rm i}0}=\tau$, $u\sqrt{\beta_4^2-{\rm i}0}=\mu$ we get:
\begin{align}
\begin{split}
{\cal F}_{(1,1,1,3)}^B&=\frac{g_s^6{\cal N}^3}{8} 
\gamma_{14}\gamma_{24}\gamma_{34}
\int_0^{\infty}d\sigma_1d\sigma_2 d\sigma_3 \\&\int_0^{\infty}dsdtdu \,\theta(\tau>\sigma)\theta(\mu>\tau){\rm e}^{-{\rm i}(m-{\rm i}0)\left(\sigma_1+\sigma_2+\sigma_3
+\sigma+\tau+\mu\right)}\\&
\left(-\sigma^2-\sigma_1^2+\gamma_{14}\sigma\sigma_1\right)^{\epsilon-1}
\left(-\tau^2-\sigma_2^2+\gamma_{24}\tau\sigma_2\right)^{\epsilon-1}
\left(-\mu^2-\sigma_3^2+\gamma_{34}\mu\sigma_3\right)^{\epsilon-1}
\end{split}
\end{align}
Next defining $\lambda_1=\sigma+\sigma_1$, $\lambda_2=\tau+\sigma_2$ and  $\lambda_3=\mu+\sigma_3$ and $x=\sigma/\lambda_1$,  $y=\tau/\lambda_2$ and $z=\mu/\lambda_3$ we get:
\begin{align}
\begin{split}
{\cal F}_{(1,1,1,3)}^B&=\frac{g_s^6{\cal N}^3}{8} 
\gamma_{14}\gamma_{24}\gamma_{34}
\int_0^{\infty}d\lambda_1\lambda_1 (-\lambda_1^2)^{\epsilon-1} d\lambda_2\lambda_2(-\lambda_2^2)^{\epsilon-1} d\lambda_3
\lambda_3(-\lambda_3^2)^{\epsilon-1} \\&\hspace*{-20pt}\int_0^{1}dx dy dz 
\,\theta(y\lambda_2>x\lambda_1)\theta(z\lambda_3>y\lambda_2){\rm e}^{-{\rm i}(m-{\rm i}0)\left(\lambda_1+\lambda_2+\lambda_3\right)}
P(x,\gamma_{14})\, P(y,\gamma_{24})\, P(z,\gamma_{34})
\end{split}
\end{align}
Finally parametrizing $\lambda_i$ as follows:
\begin{align}
\begin{split}
\lambda_1&=\zeta\lambda\\
\lambda_2&=(1-\zeta)(1-\omega)\lambda\\
\lambda_3&=(1-\zeta)\omega\lambda
\end{split}
\end{align}
and performing the dimensionful integral over $\lambda$ we get
\begin{align}
\label{calFB_1113}
\begin{split}
{\cal F}_{(1,1,1,3)}^B&= \Gamma(6\epsilon) \,\kappa^3
\gamma_{14}\gamma_{24}\gamma_{34}
\int_0^{1}dx dy dz \,P(x,\gamma_{14})\, P(y,\gamma_{24})\, P(z,\gamma_{34})
\\&\hspace*{-30pt}\times \,
\int_0^{1}d\zeta d\omega \zeta^{2\epsilon-1} (1-\zeta)^{4\epsilon-1} \, \omega^{2\epsilon-1}(1-\omega)^{2\epsilon-1}
\,\theta(y(1-\zeta)(1-\omega)>x\zeta)\, \theta(z\omega>y(1-\omega))\,.
\end{split}
\end{align}
Performing the integral over $\zeta$ and then the one over $\omega$ we get the following result for the singular terms:
\begin{align}
\begin{split}
{\cal F}_{(1,1,1,3)}^B&=
\frac{\Gamma(6\epsilon)}{(2\epsilon)^2}
\kappa^3
\gamma_{14}\gamma_{24}\gamma_{34}
\int_0^{1}dx dy dz \,P(x,\gamma_{14})\, P(y,\gamma_{24})\, P(z,\gamma_{34})
\\&\qquad
\left(\frac{y}{x}\right)^{2\epsilon}
\left[\frac32-\left(\frac{y}{z}\right)^{2\epsilon}-12\epsilon^2\left({\rm Li}_2\left(-\frac{y}{z}\right)+\zeta_2\right)+{\cal O}(\epsilon^3)\right]\,.
\end{split}
\end{align}
As mentioned above we can get the results for the momentum dependent part of all diagrams by using permutations.  We can write in general, for diagram $d=A..F$, 
\begin{align}
\begin{split}
{\cal F}_{(1,1,1,3); d}&=
\frac{\Gamma(6\epsilon)}{(2\epsilon)^2}
\kappa^3
\gamma_{14}\gamma_{24}\gamma_{34}
\int_0^{1}dx dy dz \,P(x,\gamma_{14})\, P(y,\gamma_{24})\, P(z,\gamma_{34}) \, f_d(x,y,z;\epsilon)
\end{split}
\end{align}
with 
\begin{subequations}
\begin{align}
\begin{split}
f_A(x,y,z;\epsilon)&= f_B(x,z,y;\epsilon)=
\frac12+\left(-\ln(x)+2 \ln(y)-\ln(z)\right)\,\epsilon\,+\,
\left(\ln(z)^2-4 \ln(y) \ln(z)+
\right.\\&\left.
4 \ln(y)^2+12 {\rm Li}_2(-y/z)+2 \ln(z) \ln(x)+\ln(x)^2-4 \ln(x) \ln(y)\right)\,\epsilon^2+\,{\cal O}(\epsilon^3)
\end{split}
\\
\begin{split}
f_B(x,y,z;\epsilon)&= f_B(x,z,y;\epsilon)=
\frac12+\left(-\ln(x)-\ln(y)+2 \ln(z)\right)\,\epsilon\,+\,
\left(\ln(y)^2-4 \ln(y) \ln(z)+
\right.\\&\left.
4\ln(z)^2-4 \ln(z) \ln(x)+\ln(x)^2+2 \ln(x) \ln(y)+12 {\rm Li}_2(-z/x)\right)\epsilon^2
\end{split}
\\
\begin{split}
f_C(x,y,z;\epsilon)&= f_B(y,x,z;\epsilon)=
\frac12+\left(-\ln(x)-\ln(y)+2 \ln(z)\right)\,\epsilon\,+
\,
\left(\ln(y)^2-4 \ln(y) \ln(z)+
\right.\\&\left.
4 \ln(z)^2-4 \ln(z) \ln(x)+\ln(x)^2+2 \ln(x) \ln(y)+12 {\rm Li}_2(-z/x)\right)\,\epsilon^2+\,{\cal O}(\epsilon^3)
\end{split}
\\
\begin{split}
f_D(x,y,z;\epsilon)&= f_A(y,x,z;\epsilon)=
\frac12+(2 \ln(x)-\ln(y)-\ln(z))\,\epsilon\,+\,
\left(\ln(z)^2-4 \ln(z) \ln(x)+
\right.\\&\left.
4 \ln(x)^2+12 {\rm Li}_2(-x/z)+2 \ln(y) \ln(z)+\ln(y)^2-4 \ln(x) \ln(y)\right)\,\epsilon^2+\,{\cal O}(\epsilon^3)
\end{split}
\\
\begin{split}
f_E(x,y,z;\epsilon)&= f_A(z,y,x;\epsilon)=
\frac12+(-\ln(x)+2 \ln(y)-\ln(z))\,\epsilon\,+\,
\left(\ln(x)^2-4 \ln(x) \ln(y)+
\right.\\&\left.
4 \ln(y)^2+12 {\rm Li}_2(-y/x)+2 \ln(z) \ln(x)+\ln(z)^2-4 \ln(y) \ln(z)\right)\,\epsilon^2+\,{\cal O}(\epsilon^3)
\end{split}
\\
\begin{split}
f_F(x,y,z;\epsilon)&= f_B(z,y,x;\epsilon)=
\frac12+(2 \ln(x)-\ln(y)-\ln(z))\,\epsilon\,+\,
\left(\ln(y)^2-4 \ln(x) \ln(y)+
\right.\\&\left.
4 \ln(x)^2+12 {\rm Li}_2( -x/y)+2 \ln(y) \ln(z)+\ln(z)^2-4 \ln(z) \ln(x)\right)\,\epsilon^2+\,{\cal O}(\epsilon^3)
\end{split}
\end{align}
\end{subequations}
Using now the results for the mixing matrix and connected colour factors of this web we get the final results quoted in eqs.~(\ref{ijklll}) and (\ref{phi3}).


\section{Subtracted webs at three loops\label{sec:combining}}

Using eq.~(\ref{Gamma_3}) we observe that for three-loop webs consisting of three gluon exchanges one gets a contribution to the three-loop soft anomalous dimension, of the form $-6\overline{w}^{(3,-1)}$ where 
$\overline{w}^{(3,-1)}$ is the 
\emph{subtracted web} given by eq.~(\ref{sub_web_3l_}).
In appendix \ref{sec:3_loop_calc} we derived the web kernels for the 1-2-2-1 and 1-1-1-3 webs, and we are now ready to combine these integrands with those corresponding to the commutator terms. All the necessary input is summarised in appendix~\ref{sec:summary_of_integrals}.
Let us consider the two webs in turn.

\subsection{Combining the 1-2-2-1 web with the corresponding commutators}

Using eq.~(\ref{sub_web_3l_}) for the 1-2-2-1 web, and selecting the commutators of lower-order webs which correspond to its subdiagrams, we obtain: 
\begin{align}
\label{w3_122331}
\begin{split}
&\overline{w}^{(3,-1)}_{(122334)} \!=\!
w^{(3,-1)}(122334)
-\frac16\bigg\{3\left[w^{(1,0)}(12),w^{(2,-1)}(2334)\right]+3\left[w^{(1,0)}(34),w^{(2,-1)}(1223)\right]
\\&
+3\left[w^{(2,0)}(2334),w^{(1,-1)}(12)\right]+3\left[w^{(2,0)}(1223),w^{(1,-1)}(34)\right]
\\& 
+\left[w^{(1,0)}(12),\left[w^{(1,-1)}(23),w^{(1,0)}(34)\right]\right]
+\left[w^{(1,0)}(12),\left[w^{(1,-1)}(34),w^{(1,0)}(23)\right]\right]
\\& 
+\left[w^{(1,0)}(34),\left[w^{(1,-1)}(23),w^{(1,0)}(12)\right]\right]
+\left[w^{(1,0)}(34),\left[w^{(1,-1)}(12),w^{(1,0)}(23)\right]\right]
\\&
+\left[w^{(1,-1)}(12),\left[w^{(1,1)}(23),w^{(1,-1)}(34)\right]\right]
+\left[w^{(1,-1)}(12),\left[w^{(1,1)}(34),w^{(1,-1)}(23)\right]\right]
\\&
+\left[w^{(1,-1)}(34),\left[w^{(1,1)}(23),w^{(1,-1)}(12)\right]\right]
+\left[w^{(1,-1)}(34),\left[w^{(1,1)}(12),w^{(1,-1)}(23)\right]\right]\bigg\}\,.
\end{split}
\end{align}
Upon substituting the webs in terms of colour times kinematic factors we get:
\begin{align}
\begin{split}
&\overline{w}^{(3,-1)}_{(122334)} \alpha_s^3 =\frac16 f^{abe}f^{cde}T_1^aT_2^bT_3^cT_4^d
\,
\bigg[6 {\cal F}_{122334}^{(3,-1)}(\alpha_{12},\alpha_{23},\alpha_{34})
\\&
+3{\cal F}_{12}^{(1,0)}(\alpha_{12})\,{\cal F}_{2334}^{(2,-1)}(\alpha_{23},\alpha_{34})
-3{\cal F}_{34}^{(1,0)}(\alpha_{34})\,{\cal F}_{1223}^{(2,-1)}(\alpha_{12},\alpha_{23})
\\&-3{\cal F}_{12}^{(1,-1)}(\alpha_{12})\,{\cal F}_{2334}^{(2,0)}(\alpha_{23},\alpha_{34})
+3{\cal F}_{34}^{(1,-1)}(\alpha_{34})\,{\cal F}_{1223}^{(2,0)}(\alpha_{12},\alpha_{23})
\\&+{\cal F}_{12}^{(1,0)}(\alpha_{12})\left({\cal F}_{23}^{(1,-1)}(\alpha_{23}){\cal F}_{34}^{(1,0)}(\alpha_{34})- {\cal F}_{34}^{(1,-1)}(\alpha_{34}){\cal F}_{23}^{(1,0)}(\alpha_{23})\right)
\\&+{\cal F}_{34}^{(1,0)}(\alpha_{34})\left({\cal F}_{23}^{(1,-1)}(\alpha_{23}){\cal F}_{12}^{(1,0)}(\alpha_{12})- {\cal F}_{12}^{(1,-1)}(\alpha_{12}){\cal F}_{23}^{(1,0)}(\alpha_{23})\right)
\\&+{\cal F}_{12}^{(1,-1)}(\alpha_{12})\left({\cal F}_{23}^{(1,1)}(\alpha_{23}){\cal F}_{34}^{(1,-1)}(\alpha_{34})- {\cal F}_{34}^{(1,1)}(\alpha_{34}){\cal F}_{23}^{(1,-1)}(\alpha_{23})\right)
\\&+{\cal F}_{34}^{(1,-1)}(\alpha_{34})\left({\cal F}_{23}^{(1,1)}(\alpha_{23}){\cal F}_{12}^{(1,-1)}(\alpha_{12})- {\cal F}_{12}^{(1,1)}(\alpha_{12}){\cal F}_{23}^{(1,-1)}(\alpha_{23})\right)
\bigg]\,.
\end{split}
\end{align}
Next, substituting the integrals for the kinematic functions ${\cal F}$ summarized in appendix \ref{sec:summary_of_integrals} we get:
\begin{align}
\begin{split}
\overline{w}^{(3,-1)}_{(122334)} \alpha_s^3 &=\frac16 f^{abe}f^{cde}T_1^aT_2^bT_3^cT_4^d\,\,
\left(\frac{\kappa}{2}\right)^3\,
\int_0^1 {\rm 
      d}x{\rm d} y {\rm d}z\, p_0(x,\alpha_{ij}) p_0(y,\alpha_{jk}) p_0(z,\alpha_{kl})\\
&
\Bigg[
\psi_3^{(1)}(x,y,z)+
\psi_3^{(0)}(x,y,z)\Big(\ln q(x,\alpha_{12})+ \ln q(y,\alpha_{23}) + \ln q(z,\alpha_{34})\Big)
\\&+ 
3\left(\phi_2^{(1)}(x,y)-\phi_2^{(1)}(y,z)\right)
+3\left(\phi_2^{(0)}(x,y)-\phi_2^{(0)}(y,z)\right) \ln q(y,\alpha_{23})
\\&+3\left(\ln q(x,\alpha_{12}) -\ln q(z,\alpha_{34})\right)\left(\phi_2^{(0)}(x,y)+\phi_2^{(0)}(y,z)\right)
\\&
+2\ln q(x,\alpha_{12}) \ln q(z,\alpha_{34}) 
-\ln q(y,\alpha_{23})  \ln q(x,\alpha_{12})
-\ln q(y,\alpha_{23})  \ln q(z,\alpha_{34})
\\&
-\frac12\ln^2 q(x,\alpha_{12})
+\ln^2 q(y,\alpha_{23}) 
-\frac12\ln^2 q(z,\alpha_{34})\Bigg]  \,.
\end{split}
\end{align}
Using the expressions for the kernels $\psi_3$ and $\phi_2$ we get after some algebra:
\begin{align}
\label{122334_factorization}
\begin{split}
\overline{w}^{(3,-1)}_{(122334)} \alpha_s^3&=\frac16 f^{abe}f^{cde}T_1^aT_2^bT_3^cT_4^d\,\,
\left(\frac{\kappa}{2}\right)^3\,
\int_0^1 {\rm 
      d}x{\rm d} y {\rm d}z\, p_0(x,\alpha_{ij}) p_0(y,\alpha_{jk}) p_0(z,\alpha_{kl})\\
&
\Bigg[
\left(-2\ln^2\frac{y}{x}-2\ln^2\frac{1-y}{z}+8\ln\frac{y}{x}\ln\frac{1-y}{z}\right)
+2\left(\ln\frac{x}{y}+\ln\frac{z}{y} \right) \ln q(y,\alpha_{23})
\\&+2\left(\ln\frac{x}{z}+\ln\frac{y}{z} \right) \ln q(x,\alpha_{12})\,+\,2\left(\ln\frac{z}{x}+\ln\frac{y}{x} \right) \ln q(z,\alpha_{34})
\\&
+2\ln q(x,\alpha_{12}) \ln q(z,\alpha_{34}) 
-\ln q(y,\alpha_{23})  \ln q(x,\alpha_{12})
-\ln q(y,\alpha_{23})  \ln q(z,\alpha_{34})
\\&
-\frac12\ln^2 q(x,\alpha_{12})
+\ln^2 q(y,\alpha_{23}) 
-\frac12\ln^2 q(z,\alpha_{34})\Bigg]  \,.
\end{split}
\end{align}
Here we note that all dilogarithms cancel in the difference between the three loop webs and the commutators involving lower-order webs. The consequence is that the integrand of eq.~(\ref{122334_factorization}) contains only logarithms, so in each term in the sum the three integrals over $x$, $y$ and $z$ factorize, and can be performed independently of each other. 
Finally, substituting the results for the integrals according to the definitions in table~\ref{tab:4leg3loop_transc_func_symb} and replacing $\frac{\kappa}{2}$ by its $\epsilon=0$ limit of $-\frac{g_s^2}{16\pi^2}$, we get:
\begin{align}
\begin{split}
\overline{w}^{(3,-1)}_{(122334)} \alpha_s^3&=-\frac16 f^{abe}f^{cde}T_1^aT_2^bT_3^cT_4^d\,\,
\left(\frac{g_s^2}{16\pi^2}\right)^3\,
G(\alpha_{12},\alpha_{23}, \alpha_{34})\,.
\end{split}
\end{align}
where
\begin{align}
\label{G1221}
\begin{split}
G(\alpha_{12},\alpha_{23}, \alpha_{34})&=\frac{1+\alpha_{12}^2}{1-\alpha_{12}^2}\,\frac{1+\alpha_{23}^2}{1-\alpha_{23}^2}\,\frac{1+\alpha_{34}^2}{1-\alpha_{34}^2}\, \Bigg[16 \Big(S_2(\alpha_{23})-2 \widetilde{S}_2(\alpha_{23})\Big)\,\ln \alpha_{12}\, \ln \alpha_{34}
\\&-2S_1(\alpha_{23}) \Big(S_1(\alpha_{12}) \ln \alpha_{34} + S_1(\alpha_{34}) \ln \alpha_{12} \Big)
\\
&+ 4\ln \alpha_{23}\Big(2S_2(\alpha_{12}) \ln \alpha_{34} +2S_2(\alpha_{34}) \ln \alpha_{12} 
+S_1(\alpha_{12})S_1(\alpha_{34})
\Big)
\\
&-4 \Bigg(R_1(\alpha_{23})
\Big(S_1(\alpha_{12})\ln \alpha_{34}+S_1(\alpha_{34})\ln \alpha_{12} \Big)
+4V_2(\alpha_{23}) \ln \alpha_{12} \ln \alpha_{34}\Bigg)
\\&+4\Bigg(2V_2(\alpha_{12}) \ln \alpha_{23} \ln \alpha_{34}
-R_1(\alpha_{12})\Big(S_1(\alpha_{23}) \ln \alpha_{34} -2 S_1(\alpha_{34})\ln \alpha_{23}\Big)\Bigg)
\\&+4\Bigg(2V_2(\alpha_{34}) \ln \alpha_{23} \ln \alpha_{12}
-R_1(\alpha_{34})\Big(S_1(\alpha_{23}) \ln \alpha_{12} -2 S_1(\alpha_{12})\ln \alpha_{23}\Big)\Bigg)
\\&
+8\ln \alpha_{23}\,\Big(2R_1(\alpha_{12})R_1(\alpha_{34}) -R_2(\alpha_{12}) \ln \alpha_{34} -R_2(\alpha_{34}) \ln \alpha_{12}  
\Big)
\\&
-8R_1(\alpha_{23}) \Big(R_1(\alpha_{12})\ln \alpha_{34} +R_1(\alpha_{34})\ln \alpha_{12}\Big)
+16 R_2(\alpha_{23}) \, \ln \alpha_{12}\, \ln \alpha_{34}\Bigg]
\end{split}
\end{align}
where $G$ has the symmetry property: $G(\alpha_{12},\alpha_{23}, \alpha_{34})= G(\alpha_{34},\alpha_{23}, \alpha_{12})$.
 We see that owing to the factorization property of the subtracted webs the result is written as a sum of factorized terms, each of which is a function of a single cusp angle. There are no multiple polylogarithms that depend on more than one cusp angle. As explained in the main text this is expected to be a general feature of multiple-gluon-exchange webs.

Eq.~(\ref{G1221}) is written in terms of the functions $R_i$, $S_i$ etcetera, defined in table~\ref{tab:4leg3loop_transc_func_symb}. As discussed in section~\ref{sec:functions}, each of these functions separately violates the alphabet conjecture, and in particular does not
display the crossing symmetry $\alpha\to -\alpha$. Yet the symmetry must be present in eq.~(\ref{G1221}), which is a subtracted web.
It can be made manifest by selecting the basis of functions of eq.~(\ref{symfunc}). Indeed, the result for $G$ can be written in this basis -- it is given in eq.~(\ref{G1221_sym}), which is much more compact than eq.~(\ref{G1221}). Besides demonstrating the conclusions of section~\ref{sec:functions}, this last step provides a powerful consistency check of the calculation.

\subsection{Combining the 1-1-1-3 web with the corresponding commutators}

Let us now consider the 1-1-1-3 web. Using eq.~(\ref{sub_web_3l_}) and selecting the relevant commutators of lower-order webs we obtain: 
\begin{eqnarray}
\label{sub_web_1113}
&& \nonumber
\overline{w}^{(3,-1)}_{(123444)} =
w^{(3,-1)}(123444)-\frac16\times
\\
&&\bigg\{3\left[w^{(1,0)}(14),w^{(2,-1)}(2344)\right]+3\left[w^{(1,0)}(24),w^{(2,-1)}(1344)\right]+3\left[w^{(1,0)}(34),w^{(2,-1)}(1244)\right]
\nonumber \\
&&+3\left[w^{(2,0)}(2344),w^{(1,-1)}(14)\right]+3\left[w^{(2,0)}(1344),w^{(1,-1)}(24)\right]+3\left[w^{(2,0)}(1244),w^{(1,-1)}(34)\right]
\nonumber\\&&
+\left[w^{(1,0)}(34),\left[w^{(1,-1)}(14),w^{(1,0)}(24)\right]\right]
+\left[w^{(1,0)}(34),\left[w^{(1,-1)}(24),w^{(1,0)}(14)\right]\right]
\nonumber\\&&
+\left[w^{(1,0)}(24),\left[w^{(1,-1)}(14),w^{(1,0)}(34)\right]\right]
+\left[w^{(1,0)}(24),\left[w^{(1,-1)}(34),w^{(1,0)}(14)\right]\right]
\nonumber \\&&
+\left[w^{(1,0)}(14),\left[w^{(1,-1)}(34),w^{(1,0)}(24)\right]\right]
+\left[w^{(1,0)}(14),\left[w^{(1,-1)}(24),w^{(1,0)}(34)\right]\right]
\nonumber\\&&
+\left[w^{(1,-1)}(34),\left[w^{(1,1)}(14),w^{(1,-1)}(24)\right]\right]
+\left[w^{(1,-1)}(34),\left[w^{(1,1)}(24),w^{(1,-1)}(14)\right]\right]
\nonumber\\&&
+\left[w^{(1,-1)}(24),\left[w^{(1,1)}(14),w^{(1,-1)}(34)\right]\right]
+\left[w^{(1,-1)}(24),\left[w^{(1,1)}(34),w^{(1,-1)}(14)\right]\right]
\nonumber\\&&
+\left[w^{(1,-1)}(14),\left[w^{(1,1)}(34),w^{(1,-1)}(24)\right]\right]
+\left[w^{(1,-1)}(14),\left[w^{(1,1)}(24),w^{(1,-1)}(34)\right]\right]\bigg\}\,.
\end{eqnarray}
Upon substituting the webs in terms of colour time kinematic factors we get:
\begin{align}
\begin{split}
&\overline{w}^{(3,-1)}_{(123444)} \alpha_s^3 =
\frac16 f^{ade}f^{bce}T_1^aT_2^bT_3^cT_4^d
\Bigg[
6{\cal F}_{123444; 1}^{\,(3,-1)}(\alpha_{14},\alpha_{24},\alpha_{34})
\\&-3{\cal F}^{(1,0)}_{14}(\alpha_{14})\,{\cal F}_{2344}^{(2,-1)}(\alpha_{24},\alpha_{34})
+3{\cal F}^{(1,-1)}_{14}(\alpha_{14})\,{\cal F}_{2344}^{(2,0)}(\alpha_{24},\alpha_{34})  
\\&
+{\cal F}_{14}^{(1,0)}(\alpha_{14}) \left({\cal F}_{34}^{(1,-1)}(\alpha_{34})
{\cal F}_{24}^{(1,0)}(\alpha_{24})-{\cal F}_{24}^{(1,-1)}(\alpha_{24})
{\cal F}_{34}^{(1,0)}(\alpha_{34})\right)
\\&
+{\cal F}_{14}^{(1,-1)}(\alpha_{14}) \left({\cal F}_{34}^{(1,1)}(\alpha_{34})
{\cal F}_{24}^{(1,-1)}(\alpha_{24})-{\cal F}_{24}^{(1,1)}(\alpha_{24})
{\cal F}_{34}^{(1,-1)}(\alpha_{34})\right)
\Bigg]
\\&+\frac16 f^{ace}f^{bde}T_1^aT_2^bT_3^cT_4^d
\Bigg[
6{\cal F}_{123444; 2}^{\,(3,-1)}(\alpha_{14},\alpha_{24},\alpha_{34})
\\&
-3{\cal F}^{(1,0)}_{24}(\alpha_{24})\,{\cal F}_{1344}^{(2,-1)}(\alpha_{14},\alpha_{34})
+3{\cal F}^{(1,-1)}_{24}(\alpha_{24})\,{\cal F}_{1344}^{(2,0)}(\alpha_{14},\alpha_{34})
\\&
-{\cal F}_{24}^{(1,0)}(\alpha_{24}) \left({\cal F}_{14}^{(1,-1)}(\alpha_{14})
{\cal F}_{34}^{(1,0)}(\alpha_{34})-{\cal F}_{34}^{(1,-1)}(\alpha_{34})
{\cal F}_{14}^{(1,0)}(\alpha_{14})\right)
\\&
-{\cal F}_{24}^{(1,-1)}(\alpha_{24}) \left({\cal F}_{14}^{(1,1)}(\alpha_{14})
{\cal F}_{34}^{(1,-1)}(\alpha_{34})-{\cal F}_{34}^{(1,1)}(\alpha_{34})
{\cal F}_{14}^{(1,-1)}(\alpha_{14})\right)
\Bigg]
\\&+\frac16 f^{abe}f^{cde}T_1^aT_2^bT_3^cT_4^d
\Bigg[
-3{\cal F}^{(1,0)}_{34}(\alpha_{34})\,{\cal F}_{1244}^{(2,-1)}(\alpha_{14},\alpha_{24})
+3{\cal F}^{(1,-1)}_{34}(\alpha_{34})\,{\cal F}_{1244}^{(2,0)}(\alpha_{14},\alpha_{24})
\\&-{\cal F}_{34}^{(1,0)}(\alpha_{34}) \left({\cal F}_{14}^{(1,-1)}(\alpha_{14})
{\cal F}_{24}^{(1,0)}(\alpha_{24})-{\cal F}_{24}^{(1,-1)}(\alpha_{24})
{\cal F}_{14}^{(1,0)}(\alpha_{14})\right)
\\&-{\cal F}_{34}^{(1,-1)}(\alpha_{34}) \left({\cal F}_{14}^{(1,1)}(\alpha_{14})
{\cal F}_{24}^{(1,-1)}(\alpha_{24})-{\cal F}_{24}^{(1,1)}(\alpha_{24})
{\cal F}_{14}^{(1,-1)}(\alpha_{14})\right)
\Bigg]
\end{split}
\end{align}
where by convention ${\cal F}_{ijkk}={\cal F}_{ikkj}$.
At this point we use the fact that the three colour factors are related via the Jacobi identity:
\begin{equation}
\label{Jacobi}
f^{abe}f^{cde}T_1^aT_2^bT_3^cT_4^d= f^{ace}f^{bde}T_1^aT_2^bT_3^cT_4^d -f^{ade}f^{bce}T_1^aT_2^bT_3^cT_4^d
\end{equation}
to write the result in terms of two independent colour factors.
Substituting the functions ${\cal F}$ as summarized in appendix \ref{sec:summary_of_integrals} and inserting the explicit expressions $\phi_{3,i}$ and $\phi_2$ we get:
\begin{align*}
\begin{split}
&\overline{w}^{(3,-1)}_{(123444)} \alpha_s^3=\frac16 
\left(\frac{\kappa}{2}\right)^3\,\int_0^1 {\rm 
      d}x{\rm d} y {\rm d}z\, p_0(x,\alpha_{14}) p_0(y,\alpha_{24}) p_0(z,\alpha_{34})\,\times\\
&
\bigg\{
f^{ade}f^{bce}T_1^aT_2^bT_3^cT_4^d
\bigg[
2\ln^2 x\,-\,4\ln^2y\, +2\ln^2z +4\ln x\,\ln y \,+\,4\ln y\,\ln z -8\ln x\,\ln z
\\&
+2\ln q(x,\alpha_{14}) \, \Big(\ln\frac{z}{x}+\ln\frac{z}{y}\Big)
+2\ln q(y,\alpha_{24}) \, \Big(\ln\frac{y}{x}+\ln\frac{y}{z}\Big)
+2\ln q(z,\alpha_{34}) \, \Big(\ln\frac{x}{y}+\ln\frac{x}{z}\Big)
\\&
+\, \ln q(x,\alpha_{14}) \ln q(y,\alpha_{24})
-\,2 \ln q(x,\alpha_{14}) \ln q(z,\alpha_{34})
+\,\ln q(z,\alpha_{34})\,\ln q(y,\alpha_{24})\\&
+\frac12 \,\ln^2 q(z,\alpha_{34})- \,\ln^2 q(y,\alpha_{24})
+\frac12 \,\ln^2 q(x,\alpha_{14})
\bigg]
\\&+f^{ace}f^{bde}T_1^aT_2^bT_3^cT_4^d
\bigg[
-4\ln^2 x\,+\,2\ln^2y\, +2\ln^2z +4\ln x\,\ln y \,-\,8\ln y\,\ln z +4\ln x\,\ln z
\\&
+2\ln q(x,\alpha_{14}) \, \Big(\ln\frac{x}{y}+\ln\frac{x}{z}\Big)
+2\ln q(y,\alpha_{24}) \, \Big(\ln\frac{z}{y}+\ln\frac{z}{x}\Big)
+2\ln q(z,\alpha_{34}) \, \Big(\ln\frac{y}{x}+\ln\frac{y}{z}\Big)
\\&
+\, \ln q(y,\alpha_{24}) \ln q(x,\alpha_{14})
-\,2 \ln q(y,\alpha_{24}) \ln q(z,\alpha_{34})
+\,\ln q(z,\alpha_{34})\,\ln q(x,\alpha_{14})
\\&
+\frac12 \,\ln^2 q(z,\alpha_{34})-\,\ln^2 q(x,\alpha_{14})
+\frac12 \,\ln^2 q(y,\alpha_{24})
\bigg]
\bigg\}
\end{split}
\end{align*}
where we observe again the complete cancellation of the dilogrithms, which results in the integrals being factorisable.  Performing the three integrals, the final result reads:
\begin{align*}
\begin{split}
\overline{w}^{(3,-1)}_{(123444)} \alpha_s^3=-\frac16 
\left(\frac{g^2}{16\pi^2}\right)^3 T_1^aT_2^bT_3^cT_4^d
\bigg[f^{ade}f^{bce} F(\alpha_{14},\alpha_{24},\alpha_{34})+f^{ace}f^{bde}
 \, F(\alpha_{24},\alpha_{14},\alpha_{34})
\bigg]
\end{split}
\end{align*}
\begin{eqnarray}
\text{with} \qquad& &F(\alpha_{14},\alpha_{24},\alpha_{34})=
\frac{1+\alpha_{14}^2}{1-\alpha_{14}^2}\,\frac{1+\alpha_{24}^2}{1-\alpha_{24}^2}\,\frac{1+\alpha_{34}^2}{1-\alpha_{34}^2}
\,\,\times \\&&\nonumber\bigg[
-8 S_2(\alpha_{14}) \,\ln \alpha_{24}\, \ln \alpha_{34}
+16 S_2(\alpha_{24}) \,\ln \alpha_{14}\, \ln \alpha_{34}
-8 S_2(\alpha_{34}) \,\ln \alpha_{14}\, \ln \alpha_{24}
\\&&\nonumber
+2 S_1(\alpha_{14})\,S_1(\alpha_{24})\, \ln \alpha_{34} 
+2 S_1(\alpha_{24})\,S_1(\alpha_{34})\, \ln \alpha_{14}
-4 S_1(\alpha_{14})\,S_1(\alpha_{34})\, \ln \alpha_{24}
\\&&\nonumber
-8 V_2(\alpha_{14}) \ln \alpha_{24} \ln \alpha_{34}
+ 4 R_1(\alpha_{14})\Big(S_1(\alpha_{24}) \ln \alpha_{34} -2 S_1(\alpha_{34}) \ln \alpha_{24}\Big)
\\&&\nonumber
+16 V_2(\alpha_{24}) \ln \alpha_{14} \ln \alpha_{34} 
+4 R_1(\alpha_{24}) \Big(S_1(\alpha_{14}) \ln \alpha_{34} + S_1(\alpha_{34}) \ln \alpha_{14}
\Big) 
\\&&\nonumber
-8 V_2(\alpha_{34}) \ln \alpha_{24} \ln \alpha_{14}
+ 4 R_1(\alpha_{34})\Big(S_1(\alpha_{24}) \ln \alpha_{14} -2 S_1(\alpha_{14}) \ln \alpha_{24}\Big)
\\&&\nonumber
+8 R_1(\alpha_{14}) \, R_1(\alpha_{24}) \, \ln \alpha_{34}
+8 R_1(\alpha_{24}) \, R_1(\alpha_{34}) \, \ln \alpha_{14} 
-16 R_1(\alpha_{14}) \, R_1(\alpha_{34}) \, \ln \alpha_{24}
\\&&\nonumber +
8 R_2(\alpha_{34}) \, \ln \alpha_{14}\, \ln \alpha_{24} 
-16 R_2(\alpha_{24}) \, \ln \alpha_{14}\, \ln \alpha_{34} 
+8 R_2(\alpha_{14}) \, \ln \alpha_{24}\, \ln \alpha_{34} \bigg]\,.
\end{eqnarray}
 Proceeding as in the case of the 1-2-2-1 web to write the function $F$ above using the basis functions of eq.~(\ref{symfunc}) which admit the alphabet conjecture, we get the very elegant result of eq. (\ref{F1113_sym}). This step provides a powerful consistency check of the entire calculation.

\bibliographystyle{JHEP}
\bibliography{refs3}

\end{document}